\documentclass[12pt]{article}
\usepackage[british]{babel}
\def\AnswerYes{y}
\def\pdflatex{y}                  
\def\ShowLabelsVersion{n}         
\def\ShowChangesVersion{n}        
\def\ShowAnnotationsVersion{n}    
\def\ShowFigures{y}               
\def\feynVersion{n}               
\def\MakeArXivLinksActive{y}      
\ifx\ShowFigures\AnswerYes
   \usepackage{graphicx}          
\else
   \usepackage[draft]{graphicx}   
   \renewcommand{\includegraphics}[2][]{\fbox{#2}}
\fi

\graphicspath{{figures/}{./}}

\ifx\pdflatex\AnswerYes
   \usepackage[update,prepend]{epstopdf} 
   \usepackage{grffile} 
\fi
\ifx\pdflatex\AnswerYes
    \usepackage{color}
\else
    \usepackage[dvips]{color}
\fi
\ifx\pdflatex\AnswerYes
    \usepackage{thumbpdf}    %
\else
    \usepackage[dvips]{thumbpdf}
\fi

\definecolor{green}{rgb}{0.,0.7,0.}
\definecolor{blue}{rgb}{0.,0.,1.0}
\definecolor{red}{rgb}{1,0,0}

\ifx\pdflatex\AnswerYes
    \usepackage[ 
        final,
        breaklinks=true,
        colorlinks=true,           
        pdfborder={0 0 1},    
        citecolor={blue},linkcolor={blue},urlcolor={red},
        citebordercolor={1 0 0},linkbordercolor={0 0 1},urlbordercolor={1 0 0},
        pdfpagemode=UseOutlines,
        bookmarks=true,bookmarksopenlevel=4
        ]{hyperref}         
\else
    \usepackage[dvips,              
        final,
        breaklinks=true,
        colorlinks=true,      
        pdfborder={0 0 1},    
        citecolor={blue},linkcolor={blue},urlcolor={red},
        citebordercolor={1 0 0},linkbordercolor={0 0 1},urlbordercolor={1 0 0},
        pdfpagemode=UseOutlines,
        bookmarks=true,bookmarksopenlevel=4
         ]{hyperref}         
\fi
\usepackage[numbers,sort&compress]{natbib}

\usepackage{multirow}                   
\usepackage{amssymb}
\usepackage{amsmath}
\usepackage{bbm} 
\usepackage{xspace}                    
\usepackage{dcolumn}                    
\usepackage{bm}                        
\usepackage{cancel}	                  

\usepackage{rotating}
\usepackage{floatpag} 

\usepackage{slashed}
\ifx\feynVersion\AnswerYes
   \usepackage{feynmp} 
\fi

\usepackage[para]{footmisc} 

\usepackage{pbox}

\usepackage{epsfig}

\usepackage[textsize=scriptsize]{todonotes}

\usepackage{chngcntr} 
\counterwithin*{equation}{section} 

\ifx\ShowLabelsVersion\AnswerYes
   \usepackage[color]{showkeys} 
   \definecolor{refkey}{gray}{.5}   
   \definecolor{labelkey}{gray}{.5} 
   
\fi

\ifx\ShowAnnotationsVersion\AnswerYes
   \newcommand{\comment}[1]{{\scriptsize\sffamily\bfseries{#1}}}
   \newcommand{\margin}[1]{\marginpar{\scriptsize\sffamily\bfseries{#1}}}
   \newcommand{\drafty}{\textbf{Draft version \today} \hfill}
   \usepackage{lineno}
   \linenumbers
\else
   \newcommand{\comment}[1]{}
   \newcommand{\margin}[1]{}
   \newcommand{\drafty}{}
\fi

\ifx\ShowChangesVersion\AnswerYes
   \usepackage{ulem}
    
   \newcommand{\delete}[1]{\sout{#1}}            
   \renewcommand{\emph}[1]{\textit{#1}}           
\else

   \newcommand{\sout}[1]{}
   \newcommand{\xout}[1]{}

   \newcommand{\delete}[1]{}
\fi

\ifx\MakeArXivLinksActive\AnswerYes
\usepackage{xparse}
\NewDocumentCommand{\arxiv} {r [: u{ [} u{]]} }{[\href{http://arxiv.org/abs/#2}{arXiv:#2}~[#3]]}
\NewDocumentCommand{\arxivold} {r[]}{[\href{http://arxiv.org/abs/#1}{#1}]}
\NewDocumentCommand{\arXiv} {r [: u{ [} u{]]} }{[\href{http://arxiv.org/abs/#2}{arXiv:#2}~[#3]]}
\NewDocumentCommand{\arXivold} {r[]}{[\href{http://arxiv.org/abs/#1}{#1}]}
\else
\newcommand{\arXiv}[1][]{[#1]}
\newcommand{\arXivold}[1][]{[#1]}
\newcommand{\arxiv}[1][]{[#1]}
\newcommand{\arxivold}[1][]{[#1]}
\fi
\newcommand{\disc}{\discretionary{}{}{}}

\definecolor{green}{rgb}{0.,0.7,0.}
\definecolor{blue}{rgb}{0.,0.,1.0}
\definecolor{red}{rgb}{1,0,0}

\newcommand{\eg}{\textit{e.g.}\xspace}
\newcommand{\ie}{\textit{i.e.}\xspace}
\newcommand{\etal}{\textit{et al.}\xspace}

\newcommand{\solid}{\rule[0.5ex]{5ex}{1pt}}

\newcommand{\longdashed}{\rule[0.5ex]{1ex}{1pt}\hspace*{1ex}\rule[0.5ex]{1ex}{1pt}\hspace*{1ex}\rule[0.5ex]{1ex}{1pt}}

\newcommand{\dotted}{$\cdot$\hspace*{0.2ex}$\cdot$\hspace*{0.2ex}$\cdot$\hspace*{0.2ex}$\cdot$\hspace*{0.2ex}$\cdot$\hspace*{0.2ex}$\cdot$\hspace*{0.2ex}}

\newcommand{\dis}{\displaystyle}

\newcommand{\non}{\nonumber}

\newcommand{\hq}{\hspace{0.5em}}
\newcommand{\hqq}{\hspace{1em}}
\newcommand{\hqqq}{\hspace{2em}}

\newcommand{\hqmm}{\hspace*{-0.5em}}

\newcommand{\half}{\frac{1}{2}}
\newcommand{\e}{\mathrm{e}}
\newcommand{\ii}{\mathrm{i}}
\newcommand{\dd}{\mathrm{d}}
\newcommand{\tr}{\mathrm{tr}}
\newcommand{\T}{\mathrm{T}}

\newcommand{\ev}{\vec{e}}
\newcommand{\kv}{\vec{k}}
\newcommand{\pv}{\vec{p}}
\newcommand{\qv}{\vec{q}}

\def\lsim{\mathrel{\rlap{\lower4pt\hbox{\hskip1pt$\sim$}}
    \raise1pt\hbox{$<$}}}         
\def\gsim{\mathrel {\rlap{\lower4pt\hbox{\hskip1pt$\sim$}}
    \raise1pt\hbox{$>$}}}         

\newcommand{\bra}{\langle}
\newcommand{\ket}{\rangle}

\newcommand{\mpi}{\ensuremath{m_\pi}}

\newcommand{\MeV}{\ensuremath{\mathrm{MeV}}}

\newcommand{\ChiEFT}{$\chi$EFT\xspace}

\newcommand{\NXLO}[1]{N\ensuremath{{}^{#1}}LO\xspace}

\newcommand{\HIGS}{HI$\gamma$S\xspace}
\newcommand{\threeHe}{\ensuremath{{}^3\text{He}}\xspace}

\newcommand{\alphae}{\ensuremath{\alpha_{E1}}}
\newcommand{\betam}{\ensuremath{\beta_{M1}}}
\newcommand{\gammaee}{\ensuremath{\gamma_{E1E1}}}
\newcommand{\gammamm}{\ensuremath{\gamma_{M1M1}}}
\newcommand{\gammaem}{\ensuremath{\gamma_{E1M2}}}
\newcommand{\gammame}{\ensuremath{\gamma_{M1E2}}}

\newcommand{\alphaep}{\ensuremath{\alpha_{E1}^{(\mathrm{p})}}}
\newcommand{\betamp}{\ensuremath{\beta_{M1}^{(\mathrm{p})}}}

\newcommand{\alphaen}{\ensuremath{\alpha_{E1}^{(\mathrm{n})}}}
\newcommand{\betamn}{\ensuremath{\beta_{M1}^{(\mathrm{n})}}}
\newcommand{\gammaeen}{\ensuremath{\gamma_{E1E1}^{(\mathrm{n})}}}
\newcommand{\gammammn}{\ensuremath{\gamma_{M1M1}^{(\mathrm{n})}}}
\newcommand{\gammaemn}{\ensuremath{\gamma_{E1M2}^{(\mathrm{n})}}}
\newcommand{\gammamen}{\ensuremath{\gamma_{M1E2}^{(\mathrm{n})}}}

\newcommand{\philin}{\ensuremath{\varphi_\text{lin}}}
\newcommand{\thetan}{\ensuremath{\vartheta_{\vec{n}}}}
\newcommand{\phin}{\ensuremath{\varphi_{\vec{n}}}}

\newcommand{\rhoS}{\ensuremath{\rho^{(S)}}}
\newcommand{\rhogamma}{\ensuremath{\rho^{(\gamma)}}}

\newcommand{\PthreeHe}{\ensuremath{P^{(\text{\threeHe})}}}

\newcommand{\threej}[6]
    {\ensuremath{\left(\begin{matrix}#1&#2&#3\\#4&#5&#6\end{matrix}\right)}}

\newcommand{\piN}{\pi\mathrm{N}}

\newcommand{\MN}{\ensuremath{M_\mathrm{N}}} 
\newcommand{\MHe}{\ensuremath{M_\text{\threeHe}}} 
\newcommand{\DeltaM}{\ensuremath{\Delta_{\scriptscriptstyle M}}} 

\newcommand{\omegalab}{\ensuremath{\omega_\mathrm{lab}}}
\newcommand{\omegaprimelab}{\ensuremath{\omega_\mathrm{lab}^\prime}}
\newcommand{\omegacm}{\ensuremath{\omega_\mathrm{cm}}}
\newcommand{\thetalab}{\ensuremath{\theta_\mathrm{lab}}}
\newcommand{\thetacm}{\ensuremath{\theta_\mathrm{cm}}}

\newcommand{\calA}{\mathcal{A}} 
 \newcommand{\calD}{\mathcal{D}}

\newcommand{\calO}{\mathcal{O}}

\newcommand{\mytitle}[1]{\begin{center}\LARGE{\textbf{#1}}\end{center}}
\newcommand{\myauthor}[1]{\textbf{#1}}
\newcommand{\myaddress}[1]{\textit{#1}}
\newcommand{\mypreprint}[1]{\begin{flushright}#1\end{flushright}}

\begin{document}
%

\begin{titlepage}
  \setcounter{page}{-1} \mypreprint{
    \drafty
    3 April 2018, revised 31 May 2018 \\
  }
  
\vspace*{-2ex}
  
\mytitle{Elastic Compton Scattering from \threeHe\\ and the Role of the Delta}


\begin{center}
  \myauthor{Arman Margaryan$^{a}$}\footnote{Email: am343@duke.edu}
  \myauthor{Bruno Strandberg$^{b,c}$}\footnote{Email:
    b.strandberg@nikhef.nl} \myauthor{Harald W.\
    Grie{\ss}hammer$^{d}$}\footnote{Email:
    hgrie@gwu.edu},  \\
  \myauthor{Judith A.~McGovern$^{e}$}\footnote{Email:
    judith.mcgovern@manchester.ac.uk} 
  \myauthor{Daniel R.~Phillips$^{f}$}\footnote{Email: phillid1@ohio.edu}
  \emph{and} \myauthor{Deepshikha Shukla$^{g}$}\footnote{Email:
    dshukla@rockford.edu}
  
  \vspace*{0.3cm} \myaddress{$^a$L/EFT Group, Department of Physics,\\
    Duke University, Box 90305, Durham, NC 27708, USA}\\[1ex]
  \myaddress{$^b$School of Physics and Astronomy, University of Glasgow,\\  
    Glasgow G12 8QQ, Scotland, UK}\\[1ex]
  \myaddress{$^c$Nikhef, Science Park 105, 1098 XG Amsterdam,
    Netherlands}\\[1ex]
  \myaddress{$^d$Institute for Nuclear Studies, Department of Physics, \\
    The George Washington University, Washington DC 20052, USA}\\[1ex]
  \myaddress{$^e$School of Physics and Astronomy, The University of
    Manchester,\\ Manchester M13 9PL, UK}\\[1ex]
  \myaddress{$^f$Department of Physics and Astronomy and Institute of Nuclear
    and Particle Physics, Ohio University, Athens, OH 45701, USA}\\[1ex] 
  \myaddress{$^g$Department of Mathematics, Computer Science and Physics,\\ Rockford University, Rockford, IL
    61108, USA}
\end{center}

\vspace*{-2ex}

\begin{abstract}
  \noindent
  We report observables for elastic Compton scattering from \threeHe in Chiral
  Effective Field Theory with an explicit $\Delta(1232)$ degree of freedom
  ($\chi$EFT) for energies between $50$ and $120\;\MeV$. The
  $\gamma\,$\threeHe amplitude is complete at \NXLO{3}, $\calO(\e^2\delta^3)$,
  and in general converges well order by order. It includes the dominant
  pion-loop and two-body currents, as well as the Delta excitation in
  the single-nucleon amplitude. Since the cross section is two to three times
  that for deuterium and the spin of polarised \threeHe is predominantly
  carried by its constituent neutron, elastic Compton scattering promises
  information on both the scalar and spin polarisabilities of the neutron. We
  study in detail the sensitivities of $4$ observables to the neutron
  polarisabilities: the cross section, the beam asymmetry and two double
  asymmetries resulting from circularly polarised photons and a longitudinally
  or transversely polarised target. Including the Delta enhances those
  asymmetries from which neutron spin polarisabilities could be extracted. We
  also correct previous, erroneous results at \NXLO{2}, \ie~without an
  explicit Delta, and compare to the same observables on proton, neutron and
  deuterium targets.  An interactive \emph{Mathematica} notebook of our
  results is available from \texttt{hgrie@gwu.edu}.

\end{abstract}
\noindent
\begin{tabular}{rl}
  Suggested Keywords: &\begin{minipage}[t]{10.7cm} Compton scattering, 
    Helium-3, Effective Field Theory, neutron polarisabilities, spin polarisabilities,
    $\Delta(1232)$ resonance
  \end{minipage}
\end{tabular}

\thispagestyle{empty}
\end{titlepage}
\setcounter{footnote}{0}
\newpage



\section{Introduction}
\label{sec:introduction}

Elastic Compton scattering from a Helium-3 target has been identified as a
promising means to access neutron electromagnetic polarisabilities. In
refs.~\cite{ShuklaPhD, Choudhury:2007bh, Shukla:2008zc, Shukla:2017}, Shukla
\etal showed that the differential cross section in the energy range of $50$
to $120\;\MeV$ is sensitive to the electric and magnetic dipole
polarisabilities of the neutron, $\alphaen$ and $\betamn$, and that scattering
on polarised \threeHe provides good sensitivity to the neutron spin
polarisabilities. These calculations were carried out at ${\cal O}(Q^3)$ in
the Chiral Effective Field Theory expansion and led to proposals at MAMI and
\HIGS to exploit this opportunity to extract neutron polarisabilities from
elastic $\gamma\,$\threeHe scattering~\cite{Weller:2013zta, Annand:2015,
  HIGSPAC, Ahmed:2016, Briscoe}.

Here, we extend the calculation of refs.~\cite{ShuklaPhD, Choudhury:2007bh,
  Shukla:2008zc} by one order in the chiral counting by incorporating the
leading effects of the $\Delta(1232)$. As discussed in an erratum published
simultaneously with this paper~\cite{Shukla:2017}, these first results were
obtained from a computer code which contained mistakes, and we take the
opportunity to correct some of the results here as well.

\ChiEFT is the low-energy Effective Field Theory of QCD; see, \eg,
refs.~\cite{Bernard:1995dp,Be08,Scherer:2012xha} for reviews of the mesonic
and one-nucleon sector, and refs.~\cite{Bedaque:2002mn,Epelbaum:2010nr,
  Machleidt:2016rvv} for summaries of the few-nucleon sector.  It respects the
spontaneously and dynamically broken
$\mathrm{SU}(2)_L \times \mathrm{SU}(2)_R$ symmetry of QCD and has nucleons
and pions as explicit degrees of freedom. In this work, we consider a variant
which also includes the
Delta~\cite{Jenkins:1991ne,Hemmert:1996xg,Hemmert:1997ye,Pascalutsa:2002pi}.
Since \ChiEFT provides a perturbative expansion of observables in a small,
dimension-less parameter, one can calculate observables to a given order,
which in turns provides a way to estimate the residual theoretical
uncertainties from the
truncation~\cite{Furnstahl:2015rha,Griesshammer:2015ahu}.

More details on the \ChiEFT expansion are given in sect.~\ref{sec:ChiEFT}; for
now, we summarise our calculation as follows. We employ both the $\gamma N$
and $\gamma NN$ amplitudes of
refs.~\cite{Bernard:1991rq,Bernard:1995dp,Beane:1999uq} and supplement those
with the $\Delta$-pole and $\pi \Delta$ loop graphs~\cite{Hildebrandt:2003fm,
  Hildebrandt:2004hh, Hildebrandt:2005ix, Hildebrandt:2005iw}. The different
pieces of the photonuclear operator are organised in a perturbative expansion
which is complete to \NXLO{3} [${\cal O}(e^2 \delta^3)$]. Hence, it includes
not only the Thomson term for the protons, as well as magnetic moment
couplings and dynamical single-nucleon effects such as virtual pion loops and
the Delta excitation, but also significant contributions from the
coupling of external photons to the charged pions that are exchanged between
neutrons and protons (referred to hereafter as ``two-nucleon/two-body
currents''). The photonuclear operator is evaluated between \threeHe wave
functions, which are calculated from $NN$ and $3N$ potentials derived using
the same perturbative \ChiEFT expansion~\cite{Weinberg,
  Weinberg:1992yk,Phillips:2016mov}. 

While $\gamma \threeHe$ experiments are only in the planning
stages~\cite{Weller:2013zta, Annand:2015, HIGSPAC, Ahmed:2016, Briscoe}, the
past two decades have seen significant progress in measurements of Compton
scattering on the deuteron.  However, deuteron data only provides access to
the isoscalar polarisabilities; a \threeHe target provides complementary
information on neutron polarisabilities. A na\"ive model of the \threeHe
nucleus as two protons paired with total spin zero plus a neutron suggests
that observables should depend on the combinations $2\alphaep+\alphaen$,
$2\betamp+\betamn$, and that dependence on proton spin polarisabilities should
be minimal. Such a picture is somewhat over-simplified; see
sect.~\ref{sec:conclusion}.  Still, relative to the deuteron, \threeHe data
offers the promise of stronger signals and of cross-validation of the theory
used to subtract binding effects and extract nucleon polarisabilities.

Most recently, Myers \etal~\cite{Myers:2014ace, Myers:2015aba} measured the
differential cross section on the deuteron at energies ranging from $65$ to
$115\;\MeV$, doubling the world data for elastic $\gamma d$
scattering. In combination with the proton results quoted below, a fit using
the \ChiEFT $\gamma d$ amplitude at the same order, ${\cal O}(e^2 \delta^3)$,
as the current work, yields
\begin{equation}
\begin{split}
  \alphaen&=11.55 \pm 1.25 \mathrm{(stat.)} \pm 0.2 \mathrm{(Baldin)} \pm 0.8 \mathrm{(theory)}\;\;,\\
  \betamn&=3.65 \mp 1.25 \mathrm{(stat.)} \pm 0.2 \mathrm{(Baldin)} \mp 0.8
  \mathrm{(theory)}
           \label{eq:LundPRL}
\end{split}
\end{equation}
for the electric and magnetic dipole polarisabilities of the neutron. 
The canonical units of $10^{-4}~\mathrm{fm}^3$ are employed. Here,
the Baldin sum rule
$\alphaen+\betamn=15.2\pm0.4$~\cite{Olmos:2001,Levchuk:1999zy} was used as a
constraint, and the third error listed is an estimate of the theory
uncertainty.  Equation~\eqref{eq:LundPRL} is consistent with the extraction of
neutron polarisabilities from quasi-free Compton scattering on the neutron in
deuterium~\cite{Kossert:2002ws, Demissie:2016cye, BerhanPhD}. Further
refinement of extractions from deuterium data is expected thanks to ongoing
experiments at \HIGS~\cite{Weller:2013zta, HIGSPAC, Ahmed:2016,
  Feldman:2015dsc} and the ongoing extension of the \ChiEFT calculation to
$\mathcal{O}(e^2 \delta^4)$~\cite{dcompton-delta4}. A comprehensive review of
experimental and \ChiEFT work on Compton scattering from deuterium can be
found in ref.~\cite{Griesshammer:2012we}, which also summarises work on the
proton in \ChiEFT.

Concurrently, refs.~\cite{McGovern:2012ew, Griesshammer:2015ahu} extracted the
 electric dipole polarisabilities of the proton using the data set of $\gamma p$
elastic differential cross section data compiled in
ref.~\cite{Griesshammer:2012we} and the $\mathcal{O}(e^2 \delta^4)$ \ChiEFT
$\gamma p$ amplitude:
\begin{equation}
\begin{split}
  \alphaep&= 10.65\pm0.35(\text{stat})\pm0.2(\text{Baldin})
             \pm0.3(\text{theory})\;,\\
  \betamp &= \phantom{0}3.15\mp0.35(\text{stat})\pm0.2(\text{Baldin})
             \mp0.3(\text{theory})\;\;.
             \label{eq:protonvalues}
\end{split}
\end{equation}
Here, the
Baldin sum rule value $\alphaep+\betamp=13.8\pm0.4$ was used. It is fully
consistent with a more recent determination of
$\alphaep+\betamp=14.0\pm0.2$~\cite{Gryniuk:2015eza}.  Compared to the neutron
values, the uncertainties are much smaller, for two reasons. The proton
extraction used the \ChiEFT $\gamma p$ amplitude at
$\mathcal{O}(e^2 \delta^4)$, i.e.~at one order higher than the deuteron
extraction, leading to smaller theoretical uncertainties. The main difference
is however that the deuteron data set is of lesser quality than that of the
proton, contains fewer points, and is restricted to a much smaller energy
range. This results in statistical uncertainties which are nearly four times
those of the proton. Therefore, these extractions provide only a hint that the
proton and neutron polarisabilities may be different. Reducing the
experimental error bars is imperative to conclude whether short-range effects 
lead to small proton-neutron differences; such differences would have potentially
important implications for the
proton-neutron mass splitting, see, \eg, ref.~\cite{Griesshammer:2015ahu} and
references therein. We argue that Compton scattering on \threeHe can serve to
improve the neutron values.

In addition to the scalar polarisabilities, four spin polarisabilities
$\gammaee$, $\gammamm$, $\gammaem$ and $\gammame$ parametrise the response of
the nucleon's spin degrees of freedom to electromagnetic fields of particular
multipolarities.  Intuitively interpreted, the electromagnetic field of the
spin degrees causes bi-refringence in the nucleon, just as in the classical
Faraday-effect~\cite{Holstein:2000yj}.  The type of experiment that will be
most sensitive to the spin polarisabilities involves polarised photons and
targets. A comprehensive exploration of such sensitivities for the proton was
recently published by some of the current authors~\cite{Griesshammer:2017txw}.
Some data exist, and in Ref.~\cite{Martel:2014pba} it was used to fit a subset
of the spin polarisabilities.  The values obtained agree well within the
respective uncertainties with \ChiEFT predictions~\cite{Griesshammer:2015ahu}.
However, it was argued in Ref.~\cite{Griesshammer:2017txw} that much of the data 
is at energies that are too high for any
  extraction to be independent of the theoretical framework
  employed.
  
No experiments have yet probed the individual neutron spin polarisabilities,
which are also predicted by $\chi$EFT at the same
order~\cite{Griesshammer:2015ahu}. They can be measured with a polarised
deuteron target~\cite{Choudhury:2004yz, Griesshammer:2010pz,
  Griesshammer:2013vga, erratum2}, but that has not yet been
attempted. Refs.~\cite{ShuklaPhD, Choudhury:2007bh,Shukla:2008zc} identified
polarised \threeHe as a promising candidate because the dominant
($\approx 95$\%) wave function component in \threeHe consists of two protons
in a spin-singlet pair. The spin of the nucleus is then carried by the neutron
and observables are about $20$ times more sensitive to neutron spin
polarisabilities than to their proton counterparts. Indeed, we will confirm
again in sects.~\ref{sec:Tcirc11} and~\ref{sec:Tcirc10} that \ChiEFT predicts
only small corrections to this expectation, which are slightly different for
each observable and both energy- and angle-dependent.
Even if Compton scattering from a free neutron were feasible, cross sections
for coherent Compton scattering from \threeHe are markedly larger than those
for on a (quasi-)free neutron in this energy range. This is because in
\threeHe, the neutron contributions interfere with the proton ones and with
those from two-body currents; see sect.~\ref{sec:conclusion} for comparisons
to proton, neutron and deuteron targets.

In this presentation, we therefore examine the influence of the Delta
on the cross section and asymmetries, using the same photonuclear kernel as in
the extraction of the scalar polarisabilities of the neutron from $\gamma$d data, 
eq.~\eqref{eq:LundPRL}~\cite{Griesshammer:2012we, Myers:2014ace,
  Myers:2015aba}. The Delta degree of freedom plays a significant role in some
of the spin polarisabilities, especially in $\gammamm$. In $\gamma d$
scattering, its inclusion markedly enhances the pertinent
asymmetries~\cite{Griesshammer:2010pz, Griesshammer:2013vga, erratum2}.  It is
therefore important to examine its impact on the \threeHe asymmetries.

All observables presented are available via a \emph{Mathematica} notebook from
\texttt{hgrie@gwu.edu}. It contains results for cross sections, rates and
asymmetries from $50$ to about $120$~MeV in the lab frame, and allows the
scalar and spin polarisabilities to be varied, including variations
constrained by sum rules.

This article is organised as follows. In sect.~\ref{sec:formalism}, we summarise
the ${\cal O}(e^2 \delta^3)$ amplitude that constitutes the Compton-scattering
operator in our calculation and sketch details of computing matrix elements of
that operator between \threeHe wave functions. Section~\ref{sec:catalog} then
defines the different observables, before sect.~\ref{sec:results} presents the
results of our study. Section~\ref{sec:conclusion} offers a summary and
comparisons of \ChiEFT predictions for Compton scattering on the proton,
neutron, deuteron, and \threeHe.

\section{\texorpdfstring{\threeHe}{3He} Compton Scattering in
  \texorpdfstring{\ChiEFT}{ChiEFT} with the Delta}
\label{sec:formalism}

\subsection{Chiral Effective Field Theory}
\label{sec:ChiEFT}

Compton scattering on nucleons and light nuclei in \ChiEFT has been reviewed
in refs.~\cite{Griesshammer:2012we, McGovern:2012ew}, to which we refer the
reader for notation, relevant parts of the chiral Lagrangian, and full
references to the literature. Here, we first briefly summarise the power counting;
then we sketch the content of the $\mathcal{O}(e^2 \delta^3)$ amplitude  for Compton scattering on \threeHe, which at this order only
consists of one- and two-nucleon contributions to a photonuclear kernel
evaluated between nuclear wave functions.

In \ChiEFT with a dynamical Delta, Compton scattering exhibits three
low-momentum scales: the pion mass $\mpi$, the Delta-nucleon mass splitting
$\DeltaM \approx 300\;\MeV$, and the photon energy $\omega$.  Each provides a
small, dimensionless expansion parameter when it is measured in units of the
``high'' scale $\Lambda_\chi$ at which the theory breaks down because new
degrees of freedom enter.

While the two ratios $\frac{\mpi}{\Lambda_\chi}$ and
$\frac{\DeltaM}{\Lambda_\chi}$ have very different chiral behaviour, we follow
Pascalutsa and Phillips~\cite{Pascalutsa:2002pi} and take a common breakdown
scale $\Lambda_\chi \approx 650\;\MeV$, which is consistent with the masses of
the $\omega$ and $\rho$ as the next-lightest exchange mesons, and then exploit
a numerical coincidence for the physical pion masses to define a single
expansion parameter:
\begin{equation}
  \delta\equiv\frac{\Delta_M}{\Lambda_\chi}\approx
  \left(\frac{m_\pi}{\Lambda_\chi}\right)^{1/2}\approx 0.4\ll1\;\;.
\end{equation}
We also count $\MN\sim\Lambda_\chi$.  Since $\delta$ is not very small,
order-by-order convergence must be verified carefully; see our discussion for each observable
in sect.~\ref{sec:results}.

The treatment of the scale $\omega$ depends on the
experiment~\cite{Griesshammer:2012we, McGovern:2012ew}.  In this paper, we
concentrate on energies for which
$\omega\lesssim\mpi\sim\delta^2\Lambda_\chi\ll\Delta_M$ counts like a chiral
scale and pion-cloud physics therefore dominates. As reviewed below,
pion-cloud effects enter at $\calO(e^2\delta^2)$ [\NXLO{2}{}], while the Delta
appears at $\calO(e^2\delta^3)$ [\NXLO{3}{}]. Including it is simply
equivalent to adding one order. We note that the version of \ChiEFT without a
dynamical Delta (often called Heavy Baryon Chiral Perturbation Theory) is
limited to momenta well below the resonance. We denote its expansion parameter
by $Q$, and its \NXLO{2} is identical to ours,
$\calO(Q^3)\equiv\calO(e^2\delta^2)$; see
ref.~\cite[sect.~4.2.7]{Griesshammer:2012we} for more details.
The counting changes at both higher photon energies, where
$\omega \sim \Delta_M$, and also at lower energies, where $\omega\sim\mpi^2/\MN$ is
comparable to the typical nuclear binding momentum. The latter limit is more
relevant to the current paper, since, as we will see in
sect.~\ref{sec:wavefunctions}, it sets a lower limit on the energy for which
our calculations may be considered reliable. In particular, calculations
consistent with the low-energy counting are required to have the correct
nuclear Thomson limit; the current calculation would overshoot considerably as
\linebreak$\omega\to0$. We will address both extensions to higher and lower energies in
future publications~\cite{future3He}.

Our focus is therefore on energies in the overlap of the single-nucleon regime
(i) of refs.~\cite{Griesshammer:2012we, McGovern:2012ew} and the few-nucleon
regime (II) of ref.~\cite{Griesshammer:2012we} : $50\;\MeV\lesssim\omega\lesssim120\;\MeV$. 

\subsection{Compton Kernel: Single-Nucleon Contributions}
\label{sec:onenucleon}

Contributions to the single-nucleon amplitude up to $\calO(e^2\delta^3)$ or
\NXLO{3} in the regime \linebreak$50\;\MeV\lesssim\omega\lesssim120\;\MeV$ are sketched
in fig.~\ref{fig:1Bdiagrams}:
\begin{figure}[!htb]
  \begin{center}
    \includegraphics[width=0.8\textwidth]{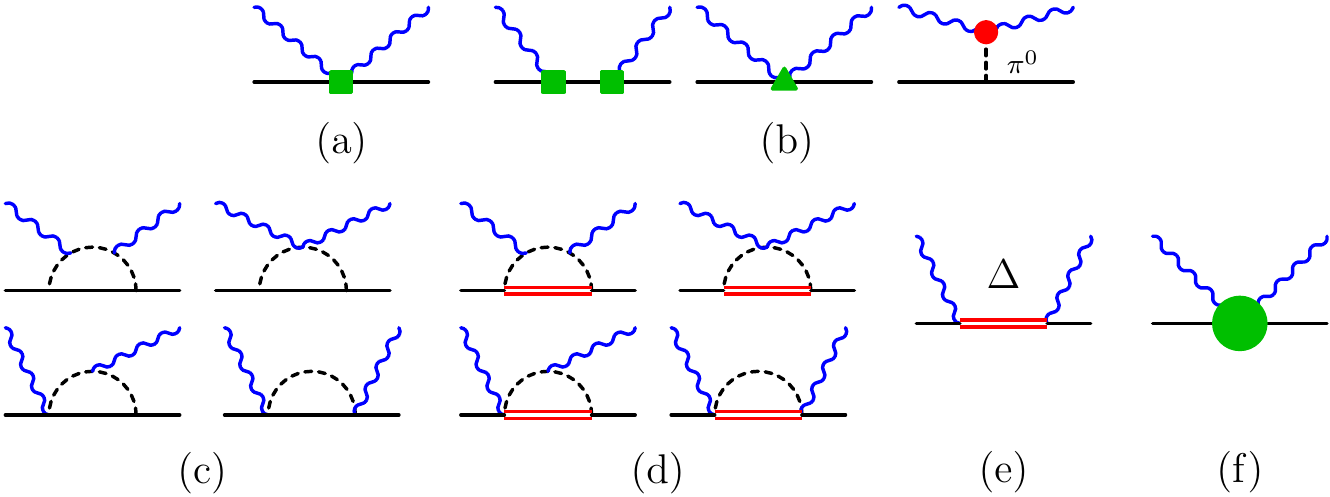}
    \caption{(Colour on-line) The single-nucleon contributions in \ChiEFT up
      to $\calO(e^2\delta^3)$ for
      $50\;\MeV\lesssim\omega\lesssim120\;\MeV$. The vertices are from:
      ${\cal L}_{\piN}^{(1)}$ (no symbol), ${\cal L}_{\piN}^{(2)}$ (green
      square), ${\cal L}_{\piN}^{(3)}$ (green triangle),
      ${\cal L}_{\pi\pi}^{(4)}$ (red disc)~\cite{Bernard:1995dp}; the green
      disc of graph (f) stands for variations of the polarisabilities,
      cf.~eq.~\eqref{eq:fit}. Permuted and crossed diagrams are not
      displayed.}
    \label{fig:1Bdiagrams}
  \end{center}
\end{figure}
\begin{itemize}

\item[(a)] LO [$\calO(e^2\delta^0=Q^2)$]: The single-nucleon (proton) Thomson
  term.
   
\item[(b)] \NXLO{2} [$\calO(e^2\delta^2=Q^3)$] non-structure/Born terms:
  photon couplings to the nucleon charge beyond LO, to its magnetic moment, or
  to the $t$-channel exchange of a $\pi^0$ meson. $\gamma\,$\threeHe
  scattering is sensitive to the latter, in contradistinction to $\gamma d$
  scattering, where its expectation value is zero, as the deuteron is
  isoscalar.

\item[(c)] \NXLO{2} [$\calO(e^2\delta^2=Q^3)$] structure/non-Born terms:
  photon couplings to the pion cloud around the nucleon, the source of the LO
  contributions to the polarisabilities as first reported in
  refs.~\cite{Bernard:1991rq, Bernard:1995dp}.
    
\item[(d/e)] \NXLO{3} [$\calO(e^2\delta^3)$] structure/non-Born terms: photon
  couplings to the pion cloud around the Delta (d) or directly exciting the
  Delta (e), as calculated in refs.~\cite{Hemmert:1996rw, Hemmert:1997tj};
  these give NLO contributions to the polarisabilities. As detailed in
  ref.~\cite{Griesshammer:2012we}, we use a heavy-baryon propagator and a zero
  width, with $\Delta_M=293\;\MeV$, $g_{\piN\Delta}=1.425$.  The non-relativistic version of the
  $N\Delta\gamma$ coupling is $b_1=5$ and tuned to give the same effective
  strength as the relativistic value of $g_M=2.9$ found by fitting to the
  proton data set up to the Delta resonance~\cite{Griesshammer:2012we}. In
  practice, the $\pi \Delta$ loops produce almost-energy-independent
  contributions to $\alphae$ and $\betam$ for $\omega < 200\;\MeV$.

\item[(f)] Short-distance/low-energy coefficients (LECs), which encode those
  contributions to the nucleon polarisabilities which stem from neither
  pion-cloud nor Delta effects; see eq.~\eqref{eq:fit} for the precise
  form. These offsets to the polarisabilities are formally of higher order.
  For the proton, we fix these LECs to reproduce the $\calO(e^2 \delta^2)$
  values of the scalar polarisabilities.  For the neutron, we exploit the
  freedom to vary the LECs in order to assess the sensitivity of cross
  sections and asymmetries to the neutron scalar polarisabilities. We refer to
  sect.~\ref{sec:varypols} for the detailed procedure and values.

\end{itemize}
There is no contribution at NLO [$\calO(e^2\delta^1)$], and only Delta
contributions at \NXLO{3} [$\calO(e^2\delta^3)$]. The difference between the
previous $\calO(e^2\delta^2)$~\cite{ShuklaPhD, Choudhury:2007bh,Shukla:2008zc}
and our new $\calO(e^2\delta^3)$ calculation is hence the inclusion of
Delta effects. Covariant kinematics for the fermion propagators, a nonzero
Delta width, vertex corrections, etc.~are just some examples of corrections
which are of higher order than the last one we retain, \NXLO{3}
[$\calO(e^2\delta^3)$]; they are parametrically small.

\subsection{Compton Kernel: Two-Nucleon Contributions}
\label{sec:twonucleon}

The leading two-body currents in \ChiEFT occur at $\mathcal{O}(e^2 \delta^2)$
and do not involve Delta excitations. They are the two-body analogue of the
$\pi N$ loop graphs depicted in fig.~\ref{fig:1Bdiagrams} and thus denoted as
$\mathcal{O}(Q^3)$ in \ChiEFT without dynamical Delta. They were
first computed in ref.~\cite{Beane:1999uq}, where full expressions can be
found. We depict them in fig.~\ref{fig:2Bdiagrams} below.

We note that the $\mathcal{O}(e^2 \delta^2)$ diagrams only contribute for $np$
pairs, \ie~they all contain an isospin factor of
$\tau^{(1)} \cdot \tau^{(2)} - \tau^{(1)}_{z} \tau^{(2)}_{z}$.  However, one
distinction between \threeHe and the deuteron is that in the tri-nucleon case
both isospin zero and one pairs are present.

No corrections enter at the next order, $\calO(e^2\delta^3)$.  Boost
corrections and corrections with a nucleon propagating between the
pion-exchange and a photon-nucleon interaction only enter at
$\calO(e^2\delta^4)$~\cite{Beane:1999uq, Beane:2004ra, Griesshammer:2012we},
and those with an intermediate Delta at
$\calO(e^2\delta^5)$~\cite{Griesshammer:2012we}.  Note that the pieces of the
pion-exchange currents that are $1/M$ suppressed and must be derived
consistently with the $NN$ potential~\cite{Pastore:2009is,
  Kolling:2009iq,Kolling:2011mt} only enter at orders higher than we consider
here; see discussion in sect.~\ref{sec:wavefunctions}.

\begin{figure}[!htb]
  \begin{center}
    \includegraphics[width=0.5\textwidth]{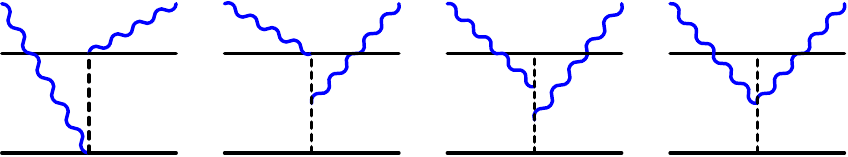}
    \caption{(Colour on-line) $\mathcal{O}(e^2 \delta^2)$ contributions to the
      (irreducible) $\gamma NN \rightarrow \gamma NN$ amplitude (no additional
      contributions at $\mathcal{O}(e^2 \delta^3)$).  Notation as in
      fig.~\ref{fig:1Bdiagrams}. Permuted and crossed diagrams not displayed.}
    \label{fig:2Bdiagrams}
  \end{center}
\end{figure}


\subsection{From the Compton Kernel to Compton Amplitudes}
\label{sec:threebody}

Here, we provide a brief description of the computation of the integrals for \threeHe matrix
elements of the Compton operators in the centre-of-mass frame;
full details can be found in refs.~\cite{Shukla:2008zc,ShuklaPhD}.

\subsubsection{Formulae for Matrix Elements}
\label{sec:MEs}

We seek a \threeHe-Compton amplitude which depends on the spin projections
$M_{i/f}$ of the incoming and outgoing \threeHe nucleus onto a common $z$ axis
(which we define below to be the beam direction) and on the helicities
$\lambda_{i/f}$ of the incident and outgoing photon.  Using permutations and
symmetries, this amplitude can be written as
\begin{equation} 
   \label{eq:T}
  A_{M_i\lambda_i}^{M_f\lambda_f}:=\langle M_f,\lambda_f|T| M_i,\lambda_i
  \rangle=3\,\bra M_f|\left(\frac{1}{2} \left[ {\hat
    O}_{\lambda_f \lambda_i}^{1B}(1)+{\hat O}_{\lambda_f \lambda_i}^{1B}(2) \right]+{\hat O}_{\lambda_f \lambda_i}^{2B}(1,2)\right)|M_i \ket\;,
\end{equation}
where nucleons are numbered and $|M \ket$ is an anti-symmetrised state of
\threeHe.  Since we are concerned with the \threeHe nucleus, we also take this
state to have isospin quantum numbers $T=M_T=1/2$.  The operator
${\hat O}_{\lambda_f \lambda_i}^{1B}(a)$ with $a=1,\dots,3$ represents the
one-body amplitude of sect.~\ref{sec:onenucleon}, where external photons of
incoming (outgoing) helicity $\lambda_i$ ($\lambda_f$) interact with nucleon
``$a$''. ${\hat O}_{\lambda_f \lambda_i}^{2B}(a,b)$ represents the
corresponding two-body current of sect.~\ref{sec:twonucleon}, where the
interaction is with the nucleon pair ``$(a,b)$''. Thus in eq.~\eqref{eq:T},
nucleon ``$3$'' serves as a spectator to the $\gamma NN \rightarrow \gamma NN$
scattering process. Three-nucleon currents (\ie~contributions of instantaneous
interactions between three nucleons and at least one photon) do not enter
before chiral order $e^2\delta^6$.

\begin{figure}[!htbp]
\begin{center}
  \includegraphics[height=6cm]{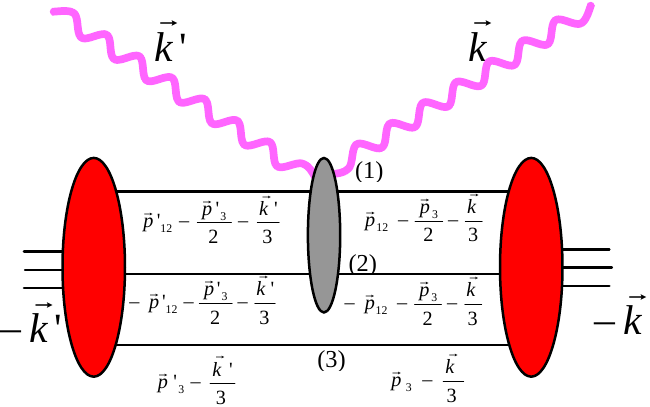}

  \caption{(Colour on-line) Kinematics of the calculation in the
    $\gamma$\threeHe centre-of-mass frame. The vectors $\vec{k}$ ($-\vec{k}$)
    and $\vec{k}'$ ($-\vec{k}'$) are the three-momenta of the incoming and
    outgoing photon (\threeHe). The nucleon loop momenta $\vec{p}_i$,
    $\vec{p}_{ij}$ and primed versions are defined and numbered as in the
    text.}
  \label{fig:kin}
\end{center}
\end{figure}

We use the approach of Kotlyar \etal~\cite{Ko00} to calculate matrix elements
between three-nucleon basis states. The loop momenta $\vec{p}_{12}$ and
$\vec{p}_3$ are the Jacobi momenta of the pair ``$(1,2)$'' and the spectator
nucleon ``$3$'' respectively; see fig.~\ref{fig:kin} for kinematics. By introducing a
complete set of states, we can write the matrix element as
\begin{equation}
\begin{split}
  A_{M_i\lambda_i}^{M_f\lambda_f} =& 3 \int d^3p_{12}' \int d^3p_3' \int d^3p_{12} \int d^3p_3 \sum \limits_{\alpha , \alpha', m_{12}, m_{12}'} \varphi (p_{12}',p_3',\alpha') \varphi (p_{12},p_3,\alpha) \\
  & \times\;{\mathcal Y}_{l_{12}',s_{12}',j_{12}', m_{12}'}^* (\hat p_{12}')\, 
  {\mathcal Y}_{l_3',{1\over 2},j_3', M_f-m_{12}'}^* (\hat p_3')\, {\mathcal
    Y}_{l_{12},s_{12},j_{12},
    m_{12}} (\hat p_{12})\,  {\mathcal Y}_{l_3,{1\over 2},j_3, M_i-m_{12}} (\hat p_3)\\
  & \times\;\bra {\vec p}_{12}^{\, \prime} \vec p_3^{\, \prime}; \alpha_T'| {\hat O}_{\lambda_f
    \lambda_i}(1,2)| \vec p_{12} \vec p_3; \alpha_T \ket\; \delta^{(3)} (\pv_3^{\, \prime} -
  \pv_3 - \qv/3) \;\;, \label{eq9}
\end{split}
\end{equation}
where
$\hat{O}_{\lambda_f \lambda_i}(1,2) \equiv \frac{1}{2} ({\hat O}_{\lambda_f
  \lambda_i}^{1B}(1)+{\hat O}_{\lambda_f \lambda_i}^{1B}(2) \big)+{\hat
  O}_{\lambda_f \lambda_i}^{2B}(1,2)$
accounts for both one- and two-nucleon currents. Note that the helicity
dependence is entirely carried by this operator.  The total angular momentum
of the nucleus $J$ is a result of coupling between $j_{12}$ and $j_3$, the
total angular momentum of the ``$(1,2)$'' subsystem, and the spectator nucleon
``$3$'', respectively.  The orbital angular momentum $l_{12}$ and the spin
$s_{12}$ of ``$(1,2)$'' combine to give $j_{12}$.  Similarly, $l_3$ and
$s_3={1\over 2}$ combine to give $j_3$. Hence, in eq.~\eqref{eq9}, we defined
\begin{equation} {\mathcal Y}_{l,s,j,m}(\hat p) = \sum \limits_{m_s}
  (l,m-m_s,s,m_s|j,m)\,  Y_{l,m-m_s} (\hat p) |s m_s \ket
  \label{eq10}
\end{equation}
and
\begin{equation}
  \varphi (p_{12},p_3,\alpha) =
  (j_{12},m_{12},j_{3},M_i-m_{12}|\frac{1}{2},M_i)\left(t_{12},mt_{12},{1 \over 2},mt_3\right|\left.{1 \over 2},{1
  \over 2}\right)\bra p_{12}p_3 \alpha |\Psi \ket\;\;,\label{eq10b}
\end{equation}
where the component $\bra p_{12}p_3\alpha|\Psi \ket$ of the \threeHe wave
function is parametrised by the quantum numbers
$\alpha=|(l_{12}s_{12})j_{12}(l_3\half)j_3(j_{12}j_3)JM_J\ket\;|\alpha_T\ket$.
Similarly, the isospin quantum numbers are $\alpha_T=|(t_{12}\half)TM_T\ket$,
where the isospin of the two-nucleon subsystem is $t_{12}$ (projection
$mt_{12}$) and combines with the isospin of the spectator nucleon to give the
total isospin and its projection of the \threeHe nucleus. The Pauli principle
guarantees that $l_{12} + s_{12} + t_{12}$ is odd.

The $\delta$ function can be used to perform the integral over $\hat{p}_3'$, and so
eq.~\eqref{eq9} can be recast as
\begin{equation}
\begin{split}
  A_{M_i\lambda_i}^{M_f\lambda_f}
     =3& \sum \limits_{j_{12}', j_{12}, m_{12}',m_{12}, s_{12}',
     s_{12}, l_{12}', l_{12},
     j_3', j_3, l_3', l_3,mt_{12}}  \int p_{12}'^2 dp_{12}'  \int p_{12}^2
     dp_{12} \int p_{3}^2 dp_{3}\\
  &\varphi (p_{12},p_3,\alpha)\,  {\mathcal I}_2 (p_{12}, p_{12}'; l_{12}, l_{12}', s_{12},
     s_{12}', j_{12}, j_{12}', m_{12},
     m_{12}', mt_{12})\\
  &\times\;\int dp_3' \varphi (p_{12}',p_3',\alpha')\,  {\mathcal
     I}_3(p_{3}, p_{3}'; l_{3}, l_{3}', j_{3}, j_{3}', m_{12}, m_{12}',
     M_i, M_f) \;\;.\label{eq11}
\end{split}
\end{equation}
In this expression, the integral
\begin{equation}
\begin{split}
  &{\mathcal I}_2(p_{12}, p_{12}';
     l_{12}, l_{12}', s_{12}, s_{12}', j_{12}, j_{12}', m_{12}, m_{12}',
     mt_{12})= \\
  &\hq \int d{\hat p_{12}'} \int d{\hat p_{12}}\, {\mathcal
     Y}_{l_{12}',s_{12}',j_{12}', m_{12}'}^* (\hat p_{12}')\bra \vec{p}_{12}^{\, \prime}; t_{12}' mt_{12}|{\hat O}(1,2)| \vec p_{12};
     t_{12} mt_{12}\ket\,  {\mathcal Y}_{l_{12},s_{12},j_{12}, m_{12}} (\hat
     p_{12})
\end{split}
\end{equation}
is computed numerically, as is
\begin{equation}
\begin{split}
  &{\mathcal I}_3(p_{3}, p_{3}'; l_{3}, l_{3}', j_{3}, j_{3}',
     m_{12}, m_{12}', M_i, M_f)= \\
  & \qquad  \int d{\hat p_{3}}\, {\mathcal Y}_{l_3',{1\over 2},j_3',
     M_f-m_{12}'}^* \left(\widehat {p_3+{q \over 3}}\right) \delta
     \left( p_3' - \left|\vec{p_3} + \vec {{ q \over 3}}\right|\right){\mathcal
     Y}_{l_3,{1\over 2},j_3, M_i-m_{12}} (\hat p_3)\;\;.
\end{split}
\end{equation}

It turns out that a rather small number of partial waves is sufficient to
achieve convergence. We test this by comparing to results with one more unit
of $j_{12}$. The slowest convergence is at the extremes of energies and
momentum transfer ($\omegacm\approx120\;\MeV$,
$\thetacm\gtrsim165^\circ$). When one includes all partial waves with
$j_{12} \le2$, the one-nucleon matrix elements are converged to within $0.5\%$
there, and better at lower energies and less-backward angles. The large and
medium-sized two-nucleon matrix elements are converged to better than $0.7\%$
for $j_{12} \le1$, and better than that for lower $\omega$ and
$\theta$. Higher numerical accuracy is only limited by computational cost:
two-nucleon runs with $j_{12} \le2$ take about $10$ times longer. Radial and
angular integrations are converged at the level of a few per-mille. With these
parameters, at $\omegacm\approx120\;\MeV$, $\thetacm\gtrsim165^\circ$ the
cross section is numerically converged at about $1.2\%$ or $0.35$~nb/sr. At
lower energies and smaller angles (and hence smaller momentum transfers),
convergence is substantially better. Results for the other observables are
similar. Increased numerical accuracy is not really useful here, since the dominant uncertainty comes instead from the
truncation of the \ChiEFT expansion at $\calO(e^2\delta^3)$, translating
roughly into a truncation error of $\delta^4\approx\pm3\%$ of the LO result
(see sect.~\ref{sec:uncertainties}).

\subsubsection{Choice of Wave Functions}
\label{sec:wavefunctions}

Following Weinberg's ``hybrid approach''~\cite{Weinberg}, we finally convolve
the \ChiEFT photonuclear kernels with wave functions which are obtained from
three choices for the nuclear interaction: the chiral Idaho \NXLO{3}
interaction for the $NN$ system at cutoff $500\;\MeV$~\cite{En03} with the
$\mathcal{O}(Q^3)$ \ChiEFT $3N$ interaction of variants ``b'' (our
``standard'') and ``a'' as described in refs.~\cite{No06a, No06b}, and the
AV18 $NN$ model interaction~\cite{Wi95}, supplemented by the Urbana-IX $3N$
interaction (3NI)~\cite{Pu95}. (Unlike refs.~\cite{ShuklaPhD,
  Choudhury:2007bh, Shukla:2008zc}, we do not consider wave functions found
without $3N$ interactions.)  All choices capture the correct long-distance
physics of one-pion exchange and reproduce the $NN$ scattering data to a
degree that is superior to the accuracy aimed for in this article. They also
all reproduce the experimental value of the triton and \threeHe binding
energies.  The two \ChiEFT variants are parametrised differently and lead to
different spectra in other light nuclei. All wave functions are fully
anti-symmetrised and obtained from Faddeev calculations in momentum
space~\cite{No97, NoggaPrivComm}.

The chiral wave functions claim a higher accuracy than that of our Compton
kernels. For the purposes of this article, it is not necessary to enter the
ongoing debate about correct implementations of the chiral power counting or
the range of cutoff variations, etc.; see ref.~\cite{Phillips:2013fia} for a
concise summary. Similarly, even though the Compton Ward identities are
violated because the one-pion-exchange $NN$ potentials are regulated, any
inconsistencies between currents, wave functions and nuclear potentials will
be compensated by operators which enter at higher orders in \ChiEFT than the
last order we fully retain, namely $\calO(e^2\delta^3)$ or \NXLO{3}. In
addition, the potentials do not include explicit Delta contributions while the
kernel does.  However, it is easy to see that, for real Compton scattering
around $120\;\MeV$, a Delta excited directly by the incoming photon is more
important than one that occurs virtually between exchanges of virtual pions.
For our purposes such Delta excitations in the $NN$ potential are
well
approximated by the $\pi N$ seagull LECs that enter the \NXLO{3} interaction. 


We therefore take the difference between results with the three wave functions
as indicative of the present residual theoretical uncertainties. These do not 
affect the conclusions of our sensitivity studies, but it is
undoubtedly true that better extractions of polarisabilities from \threeHe
data will need a reduced wave function spread.

In particular, we expect that including terms in the amplitude which restore
the Thomson limit should significantly reduce the wave-function and
interaction dependence even at nonzero energies.  For the deuteron, this was
seen at energies as high as $\omega\approx120\;\MeV$~\cite{Hildebrandt:2005ix,
  Hildebrandt:2005iw, Griesshammer:2012we, Griesshammer:2013vga, erratum2}.
It is also likely to decrease the cross section at the low-energy end of our
region of interest.  As detailed in refs.~\cite{Beane:1999uq, Beane:2004ra}
and~\cite[sect.~5.2]{Griesshammer:2012we}, the coherent-propagation process
necessary to restore the Thomson limit becomes important for photon energies
lower than the inverse target size. For \threeHe, that scale is
$\gamma_3\sim\sqrt{2\MN\;B(\threeHe)/3}\lesssim70\;\MeV$~\cite{Konig:2016}.
Refs.~\cite{Hildebrandt:2005ix, Hildebrandt:2005iw, Griesshammer:2012we}
discuss in detail how the power counting for low energies,
$\omega \sim m_\pi^2/M$, leads to the restoration of the Thomson limit by
inclusion of coherent propagation of the \threeHe system in the intermediate
state between absorption and emission of photons.

However, the present formulation of \threeHe Compton scattering is not
applicable in the Thomson-limit region, since it organises contributions
under the assumption $\omega \sim m_\pi\gtrsim\gamma_3$.  If 
used for $\omega\ll\mpi$, where it does not apply,  it would not yield the
Thomson limit for \threeHe but would be several times too large.  Indeed, at
the energies we study here, $50\;\MeV\lesssim\omega\lesssim120\;\MeV$, the
power counting predicts that incoherent propagation of the intermediate
three-nucleon system dominates. This is supported by the deuteron case, where
including the effects that restore the nuclear Thomson limit leads to a
reduction of $10-20\%$ at $\omega=50$ MeV~\cite{Hildebrandt:2005ix,
  Hildebrandt:2005iw}.  For \threeHe, it is plausible that the corrections by
coherent-nuclear effects may suppress the cross section at the low end of our
energy range somewhat more: the mismatch between the Thomson-limit and the
$\omega=0$ amplitude in our calculation is larger than for the deuteron, and
\threeHe has a larger binding energy, so coherent propagation of the
three-nucleon system may be important up to higher energies than in the
two-nucleon case. While work in this direction is in
progress~\cite{future3He}, the present approach suffices for reasonable rate
estimates at $\omega\lesssim80\;\MeV$.

\subsubsection{A Note on Estimates of Theory Uncertainties}
\label{sec:uncertainties}

Since the Compton amplitudes are complete at \NXLO{3} [$\calO(e^2\delta^3)$],
they carry an \emph{a-priori} uncertainty estimate of roughly
$\delta^4\approx(0.4)^4\approx\pm3\%$ of the LO result, or twice that for
observables since they involve amplitude-squares. This translates to $\pm6\%$
for generic cross sections and beam asymmetries because they are nonzero at LO. (At
lower energies, the restoration of the Thomson limit may lead to an additional
reduction of the cross section, as discussed above.) The first nonzero
contributions to the double asymmetries enter at \NXLO{2}, so their
\emph{a-priori} accuracy is estimated as $\pm2\times\delta^2\approx\pm30\%$.

Here, we do not explore convergence with a statistically rigorous
interpretation. We nonetheless briefly mention that two \emph{post-facto}
criteria (order-by-order convergence and residual wave function dependence)
shown in sect.~\ref{sec:results} are roughly commensurate with these
estimates. An exception may be at the highest energies
$\omega\approx120\;\MeV$, where the convergence pattern discussed in
sect.~\ref{sec:results} indicates that \NXLO{4} corrections might amount to
changes by roughly $\pm20\%$ of the magnitude of the cross section. 

Reassuringly, we see that the \emph{sensitivities} of observables to
variations of polarisabilities are typically much less affected by convergence
issues than are their overall magnitudes.  We therefore judge that our
sensitivity investigations are sufficiently reliable to be useful for current
planning of experiments. We reiterate that our goal here is an exploratory
study of magnitudes and sensitivities of observables to the nucleon
polarisabilities. Once data are available, a polarisability extraction will of
course need to address residual theoretical uncertainties with more diligence,
as was already done for the proton and deuteron in
refs.~\cite{Griesshammer:2012we, McGovern:2012ew, Myers:2014ace,
  Myers:2015aba, Griesshammer:2015ahu}.
  
\section{A Catalogue of \texorpdfstring{\threeHe}{3He} Compton-Scattering Observables}
\label{sec:catalog}

\subsection{Observables for Polarised Cross Sections}
\label{sec:obs}

This presentation follows reviews on polarised scattering by Arenh\"ovel and
Sanzone~\cite{Arenhovel:1990yg, Arenhovel:2008qy}, by Paetz~\cite{Paetz}, and
the presentation of Babusci \etal \cite{Babusci:1998ww} which addresses
Compton scattering from a spin-$\half$ target.

\begin{figure}[!htbp]
  \begin{center}
    \includegraphics[width=0.4\linewidth]{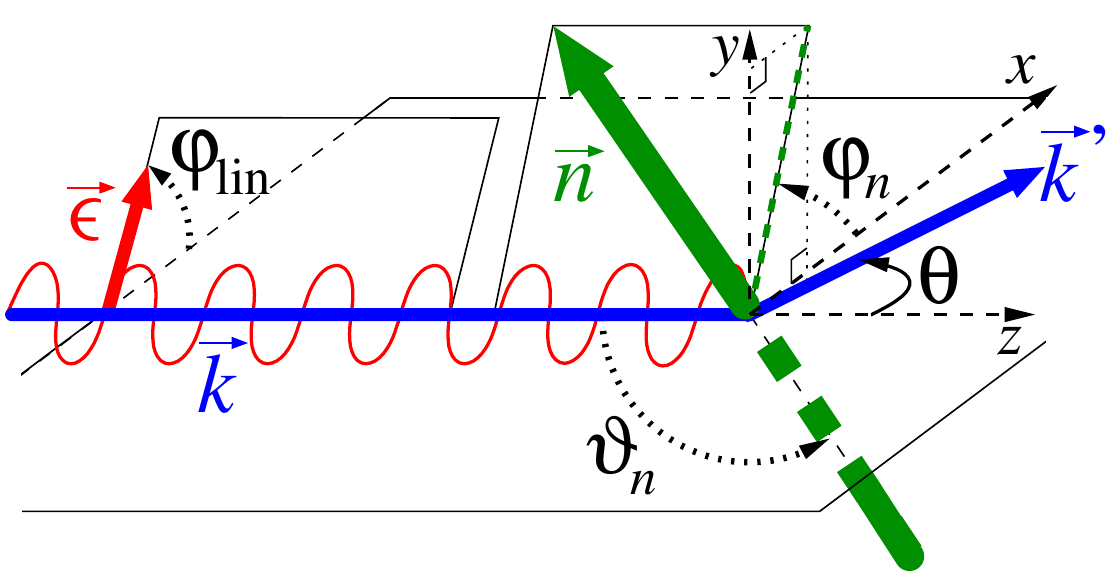}
    \caption{(Colour on-line) Kinematics of \threeHe Compton scattering. }
    \label{fig:spinkinematics}
  \end{center}
\end{figure}

We start by summarising the kinematics and coordinate system; see
fig.~\ref{fig:spinkinematics}. Unless specified otherwise, we work in the
laboratory frame. 
The incoming photon momentum, $\kv$, defines the $z$-axis. The scattered 
photon momentum, $\kv'$, lies in the $xz$-plane.
The energies of the two photons are $|\kv|=\omegalab$ and $|\kv'|=\omegaprimelab$, and
the scattering angle $\theta$ is the angle between the two momenta. The
$y$-axis is chosen to form a right-handed triplet with $\kv$ and $\kv'$, so
that finally 
\begin{equation}\kv\times\kv'= \omegalab\,\omegaprimelab\sin\theta\;\vec e_y\;\;,\;\; 
\kv\cdot\kv'=\omegalab \,\omegalab'\cos\theta\;\;,\;\; \omegaprimelab=\frac{\omegalab}{1+\omegalab(1-\cos\theta)/\MHe}\;\;.
\end{equation}
The linear-polarisation angle $0\le\philin<\pi$ of the photon is the angle
from the scattering plane to the linear photon polarisation
plane\footnote{This definition varies from that of~\cite{Arenhovel:1990yg},
  whose angle $\phi$ is measured \emph{from} the polarisation plane \emph{to}
  the normal of the scattering plane, \ie~$\philin=-\phi$.},
\ie~$\vec{\epsilon}_\text{lin}=\ev_x\cos\philin+\ev_y\sin\philin$. The
\threeHe polarisation vector is $\PthreeHe\,\vec{n}$, where
$\PthreeHe\in[0;1]$ is the degree of (vector-)polarisation and its direction
is $\vec{n}=(\sin\thetan\cos\phin,\sin\thetan\sin\phin,\cos\thetan)$ with
angle $\thetan\in[0;\pi]$ from the $z$-axis to $\vec{n}$ and angle
$\phin\in[0;2\pi)$ from the $x$-axis to the projection of $\vec{n}$ onto the
$xy$-plane.

The cross section for Compton scattering of a polarised photon beam with
density matrix $\rhogamma$ from a polarised target of spin $S$ with density
matrix $\rhoS$, without detection of the final state polarisations, is found
from the Compton tensor $T$ of eq.~\eqref{eq:T} via
\begin{equation}
  \label{eq:xsect}
  \frac{\dd\sigma}{\dd\Omega}=\Phi^2\;\tr[T\rhoS\rhogamma T^\dagger]\;\;,
\end{equation}
where the trace is taken over the polarisation states and $\Phi$ is the
frame-dependent flux factor. In the lab
frame~\cite[sect.~2.3]{Griesshammer:2012we}:
\begin{equation}
  \Phi_{\mathrm{lab}} = \frac{\omegalab^\prime}{4\pi\omegalab}\;\;.
\end{equation}
Two popular bases exist for each density matrix. In the helicity
basis, the photon polarisation is described by\footnote{We correct a
  notational inconsistency for the bra-ket notation in
  ref.~\cite{Griesshammer:2013vga} which did not have any influence on the
  final result.}
\begin{equation}
  \label{eq:photonpol}
  \left(\rhogamma\right)_{\lambda^\prime\lambda}:=
  \langle \lambda^\prime|\rhogamma|\lambda\rangle
  =\frac{1}{2}\left[
    \delta_{\lambda^\prime\lambda}\left(1+\lambda\;P^{(\gamma)}_\text{circ}
    \right)
    -\delta_{\lambda^\prime,-\lambda}\;P^{(\gamma)}_\text{lin}\;
    \e^{+2\lambda\ii\philin}\right]\;\;.
\end{equation}
Here, positive/negative helicities $\lambda,\lambda^\prime=\pm1$ are defined
by\footnote{This corrects an inconsequential misprint in
  ref.~\cite{Griesshammer:2013vga}.}
$\ev_\pm=-\ii(\ev_y\mp\ii\ev_x)/\sqrt{2}$. \linebreak$P^{(\gamma)}_\text{circ}\in[-1;1]$
is the degree of right-circular polarisation, \ie~the difference between right-
and left-circular polarisation, with $P^{(\gamma)}_\text{circ}=+1/-1$
describing a fully right-/left-circularly polarised photon.  The degree of
linear polarisation is $P^{(\gamma)}_\text{lin}\in[0;1]$.

In the Cartesian basis, the Stokes parameters $\xi_i$ parametrise
the $x$ and $y$ components of the density matrix of the incident photon. With
$i,j\in\{x;y\}$:
\begin{equation}
  \left(\rhogamma\right)_{ij}=\half\begin{pmatrix}1+\xi_3&\xi_1-\ii\xi_2\\
    \xi_1+\ii\xi_2&1-\xi_3\end{pmatrix}_{ij}\;\;.
\end{equation}
The degree of right-circular polarisation is then
$P^{(\gamma)}_\text{circ}=\xi_2$, and that of linear polarisation is
$P^{(\gamma)}_\text{lin}=\sqrt{\xi_1^2+\xi_3^2}$, with angle
$\cos[2\philin]=\xi_3/\sqrt{\xi_1^2+\xi_3^2}$ and
$\sin[2\philin]=\xi_1/\sqrt{\xi_1^2+\xi_3^2}$, so that $\xi_i\in[-1;1]$. The
combination $\xi_3=\pm1$, $\xi_1=0$ describes a beam which is linearly
polarised in/perpendicular to the scattering plane, and $\xi_1=\pm1$ with
$\xi_3=0$ a beam which is linearly polarised at angle $\philin=\pm\pi/4$
relative to the scattering plane.

The multipole-decomposition of the density matrix of a polarised spin-$S$
target is~\cite{Arenhovel:2008qy}:
\begin{equation}
  \label{eq:targetpol}
  \rhoS_{m^\prime m}:=\langle m^\prime|\rhoS|m\rangle=
  \frac{(-)^{S-m}}{\sqrt{2S+1}}\sum\limits_{J=0}^{2S}\sqrt{2J+1}\;P^{S}_J
  \sum\limits_{M=-J}^J \threej{S}{S}{J}{m}{-m^\prime}{-M}\e^{\ii M\phin}\;
  d^J_{M0}(\thetan)\;\;,
\end{equation}
where $P^{S}_0\equiv1$ and $P^{\half}_1=\PthreeHe$.  The conventions for
$3j$-symbols and reduced Wigner-$d$ matrices are those of Rose~\cite{Rose},
also listed in the Review of Particle Physics~\cite{Patrignani:2016xqp}.

Another variant for spin-$\half$ particles in the same basis uses
Pauli spin matrices $\vec{\sigma}$:
\begin{equation}
  \rho^{(\half)}=\half\bigl(1+\PthreeHe\;\vec\sigma\cdot\vec n\bigr)\;\;.
\end{equation}


These variants also lead to different parametrisations of the differential
cross section with definite beam and target polarisations (and no detection of
the final-state polarisations), after parity invariance has been taken into
account. In our analysis, both variants will be used side-by-side:
\begin{eqnarray}
    \label{eq:arencross}
    \frac{\dd\sigma}{\dd\Omega}=\hqmm&\dis\left.\frac{\dd\sigma}{\dd\Omega}
    \right|_\text{unpol}\;\bigg[\;1\hqmm
  &+\;\Sigma^\text{lin}(\omega,\theta)\;P^{(\gamma)}_\text{lin}\;\cos2\philin
    \;+\;T_{11}(\omega,\theta)\;
    \PthreeHe\;d^1_{10}(\thetan)\;\sin[\phin]\non\\
    &&+ \sum\limits_{M=0,1}\;T_{1M}^\text{circ}(\omega,\theta)\;
    \PthreeHe\;P^{(\gamma)}_\text{circ}\;d^1_{M0}(\thetan)\;
    \cos[M\phin]\\
    &&+ \sum\limits_{-1\le M\le1}\;T_{1M}^\text{lin}(\omega,\theta)\;
    \PthreeHe\;P^{(\gamma)}_\text{lin}\;d^1_{M0}(\thetan)\;
    \sin[M\phin-2\philin] \bigg]\non\\
   =\hqmm&\dis\left.\frac{\dd\sigma}{\dd\Omega}
    \right|_\text{unpol}\;\bigg[\;1\hqmm
  &+\;\xi_3\;\Sigma_3(\omega,\theta)\;
    +\;\PthreeHe\,n_y\;\Sigma_y(\omega,\theta)\,\nonumber\\
    \label{eq:babcross}
  &&
  +\;\PthreeHe\,\xi_1\Big(n_x\;\Sigma_{1x}(\omega,\theta)+
  n_z\;\Sigma_{1z}(\omega,\theta)\Big)\\
  &&+\;\PthreeHe\,\xi_2\Big(n_x\;\Sigma_{2x}(\omega,\theta)
  +n_z\;\Sigma_{2z}(\omega,\theta)\Big)
  \;+\;\PthreeHe\,\xi_3\;n_y\;\Sigma_{3y}(\omega,\theta)
  \bigg]\, .\non
\end{eqnarray}
The first one, eq.~\eqref{eq:arencross}, is based on the general multipole
decomposition, and was adapted to deuteron Compton scattering in
refs.~\cite{Griesshammer:2013vga, erratum2}. Its independent observables
$T_{JM}^X$ carry the target multipolarity $(J,M)$ as subscript and the beam
polarisation $X\in\{\mbox{none, circ, lin}\}$ as superscript, and naturally
extend to arbitrary-spin targets.

The second one, eq.~\eqref{eq:babcross}, by Babusci \etal
\cite{Babusci:1998ww}, uses a Cartesian basis and the
components\footnote{Babusci \etal denote them by
  $\zeta_i$~\cite{Babusci:1998ww}.}  $n_i$ of the polarisation vector
$\vec{n}$. It uses the indices of the Stokes parameter $\xi_i$ and the target
polarisation direction $n_\alpha$ as the labels of the asymmetries
$\Sigma_{i\alpha}$. This form is more convenient to translate rate-difference
experiments since typically only one of the parameters $\xi_i$ is nonzero.

In either case, the cross section for Compton processes on spin-$\half$
targets without detecting final-state polarisations is fully parametrised by
$8$ linearly independent functions listed below. Here, $\dd \sigma$ is
shorthand for ${\dd\sigma}/{\dd\Omega}$; superscripts refer to photon
polarisations (``$\parallel$'' for polarisation in the scattering plane,
``$\perp$'' for perpendicular to it); subscripts to target polarisations; and
the absence of either means unpolarised. The observables are:
\begin{itemize}

\item $1$ differential cross section
  $\dis\left.\frac{\dd\sigma}{\dd\Omega}\right|_\text{unpol}$ of unpolarised
  photons on an unpolarised target.

\item $1$ beam asymmetry of a linearly polarised beam on an unpolarised target:
\begin{equation} 
  \label{eq:beamasym}
  \Sigma^\text{lin}=\Sigma_3=\frac{\dd \sigma^{||}-\dd \sigma^{\perp}}{\dd \sigma^{||}+\dd \sigma^{\perp}}\, .
\end{equation}

\item $1$ vector target asymmetry for a target polarised out of the scattering
  plane along the $\pm y$ direction and an unpolarised beam:
\begin{equation} 
  \label{eq:T11def}
  T_{11}=-\sqrt{2}\;\Sigma_y=-\sqrt{2}\;\frac{\dd \sigma_{y}-\dd
    \sigma_{-y}}{\dd \sigma_{y}+\dd \sigma_{-y}}\;\; .
\end{equation}

\item $2$ double asymmetries of right-/left-circularly-polarised photons on a
  target polarised along the $\pm x$ or $\pm z$ directions:
\begin{equation}
  \label{eq:T1Xcirc}
  T^\text{circ}_{11}=-\sqrt{2}\;\Sigma_{2x}=-\sqrt{2}\;
  \frac{\dd\sigma^{R}_{x}-\dd\sigma^{L}_{x}}{\dd\sigma^{R}_x+\dd\sigma^{L}_{x}}
  \;\;,\;\;
  T^\text{circ}_{10}=\Sigma_{2z}=
  \frac{\dd\sigma^{R}_{z}-\dd\sigma^{L}_{z}}{\dd\sigma^{R}_z+\dd\sigma^{L}_{z}} \;\;.
\end{equation}

\item $3$ double asymmetries of linearly-polarised photons on a
  vector-polarised target:
\begin{equation}
  \begin{split}
  \label{eq:T1Xlin}
  T^\text{lin}_{1\pm1}=&(\pm\Sigma_{1x}-\Sigma_{3y})/\sqrt{2}\\
  &\mbox{ with }\Sigma_{1x}=
  \frac{\dd\sigma^{\pi/4}_{x}-\dd\sigma^{-\pi/4}_{x}}
  {\dd\sigma^{\pi/4}_x+\dd\sigma^{-\pi/4}_{x}}\;\;,\;\;
  \Sigma_{3y}=\frac{(\dd \sigma^{||}_y-\dd \sigma^{\perp}_y)-
    (\dd \sigma^{||}_{-y}-\dd \sigma^{\perp}_{-y})}
  {\dd \sigma^{||}_y+\dd \sigma^{\perp}_y+\dd \sigma^{||}_{-y}+\dd \sigma^{\perp}_{-y}}\;\;,
  \\T^\text{lin}_{10}=&-\Sigma_{1z}=
  -\frac{\dd\sigma^{\pi/4}_{z}-\dd\sigma^{-\pi/4}_{z}}
  {\dd\sigma^{\pi/4}_z+\dd\sigma^{-\pi/4}_{z}}\;\; .
\end{split}
\end{equation}
\end{itemize}
The decompositions of eqs.~\eqref{eq:arencross} and~\eqref{eq:babcross} hold
in any frame, but the functions are frame-dependent.

The differences of the rates, $\Delta_\alpha$, for the different orientations
associated with each asymmetry are important to facilitate run-time
estimates. These are the numerators in eqs.~\eqref{eq:beamasym}
to~\eqref{eq:T1Xlin} and can most conveniently be expressed in the Babusci
basis:
\begin{equation}
  \label{eq:deltamatrix}
  \Delta_\alpha
  =g\;\Sigma_\alpha\;\left.\frac{\dd\sigma}{\dd\Omega}\right|_{\text{unpol}}\;\;,
\end{equation}
with $g=4$ for $\Delta_{3y}$ and $g=2$ for all other
asymmetries~\cite{Griesshammer:2017txw}.

Finally, we provide the translations to the notations used by Hildebrandt et
al.~\cite{Hildebrandt:2005ix, Hildebrandt:2003md} and Shukla et
al.~\cite{ShuklaPhD, Choudhury:2007bh,Shukla:2008zc}. The asymmetries are
\begin{equation}
  \begin{split}
    \Sigma_{2x}&=-\frac{T^\text{circ}_{11}}{\sqrt{2}}=
    \Sigma_x^\text{\cite{Hildebrandt:2005ix, Hildebrandt:2003md,ShuklaPhD, Choudhury:2007bh,Shukla:2008zc}}\;\;,\\
    \Sigma_{2z}&=T^\text{circ}_{10}=
    \Sigma_z^\text{\cite{Hildebrandt:2005ix, Hildebrandt:2003md,ShuklaPhD, Choudhury:2007bh,Shukla:2008zc}}\;\;,
  \end{split}
\end{equation} 
and the corresponding rates are
\begin{equation}
  \begin{split}
    \Delta_{2x}=2\Sigma_{2x}\left.\frac{\dd\sigma}{\dd\Omega} \right|_\text{unpol}&=
    -\sqrt{2}\;T^\text{circ}_{11}\left.\frac{\dd\sigma}{\dd\Omega}\right|_\text{unpol}=
    2\;\calD_x^\text{\cite{Hildebrandt:2005ix, Hildebrandt:2003md}}=
    \Delta_x^\text{\cite{ShuklaPhD, Choudhury:2007bh,Shukla:2008zc}}\;\;,
    \\
    \Delta_{2z}=2\Sigma_{2z}\left.\frac{\dd\sigma}{\dd\Omega} \right|_\text{unpol}&=2
    T^\text{circ}_{10}\left.\frac{\dd\sigma}{\dd\Omega}\right|_\text{unpol}=
    2\;\calD_z^\text{\cite{Hildebrandt:2005ix, Hildebrandt:2003md}}=
    \Delta_z^\text{\cite{ShuklaPhD, Choudhury:2007bh,Shukla:2008zc}}\;\;.
  \end{split}
\end{equation}

\subsection{Translating Amplitudes into Observables}
\label{sec:matching}

The Compton matrix elements of sect.~\ref{sec:MEs} are provided in the basis
of spin projections and photon helicities (dependencies on $\omega,\theta$ and
other parameters are dropped for brevity in this section). For ease of
presentation, we abbreviate the sum over all polarisations of the squared
amplitude:
\begin{equation}
  \label{eq:cala}
  |\calA|^2 \equiv \sum\limits_{M_i,\lambda_i} |A_{M_i\lambda_i}|^2\equiv
  \sum\limits_{M_f,\lambda_f; M_i,\lambda_i}
  |A^{M_f\lambda_f}_{M_i\lambda_i}|^2\;\;.
\end{equation} 
Note that we also suppress the
indices and summations for straightforward final-state-polarisation sums, as
  indicated in the second half of Eq.~(\ref{eq:cala}).

By inserting the density matrices of eqs.~\eqref{eq:photonpol}
and~\eqref{eq:targetpol} into eq.~\eqref{eq:xsect}, one obtains the cross
section in terms of the amplitudes, as a function of the photon polarisations
$P^{(\gamma)}_\text{circ}$ and $P^{(\gamma)}_\text{lin}$ with polarisation
angle $\philin$ and \threeHe polarisation $\PthreeHe$ with
orientation $(\thetan,\phin)$. The functional dependence of the result on
these parameters is easily matched to the parametrisation in
eq.~\eqref{eq:arencross}. For the unpolarised part, the result is:
\begin{equation}
  \label{eq:cross}
  \left.\frac{\dd\sigma}{\dd\Omega}\right|_\text{unpol}=\frac{\Phi^2}{4}
  |\calA|^2\; .
\end{equation}
The factor $\frac{1}{4}$ is familiar from averaging over initial spins and
helicities. The asymmetries\footnote{We correct a factor of $2$ in
  $T_{JM}$ and $T^\text{circ}_{JM}$ for $M>0$ in the corresponding spin-$1$
  results of ref.~\cite{Griesshammer:2013vga}; see
  erratum~\cite{erratum2}.}
\begin{align}
  \label{eq:sigmalin}
  \Sigma^\text{lin}\;|\calA|^2&=-\sum\limits_{M_i,\lambda_i}
                                  A_{M_i\lambda_i}A^*_{M_i,-\lambda_i}\; ,\\
  \label{eq:T11}
  T_{11}\;|\calA|^2&=2\sqrt{6}\;\ii
                       \sum\limits_{M_i,M_i^\prime,\lambda_i}  (-)^{\half-M_i}
                       \threej{\half}{\half}{1}{M_i}{-M_i^\prime}{-1}
                       A_{M_i^\prime\lambda_i}A^*_{M_i\lambda_i}\; ,\\
  \label{eq:Tcirc}
  T_{1M}^\text{circ}\;|\calA|^2&=(2-\delta_{M0})\;\sqrt{6}\;
                                   \sum\limits_{M_i,M_i^\prime,\lambda_i}  (-)^{\half-M_i}\;\lambda_i
                                   \threej{\half}{\half}{1}{M_i}{-M_i^\prime}{-M}
                                   A_{M_i^\prime\lambda_i}A^*_{M_i\lambda_i}\; ,\\
  \label{eq:Tlin}
  T_{1M}^\text{lin}\;|\calA|^2&=-\sqrt{6}\;\ii
                                  \sum\limits_{M_i,M_i^\prime,\lambda_i}
                                  (-)^{\frac{1}{2}-M_i}\;
                                  \lambda_i^{M+1}
                                  \threej{\half}{\half}{1}{M_i}{-M_i^\prime}{-\lambda_iM}
                                  A_{M_i^\prime\lambda_i}A^*_{M_i,-\lambda_i}\; ,
\end{align}
can be translated into the Babusci basis using eqs.~\eqref{eq:beamasym} to
\eqref{eq:T1Xlin}.

Since the amplitudes are real below the first inelasticity and
$M_i,M_i^\prime\in\{\pm\half\}$, the occurrence of the imaginary unit in four
of the observables in eqs.~\eqref{eq:sigmalin} to~\eqref{eq:Tlin} indicates
that they are zero there, \ie below the first inelasticity,
\begin{equation}
  \label{eq:zerobelow}
  T_{11}\equiv0\;\;,\;\;
  T_{1(0,\pm1)}^\text{lin}\equiv0\;\;.
\end{equation}
This is equivalent to the statement that
 $\Sigma_y$, $\Sigma_{1x/z} $ and $\Sigma_{3y}$ are zero in this kinematic region.  For \threeHe, the first strong inelasticity starts with the knock-out
reaction $\gamma\,\threeHe\to\gamma pd$. However, in 
the regime we are concerned with, nuclear dissociation processes are
relatively small and our approach does not include them. In contradistinction,
the first appreciable strong-interaction inelasticity on the proton starts at
the one-pion production threshold. Hence, in what follows, we study the four observables for which our approach yields non-zero results below the pion-production threshold:
the differential cross section, $\Sigma^\text{lin}=\Sigma_3$,
$T_{11}^\text{circ}=-\sqrt{2}\;\Sigma_{2x}$, and
$T_{10}^\text{circ}=\Sigma_{2z}$.

\section{Results With Explicit \texorpdfstring{Delta}{Delta(1232)}}
\label{sec:results}

\subsection{Central Values and Variations of the Nucleon Polarisabilities}
\label{sec:varypols}

In this section, we present the results of our calculations, including
the sensitivity of several of the observables defined in the
previous section to neutron scalar and spin polarisabilities. 

Unless otherwise
stated, we use the $\calO(e^2\delta^3)$ amplitude described above, with the
\threeHe wave function in \ChiEFT calculated using the Idaho potential at
\NXLO{3} and the ``b'' version of the 3NI provided by A.~Nogga~\cite{No97,
  NoggaPrivComm}.  
We will, however, show some results with the
$\calO(e^2\delta^2)$ amplitude in order to exhibit the impact of adding explicit Deltas
to the calculation. 
At both $\calO(e^2\delta^2)$ and $\calO(e^2\delta^3)$ we set the
proton and neutron scalar polarisabilities to the central values of
eqs.~\eqref{eq:protonvalues} and~\eqref{eq:LundPRL}, respectively. Differences between the orders 
are thus guaranteed not to be contaminated by spurious dependencies on the
well-established Baldin sum rules. Indeed, the sensitivity of observables to
varying polarisabilities is nearly unaffected by the exact choice of their
central values.
However, for the central values of the spin polarisabilities for both proton and neutron, we use the values that are predicted by \ChiEFT
at the two orders under consideration~\cite{Griesshammer:2012we,
  Griesshammer:2015ahu}:
 \begin{align}
  \label{eq:LOvaluesspin}
  \calO(e^2\delta^2) \mbox{ (no Delta):}&\hq\gammaee=5\gammamm=-5\gammaem=-5\gammame=-5.6\;\;,\\
  \label{eq:NLOvalues}
  \calO(e^2\delta^3)\mbox{ (with Delta):}&\hq
  \gammaee=-5.1\;,\;\gammamm=3.1\;,\;\gammaem=\gammame=0.9\;\;.
\end{align}
Therefore, the amplitudes at these two orders differ by terms at
$\mathcal{O}(\omega^3)$.  At the next \ChiEFT order, $\calO(e^2\delta^4)$,
proton and neutron spin-polarisability values differ.  Results, including
theory uncertainties, can be found in ref.~\cite{Griesshammer:2015ahu}. The
corrections to the $\calO(e^2 \delta^3)$ values are less than $2$ canonical
units.

As discussed around eq.~\eqref{eq:protonvalues}, the scalar polarisabilities
of the proton are known to much better accuracy than the neutron ones.
Furthermore, given the current state of few-body theory, more
accurate values of proton polarisabilities will come from proton data than from $\gamma {}^3$He scattering.
Therefore, for our current study of $\threeHe$, we only consider variations of
the six \emph{neutron} scalar and spin polarisabilities about the baseline
values of eqs.~\eqref{eq:LundPRL} and \eqref{eq:NLOvalues}, and explore the
pattern of sensitivities to such variations across different observables.  As
in the study of deuteron asymmetries in refs.~\cite{Griesshammer:2010pz,
  Griesshammer:2013vga, erratum2}, we choose a variation of the
polarisabilities by $\pm2$ canonical units. This is roughly at the level of
the combined statistical, theoretical and Baldin sum rule induced
uncertainties of the scalar polarisabilities of the neutron, and also is about
the combined uncertainty of the spin
polarisabilities~\cite{Griesshammer:2017txw}. These changes are implemented by
adding the following term,
\begin{equation}
\begin{split}
  {T}^{\mathrm{var}}(\omega,\,z=\cos\thetacm)=& 4\pi\,\omega^2\,\bigg[
  [\delta\alpha_{E1}+z\,\delta\beta_{M1}]
  \,(\vec{\epsilon}^{\,\prime\ast}\cdot\vec{\epsilon})
  -\delta\beta_{M1}\,
  (\vec{\epsilon}^{\,\prime\ast}\cdot\hat{k})\,(\vec{\epsilon}\cdot\hat{k}^\prime)\\
  &-\ii\,[\delta\gamma_{E1E1}+z\,\delta\gamma_{M1M1}+\delta\gamma_{E1M2}
         +z\,\delta\gamma_{M1E2}]\,\omega\,\vec{\sigma}\cdot
         (\vec{\epsilon}^{\,\prime\ast}\times\vec{\epsilon})\\
  &+\ii\,[\delta\gamma_{M1E2}-\delta\gamma_{M1M1}]\,\omega\,\vec{\sigma}\cdot
         \left(\hat{k}^\prime\times\hat{k}\right)
         (\vec{\epsilon}^{\,\prime\ast}\cdot\vec{\epsilon})\\
  &+\ii\,\delta\gamma_{M1M1}\,\omega\,\vec{\sigma}\cdot
  \left[\left(\vec{\epsilon}^{\,\prime\ast}\times\hat{k}\right)
    (\vec{\epsilon}\cdot\hat{k}^\prime)-
    \left(\vec{\epsilon}\times\hat{k}^\prime\right)
    (\vec{\epsilon}^{\,\prime\ast}\cdot\hat{k})\right]\\
  &+\ii\,\delta\gamma_{E1M2}\,\omega\,\vec{\sigma}\cdot
  \left[\left(\vec{\epsilon}^{\,\prime\ast}\times\hat{k}^\prime\right)
    (\vec{\epsilon}\cdot\hat{k}^\prime)-
    \left(\vec{\epsilon}\times\hat{k}\right)
    (\vec{\epsilon}^{\,\prime\ast}\cdot\hat{k})\right]
\bigg]\;\;,
\label{eq:fit}
\end{split}
\end{equation}
to the single-nucleon amplitude in the $\gamma$N cm frame.  Note that changing
$\alphaen-\betamn$ by $+2$ units while keeping $\alphaen+\betamn$ fixed
translates to a concurrent variation of $\alphaen$ by $+1$ unit and $\betamn$
by $-1$ unit. Other variations  are quite well determined by linear
extrapolation from our results, since quadratic contributions of the
polarisability variations $\delta(\alphae,\betam,\gamma_i)$ are suppressed in
the squared amplitudes.

The sensitivities could be visualised using the ``heat maps'' employed for
proton observables in ref.~\cite{Griesshammer:2017txw}.  However, there is no
data for elastic Compton scattering on \threeHe and theory is currently
constrained to a smaller energy range,
$50\;\MeV\lesssim\omegalab\lesssim120\;\MeV$, where the rates and
sensitivities vary much less than they do in the wider energy range of the
proton study. For this exploratory study, we therefore decided to concentrate
on results at two extreme energies: $\omegalab\approx60\;\MeV$, where only
effects from varying the scalar polarisabilities are seen, and
$\omegalab\approx120\;\MeV$, where spin polarisabilities are contributing
appreciably as well.  More detailed questions about sensitivities and
constraints, such as on $\gamma_0$ and $\gamma_\pi$, are deferred to a future
study in which we extend the present formalism above the pion-production
threshold and one order further, \ie~to $\calO(e^2\delta^4)$~\cite{future3He}.

\subsection{Corrections to Previous Presentations}
\label{sec:erratum}

The analytic formulae for Feynman diagrams in
refs.~\cite{Shukla:2008zc,ShuklaPhD} are correct. However, the code that
calculated observables for Compton scattering from \threeHe contained the
following errors in the implementation of the two-nucleon Compton operator
that appears at $\calO(Q^3=e^2\delta^2)$ in the chiral expansion:
\begin{enumerate}
\item It failed to include the isospin dependence
  $(\tau^{(1)} \cdot \tau^{(2)} - \tau^{(1)}_3 \tau^{(2)}_3)/2$ of the
  two-nucleon operator; see ref.~\cite[eq.~(59)]{Shukla:2008zc}. This factor
  is $1$ for ``deuteron''-like pairs and $-1$ for isospin-$1$ $np$
  pairs. ($pp$ and $nn$ pairs do not contribute at this order.)
  
\item A factor of two was missed when coding the operator that produces
  transitions between pairs where the two-nucleon spin changes
  $s_{12}=0 \leftrightarrow s_{12}=1$.

\item There was a mistake in the implementation of the
  $s_{12}=0 \leftrightarrow s_{12}=1$ piece of the third diagram in
  fig.~\ref{fig:2Bdiagrams}.
\end{enumerate}

\begin{figure}[!htbp]
  \begin{center}
\includegraphics[width=0.45\linewidth]
{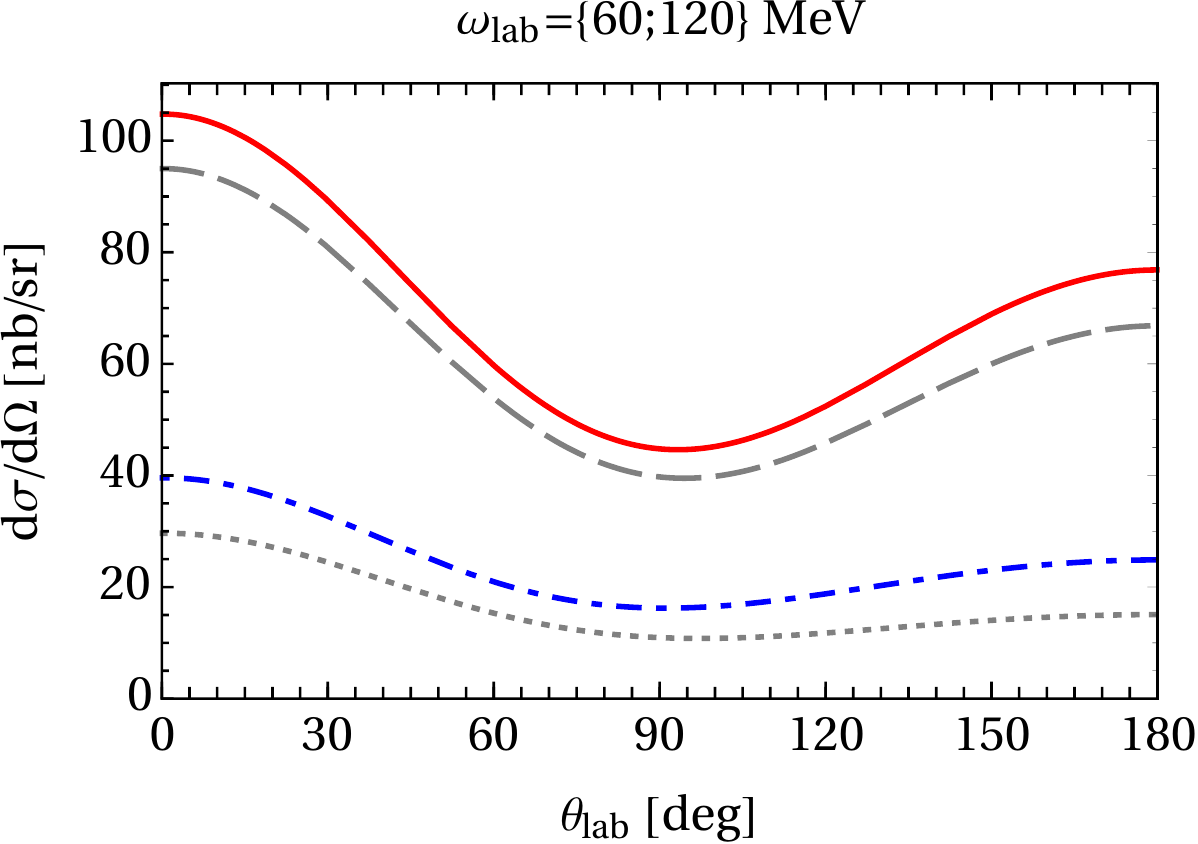}
\caption{(Colour on-line) The lab-frame \threeHe cross section at
  $\calO(e^2\delta^2)$ [\NXLO{2}, no explicit Delta] at
  $\omegalab=60\;\MeV$ after (solid red) and before (dashed gray) corrections;
  and at $\omegalab=120\;\MeV$ after (dot-dashed blue) and before (dotted
  gray) corrections. Here, and only here, we use the
  $\mathcal{O}(e^2 \delta^2)$ polarisability values for both protons and
  neutrons, $\alphaep=\alphaen=10\betamp=10\betamn=12.5$, in order to be
  consistent with the choice of refs.~\cite{ShuklaPhD,Choudhury:2007bh,Shukla:2008zc}.}
    \label{fig:erratum}
  \end{center}
\end{figure}

Of these, the first mistake had the largest consequences. The others only
affected parts of the matrix element that involve small pieces of the
\threeHe wave function. But rectifying the isospin factor increases the
prediction for the cross section markedly: by about $10\%$ at
$\omegalab\approx60\;\MeV$, and by as much as $40\%$ ($80\%$) at forward
(backward) angles at $120\;\MeV$. Figure~\ref{fig:erratum} compares the
corrected and original lab cross sections\footnote{The original papers presented cm cross
  sections, but this only makes a small difference.} at $60$
and $120\;\MeV$ at $\calO(e^2\delta^2)$ (without explicit Delta),
using the same parameters as the original
publications~\cite{Shukla:2008zc,ShuklaPhD}.  Corrections for
$\Sigma^\text{lin}=\Sigma_3$ are minimal;
$T^\text{circ}_{11}=-\sqrt{2}\,\Sigma_{2x}$ and
$T^\text{circ}_{10}=\Sigma_{2z}$ reduce by less than $5\%$ at
$\omegalab\approx60\;\MeV$ (rates increase by about $5\%$), and by less than
$14\%$ at $\omegalab\approx120\;\MeV$ (rates increase by $16\%$).

Previous investigations of deuteron Compton scattering~\cite{Beane:1999uq,
  Hildebrandt:2005ix, Hildebrandt:2004hh, Hildebrandt:2005iw,
  Choudhury:2004yz, Griesshammer:2010pz, Griesshammer:2013vga, erratum2} use
the same two-nucleon amplitudes, so it is important to stress that these
issues only affect the \threeHe calculation, and not that for the deuteron ($s_{12}=1$, $t_{12}=0$).

\subsection{Differential Cross Section}
\label{sec:crosssection}

\begin{figure}[!htbp]
  \begin{center}
    \includegraphics[height=0.33\linewidth]
    {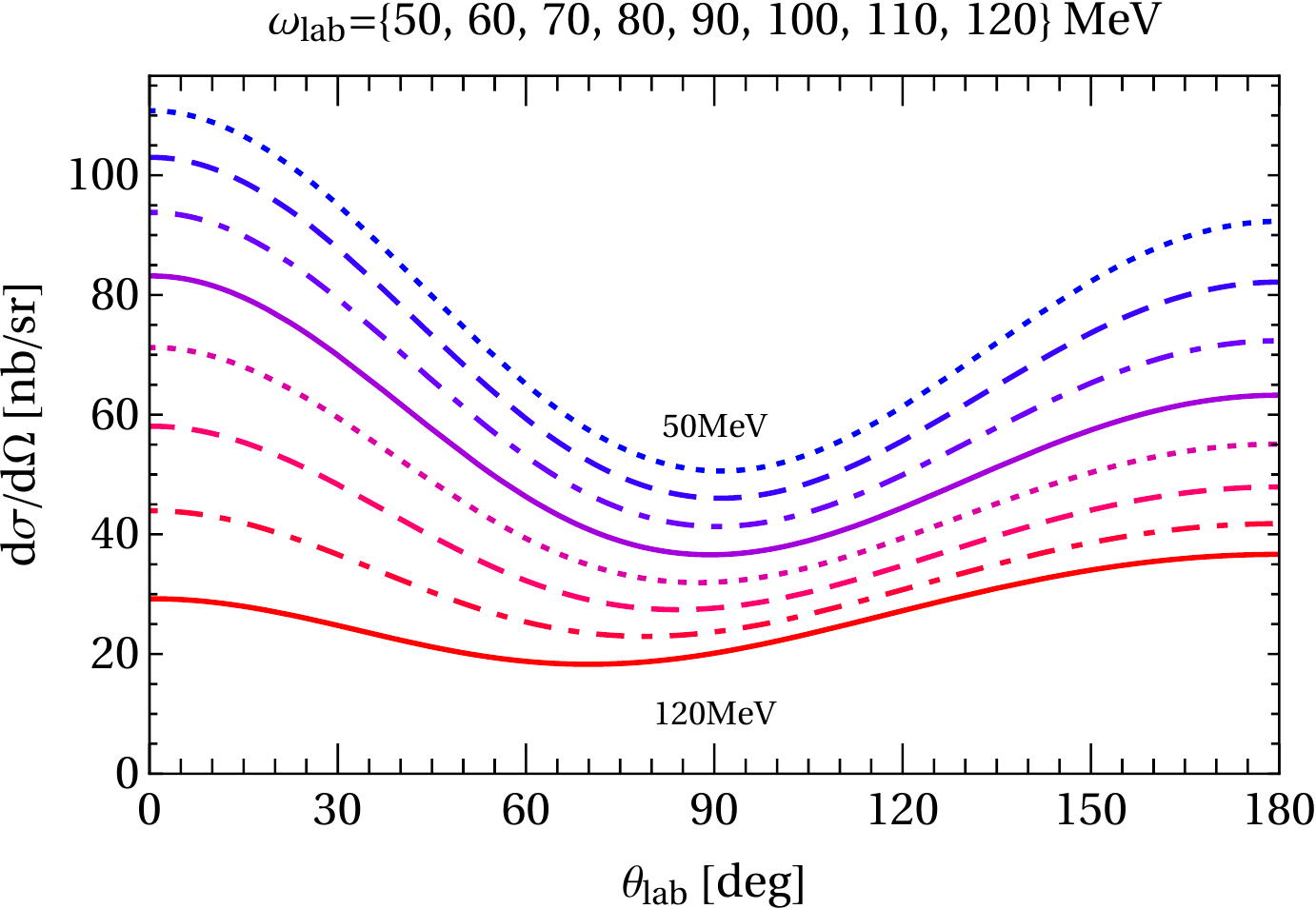}

    \caption{(Colour on-line) The \threeHe cross section at
      $\calO(e^2\delta^3)$, for $\omegalab$ between $50\;\MeV$ (top-most
      line) and $120\;\MeV$ (bottom-most line) in $10\;\MeV$ steps.}
    \label{fig:crosssect-sweep}
  \end{center}
\end{figure}

We now turn our attention to the $\calO(e^2 \delta^3)$ results for the elastic
cross section.  Figure~\ref{fig:crosssect-sweep} shows that there is a steady
decrease of the cross section between $50\;\MeV$ and $120\;\MeV$ for all
angles. Meanwhile, the forward-backward difference, which is about 20\% at
$50\;\MeV$, essentially disappears by $120\;\MeV$.

%
\begin{figure}[!htb]
  \begin{center}
    \includegraphics[height=0.33\linewidth]
    {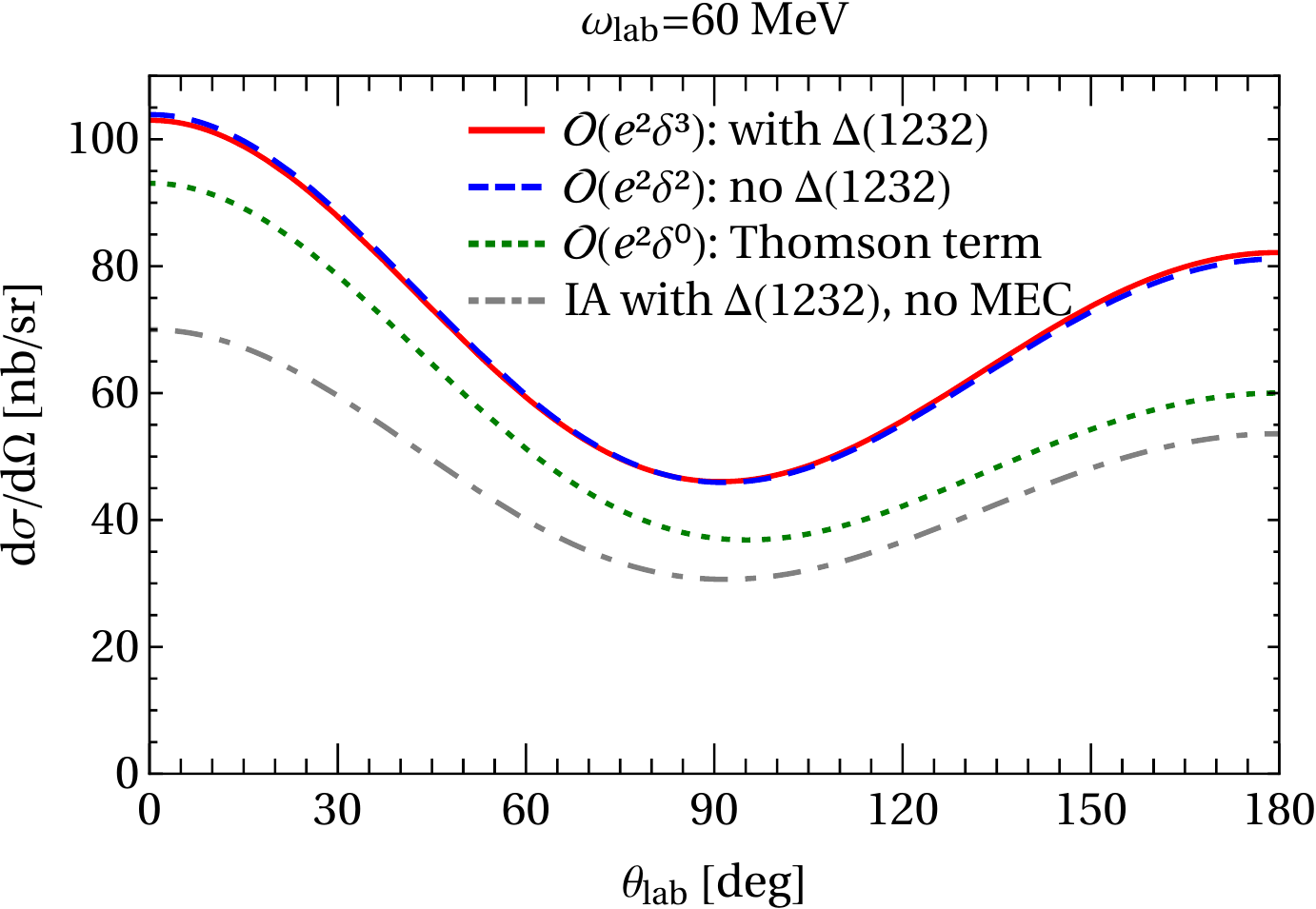}
\hq
    \includegraphics[height=0.33\linewidth]
    {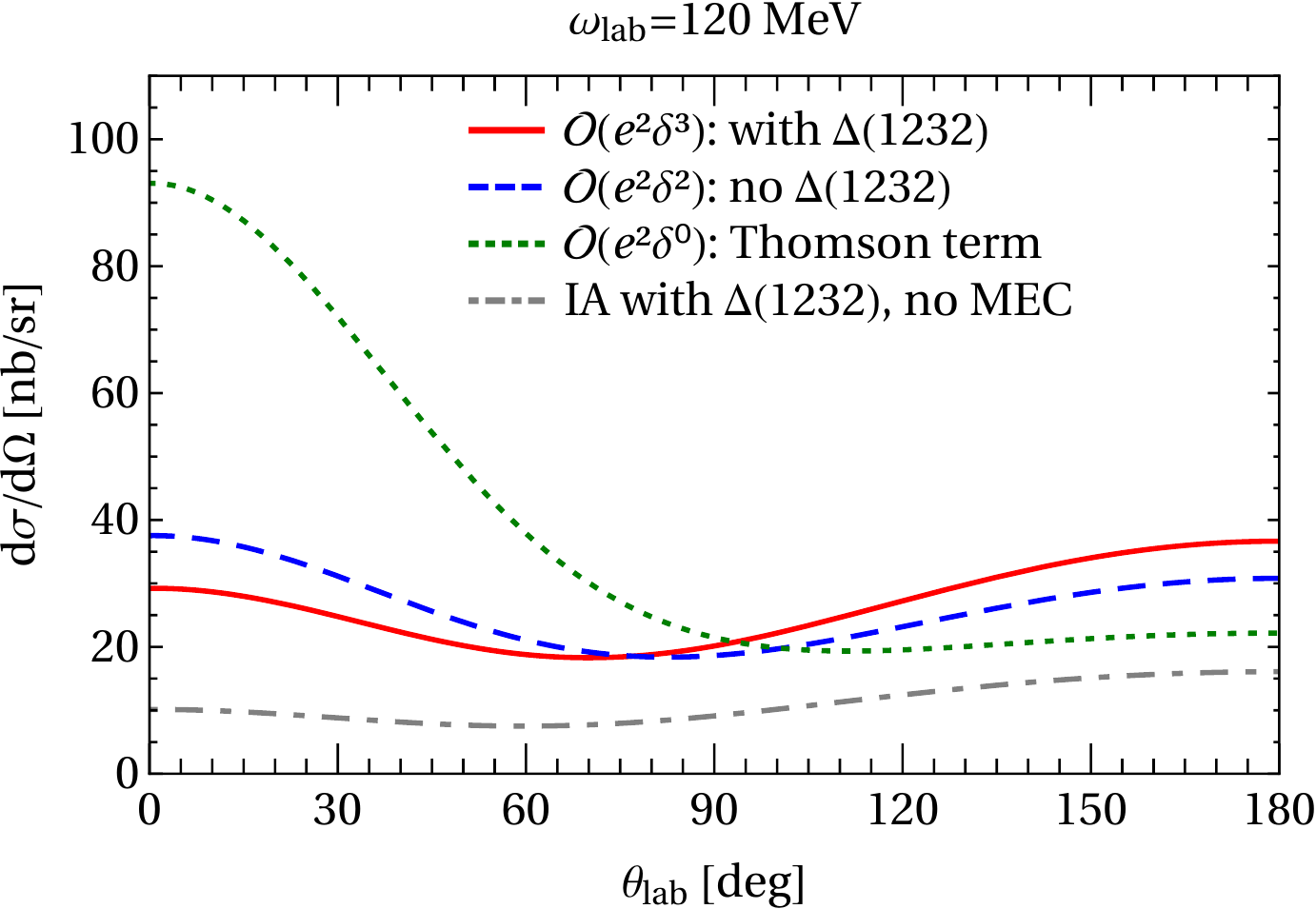}
    \caption{(Colour on-line) The \threeHe cross section at orders
      $e^2\delta^3$ (with Delta; red solid), $e^2\delta^2=Q^3$ (blue dashed)
      and $e^2\delta^0=Q^2$ (single-nucleon-Thomson term only; green dotted)
      at $\omegalab=60\;\MeV$ (left) and $120\;\MeV$ (right). We also show an
      incomplete ``Impulse Approximation'' calculation that includes
      $\calO(e^2\delta^3)$ single-nucleon (with Delta) amplitudes but not
      two-body currents (gray dot-dashed). Note that here, and in subsequent
      order-by-order comparisons, at 
      ${\cal O}(e^2\delta^2)$ and ${\cal O}(e^2\delta^3)$ the scalar polarisabilities are fixed to the same values,
      eqs.~\eqref{eq:LundPRL} and \eqref{eq:protonvalues}, but the spin
      polarisabilities differ; see eqs.~\eqref{eq:LOvaluesspin} and
      \eqref{eq:NLOvalues}.}
    \label{fig:crosssect-convergence}
  \end{center}
\end{figure}

Figure~\ref{fig:crosssect-convergence} shows that the order-by-order
convergence is good over the range of applicability. The
single-nucleon-Thomson term of fig.~\ref{fig:1Bdiagrams}(a) constitutes LO
[$\calO(e^2\delta^0)$] and indeed provides the bulk of the cross section.
Only for extreme forward scattering at the highest energies,
$\omegalab\gtrsim110\;\MeV$, does the next nonzero correction,
$\calO(e^2\delta^2)$, lead to a $60\%$ reduction. The order at which the Delta
enters [\NXLO{3}, $\calO(e^2\delta^3)$] provides a parametrically small
correction to \NXLO{2} [$\calO(e^2\delta^2)$] when the same values for the
scalar polarisabilities are chosen. At low energies, the variants with
[$\calO(e^2\delta^3)$] and without [$\calO(e^2\delta^2)$] explicit Delta are
near-indistinguishable. The difference is about $25\%$ at the highest energy
considered here, $\omegalab=120\;\MeV$. Note, though, that the angular
dependence is substantially changed once the Delta is included: backward-angle
scattering increases and forward-angle scattering decreases.

Such a strong signal from the $\Delta(1232)$ below the pion production
threshold might at first be surprising. However, even though the effect of the
Delta on the scalar polarisabilities is hidden by using their fitted values in
both orders, eqs.~\eqref{eq:LOvaluesspin} and~\eqref{eq:NLOvalues} show that
the value of $\gammamm$ differs at $\mathcal{O}(e^2 \delta^2)$ and
$\mathcal{O}(e^2 \delta^3)$, i.e., without and with the Delta. More important,
though, are the sizeable dispersive corrections to $\betam$ induced by the
$\Delta$. These cannot be absorbed into the static value of $\betam$. At
$\omega=120\;\MeV$, their impact on observables can be as large as that of
$\betam$ itself; see also the discussion and plots of ``dynamic
polarisabilities'' in refs.~\cite{Griesshammer:2012we, Griesshammer:2017txw}.

Figure~\ref{fig:crosssect-convergence} also includes the results of an
``impulse approximation'' (IA) calculation in which the two-body-current
contribution is artificially set to zero. This calculation is not consistent
with the \ChiEFT power counting, since this effect is $\calO(e^2
\delta^2)$.
Comparing the IA result to the $\mathcal{O}(e^2 \delta^0)$
(\ie~single-nucleon-Thomson only) result shows that structure effects in the
single-nucleon Compton amplitudes lead to significant suppression of the cross
section, just as for proton Compton scattering (particularly obvious in fig.~3
of ref.~\cite{Lensky:2008re}).  However, in the full $\calO(e^2 \delta^2)$
result at $60\;\MeV$, this suppression is more than compensated by two-body
currents that increase the \threeHe Compton cross section by roughly a factor
of two over the IA result. Two-body currents play an even larger fractional
role at higher energies. This confirms, yet again, the findings for Compton
scattering on \threeHe in refs.~\cite{Choudhury:2007bh, Shukla:2008zc,
  ShuklaPhD}, and on the deuteron in ref.~\cite{Beane:1999uq}, that two-body
currents must be included in any realistic description of few-nucleon Compton
scattering.

\begin{figure}[!htb]
  \begin{center}
    \includegraphics[height=0.34\linewidth]
    {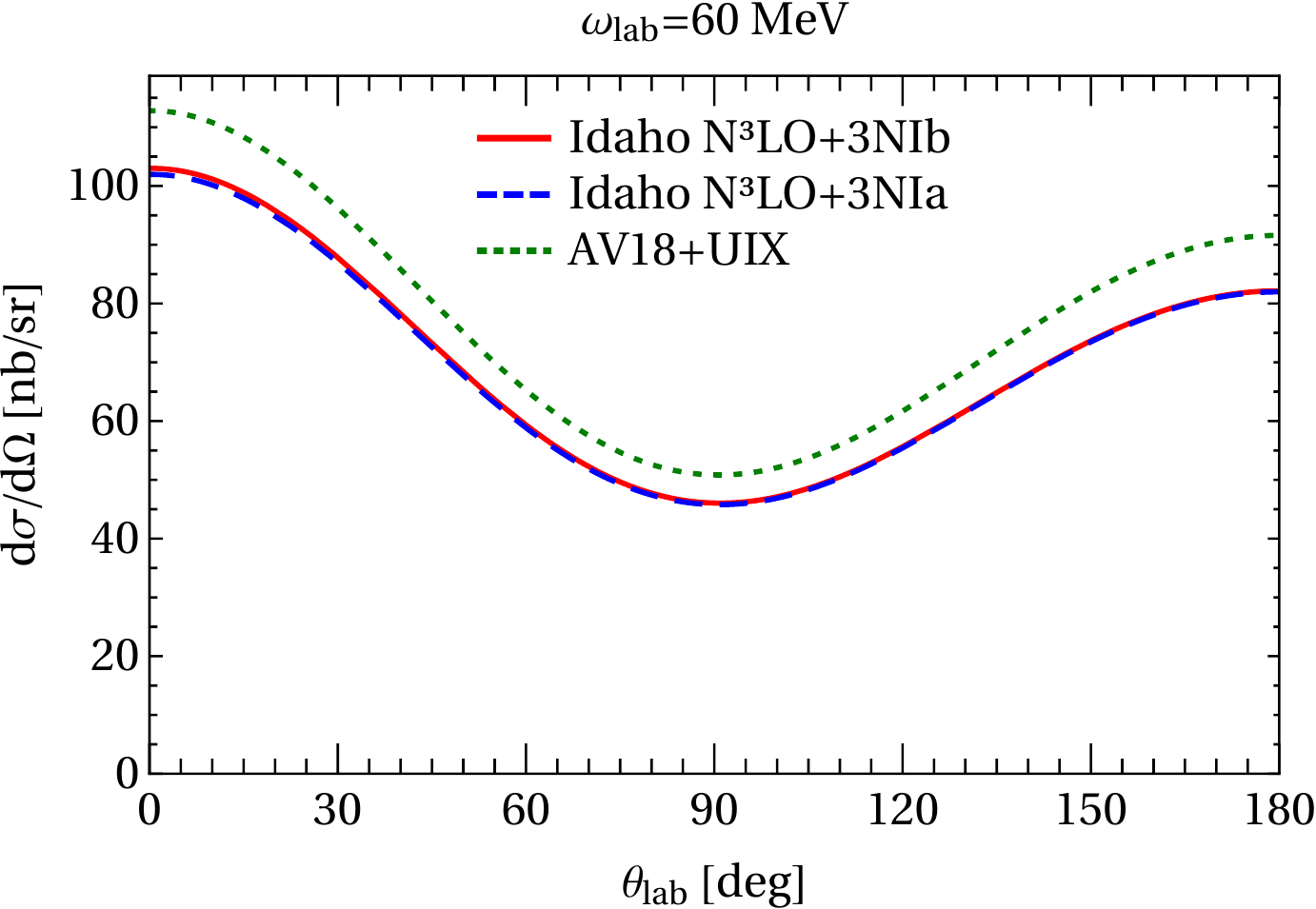}
\hq
    \includegraphics[height=0.34\linewidth]
    {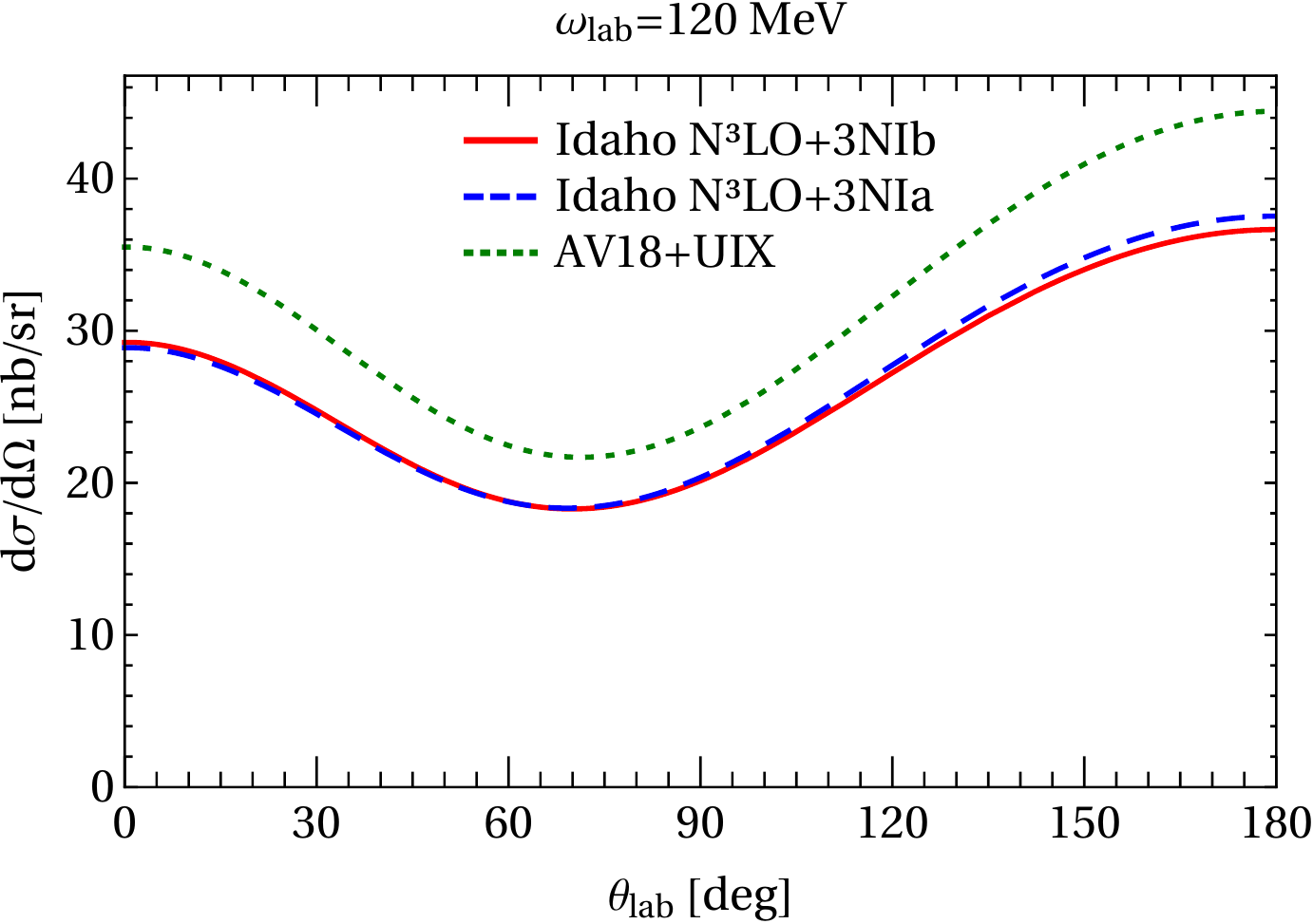}

    \caption{(Colour on-line) The \threeHe cross section for different
      \threeHe wave functions in the lab frame at $\omegalab=60\;\MeV$ (left)
      and $120\;\MeV$ (right): \ChiEFT's Idaho variant at \NXLO{3} with 3NI
      ``b'' (red solid); \ChiEFT's Idaho variant at \NXLO{3} with 3NI ``a''
      (blue dashed); AV18 with Urbana-IX 3NI (green dotted). Notice the
      difference of scales.}
    \label{fig:crosssect-wfdep}
  \end{center}
\end{figure}

\begin{figure}[!htbp]
  \begin{center}
    \includegraphics[width=0.45\linewidth]
    {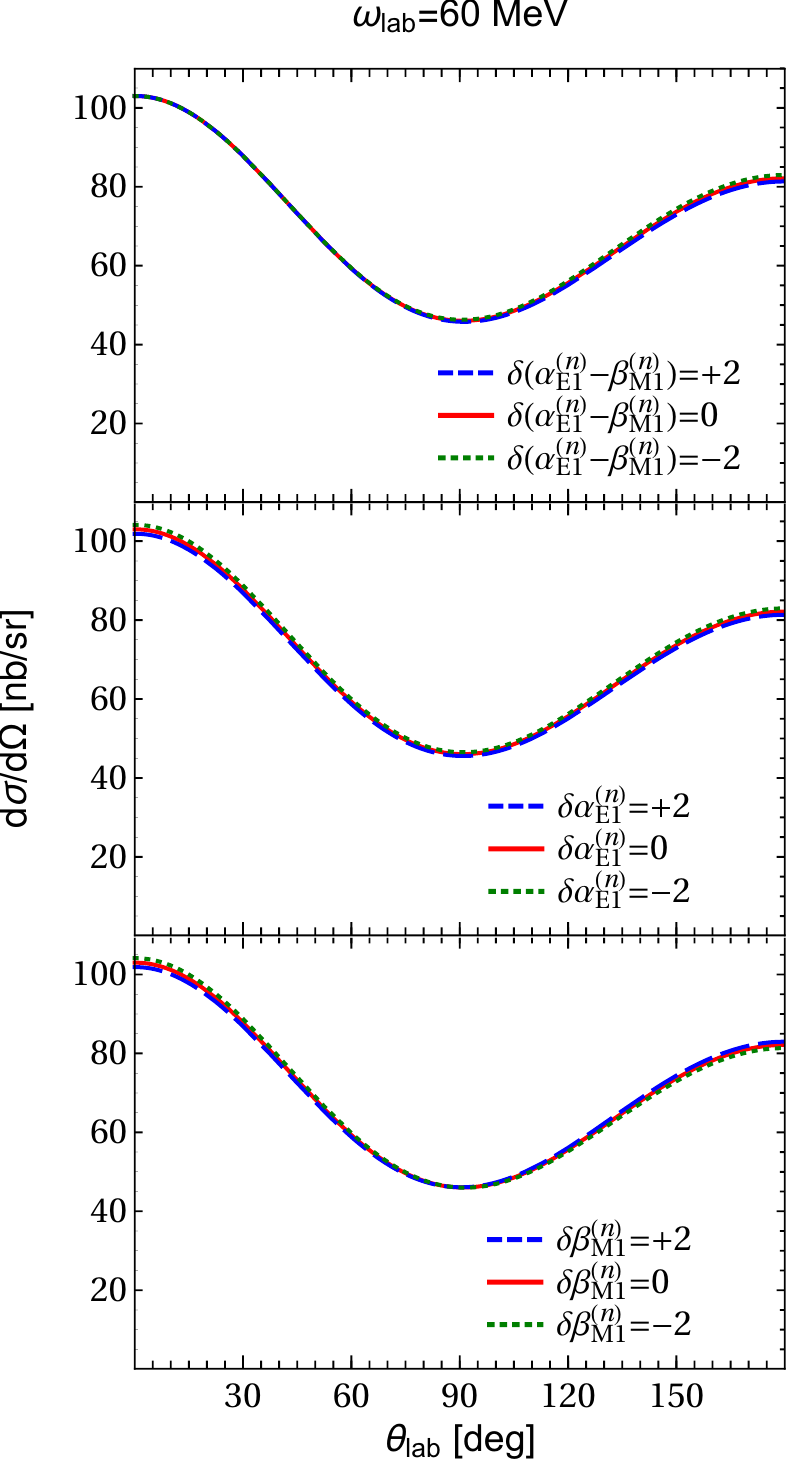}
\hqqq
    \includegraphics[width=0.45\linewidth] {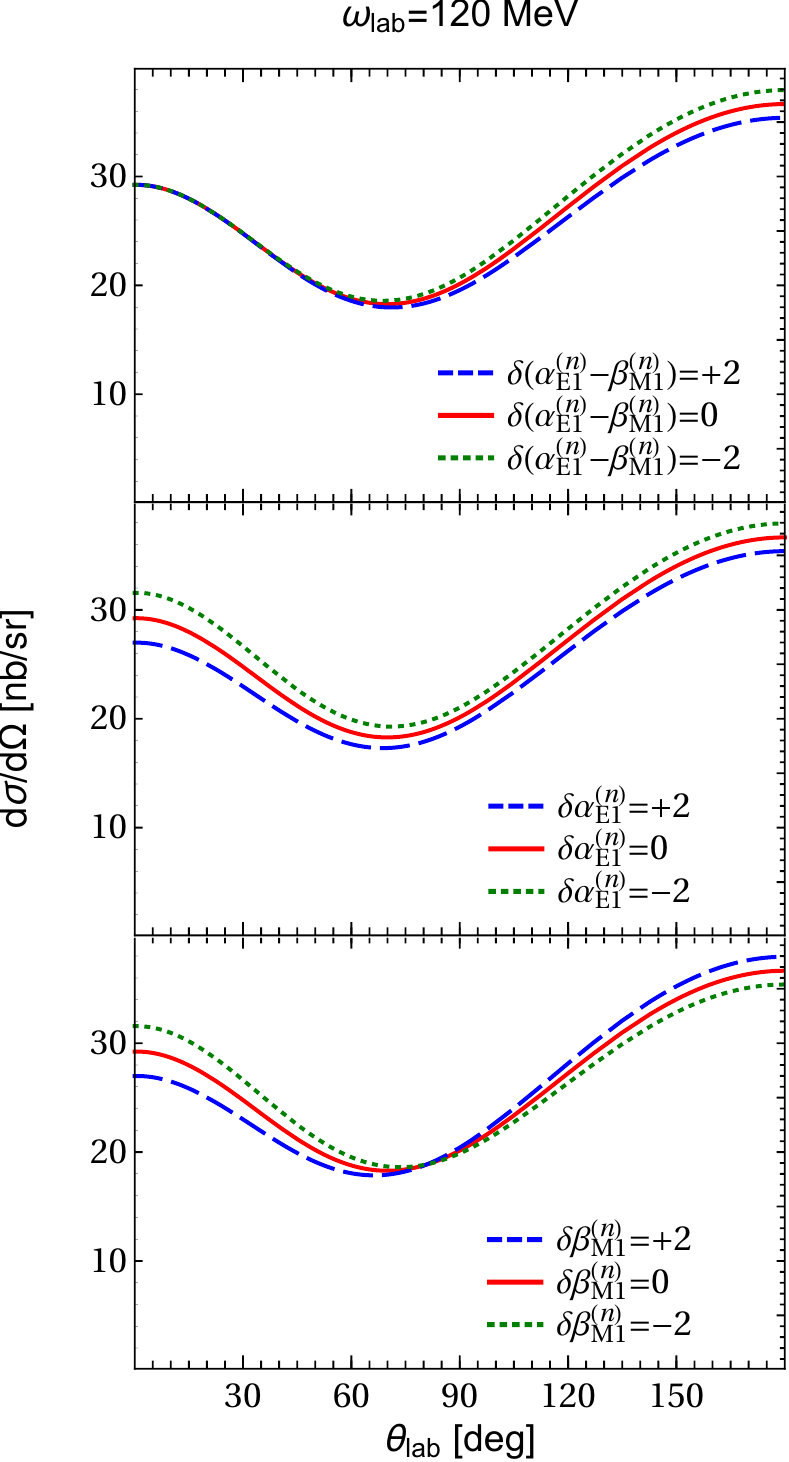}

    \caption{(Colour on-line) Sensitivity of the cross section to varying the
      scalar polarisabilities of the neutron $\alphaen-\betamn$ (top),
      $\alphaen$ (middle) and $\betamn$ (bottom) about their central values
      (\textcolor{red}{\protect\solid}) of eq.~\eqref{eq:LundPRL} by
      $+2$ (\textcolor{blue}{\protect\longdashed}) and $-2$
      (\textcolor{green}{\protect\dotted}) units, at $\omegalab=60\;\MeV$
      (left) and $120\;\MeV$ (right). Even at $\omegalab=120\;\MeV$, changes
      in the spin polarisabilities produces variations which are at most as
      large as the thickness of the line. Notice the
      difference of the scales.}
    \label{fig:crosssect-scalarpols}
  \end{center}
\end{figure}

In fig.~\ref{fig:crosssect-wfdep}, we compare the results for the three
different \threeHe wave functions discussed in sect.~\ref{sec:wavefunctions}:
\ChiEFT (Idaho formulation) at \NXLO{3} with 3NI ``b'' or ``a'', and AV18 with
the Urbana-IX 3NI. The wave functions of the two \ChiEFT variants yield
near-identical results. As anticipated, there is a larger difference between
these two and AV18 with the Urbana-IX 3NI: this wave function leads to cross
sections which are about 10\% larger at lower energies, and $\approx 20\%$
larger at $\omegalab=120\;\MeV$.    The effect of these discrepancies is
mitigated by the fact that the relative difference in the predicted cross
section in fig.~\ref{fig:crosssect-wfdep} is largely angle independent,
whereas we see in fig.~\ref{fig:crosssect-scalarpols} that the sensitivities
to the scalar polarisabilities have a rather strong angular dependence.

The difference induced by varying the scalar polarisabilities of the neutron
by $\pm2$ units at $\omegalab=60\;\MeV$ hardly exceeds the thickness of the
line. At $\omegalab=120\;\MeV$, such a variation results in cross section
variations of $\lesssim\pm8\%$ or $\pm1$nb/sr. Equivalent polarisability
variations in the deuteron case lead to changes in the cross section that are
smaller in absolute terms, but represent a larger fraction of the (much)
smaller cross section for that process.  At both energies, varying the spin
polarisabilities by up to $\pm100\%$ of their baseline values (or at least
$\pm2$ canonical units) produces variations which are at most as large as the
line thickness.

\clearpage
\subsection{Beam Asymmetry}
\label{sec:beamasymmetry}

\begin{figure}[!htbp]
  \begin{center}
    \includegraphics[height=0.33\linewidth]
{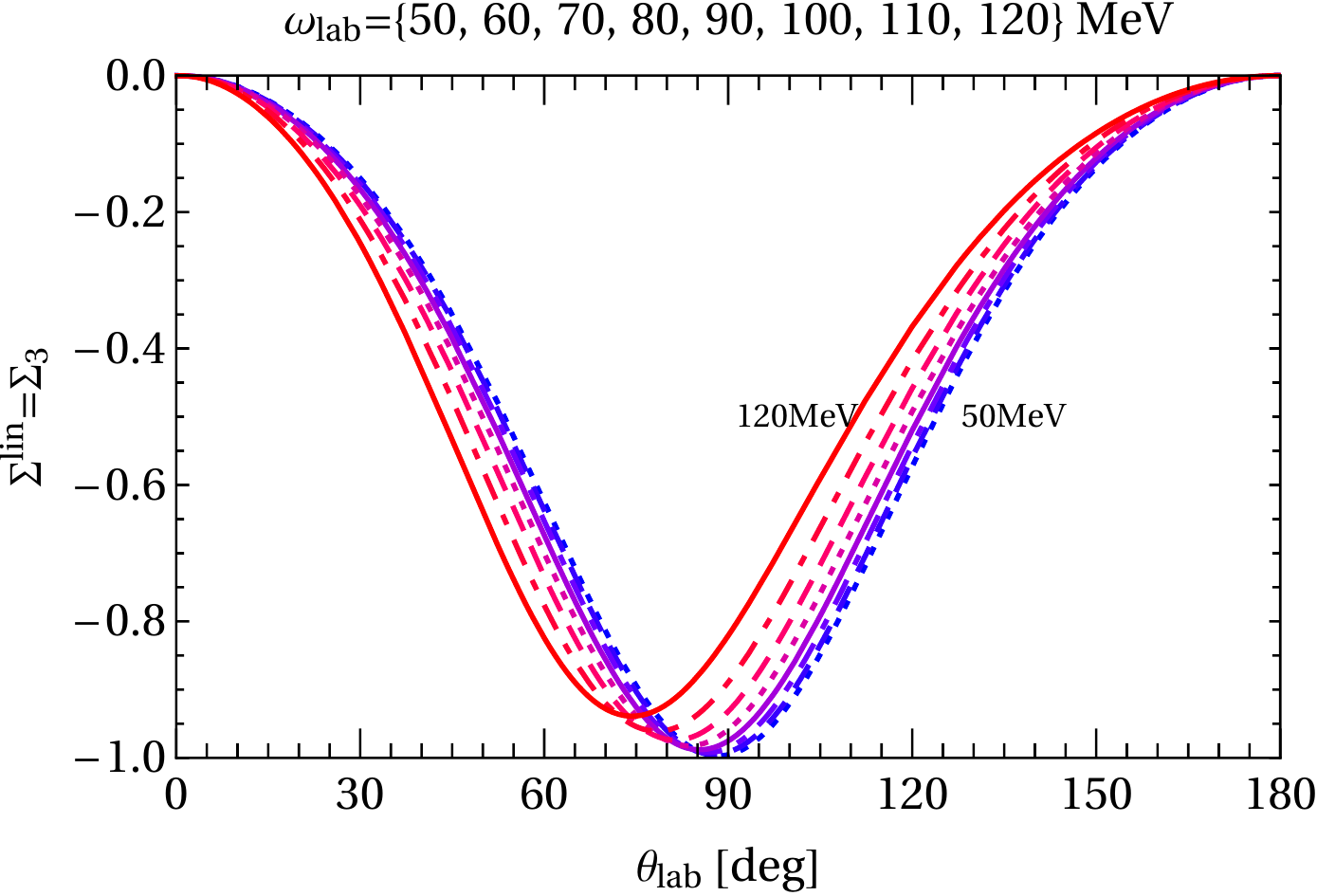}
    \includegraphics[height=0.33\linewidth]
{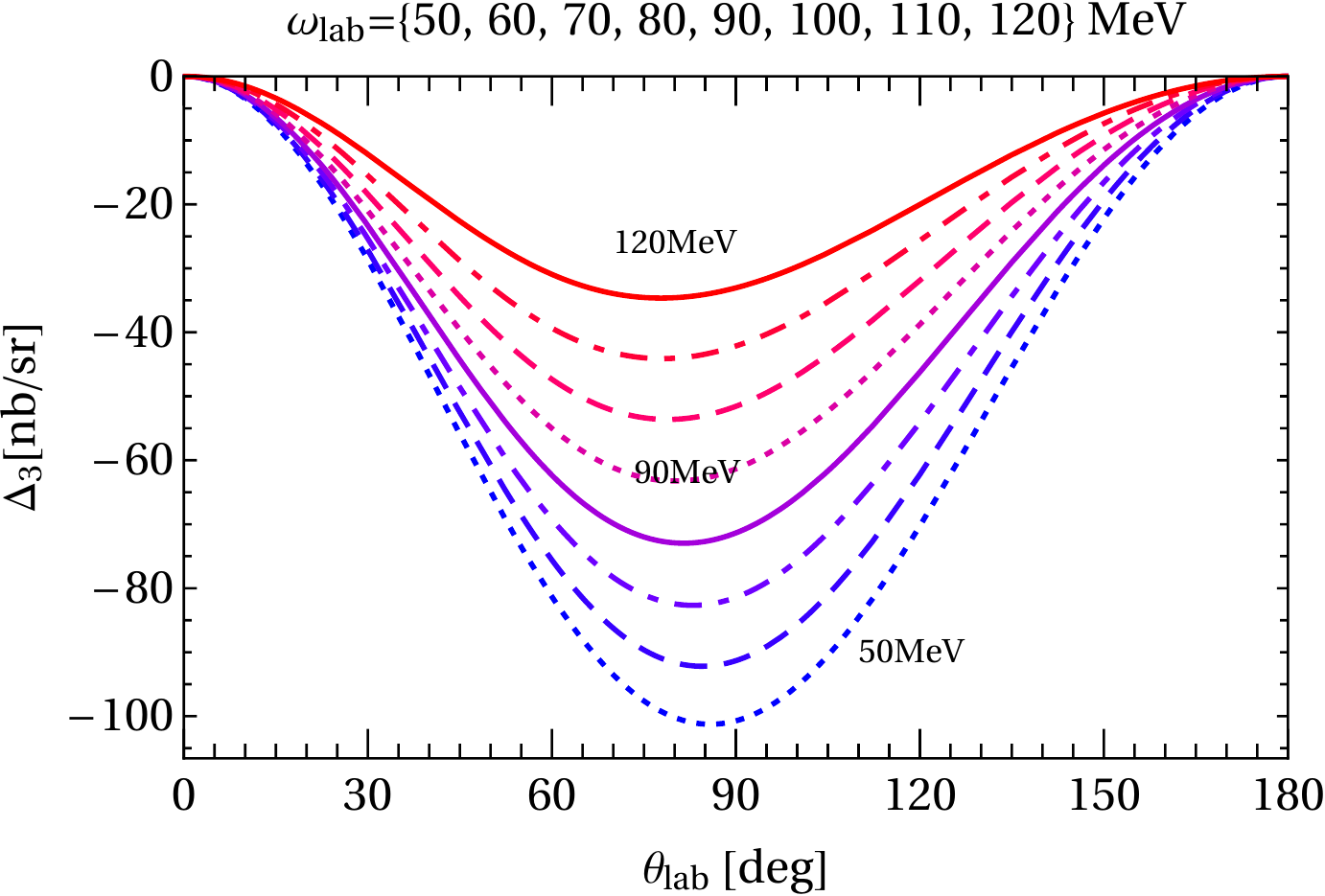}
\\[2ex]
    \includegraphics[height=0.33\linewidth]
    {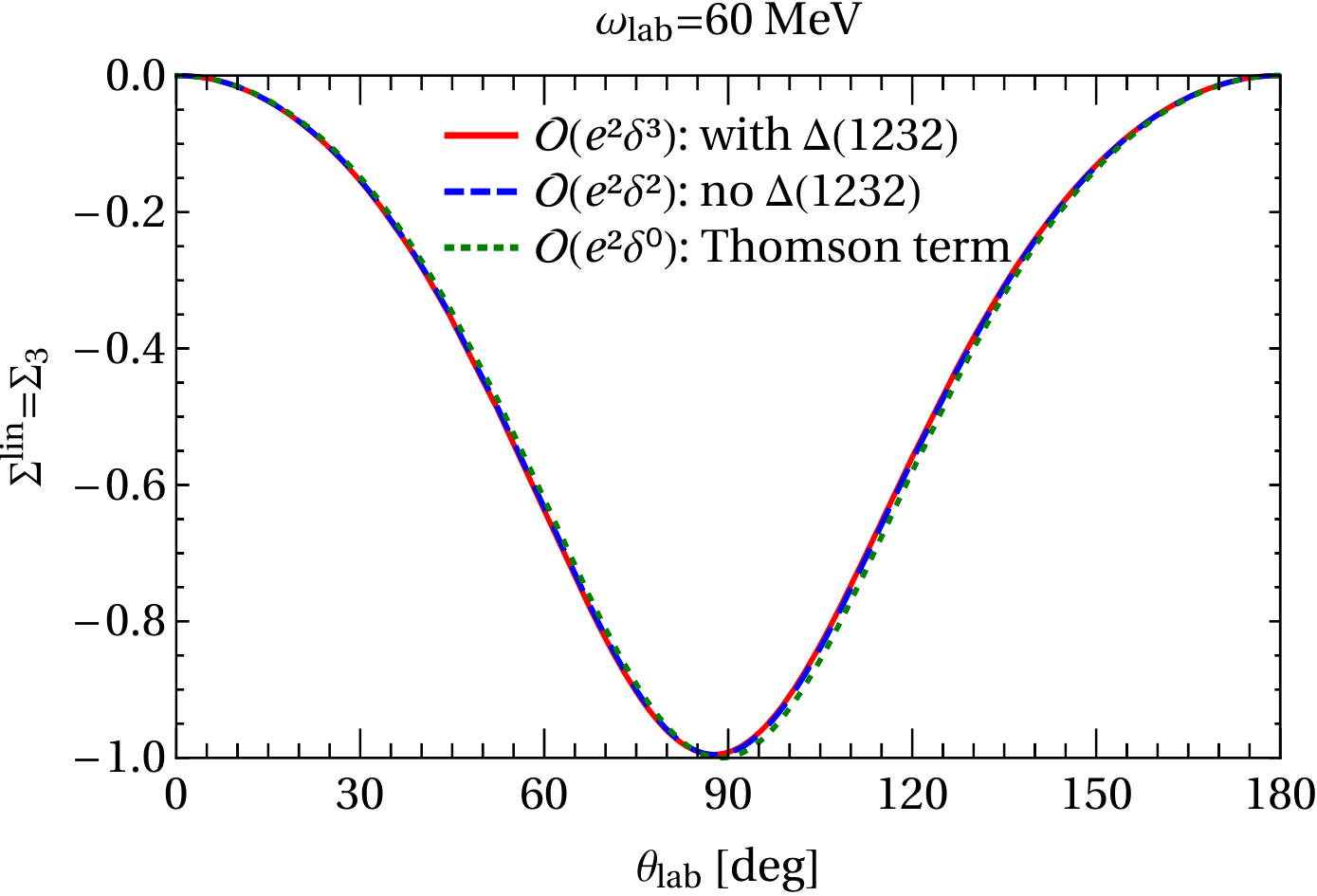}
    \includegraphics[height=0.33\linewidth]
    {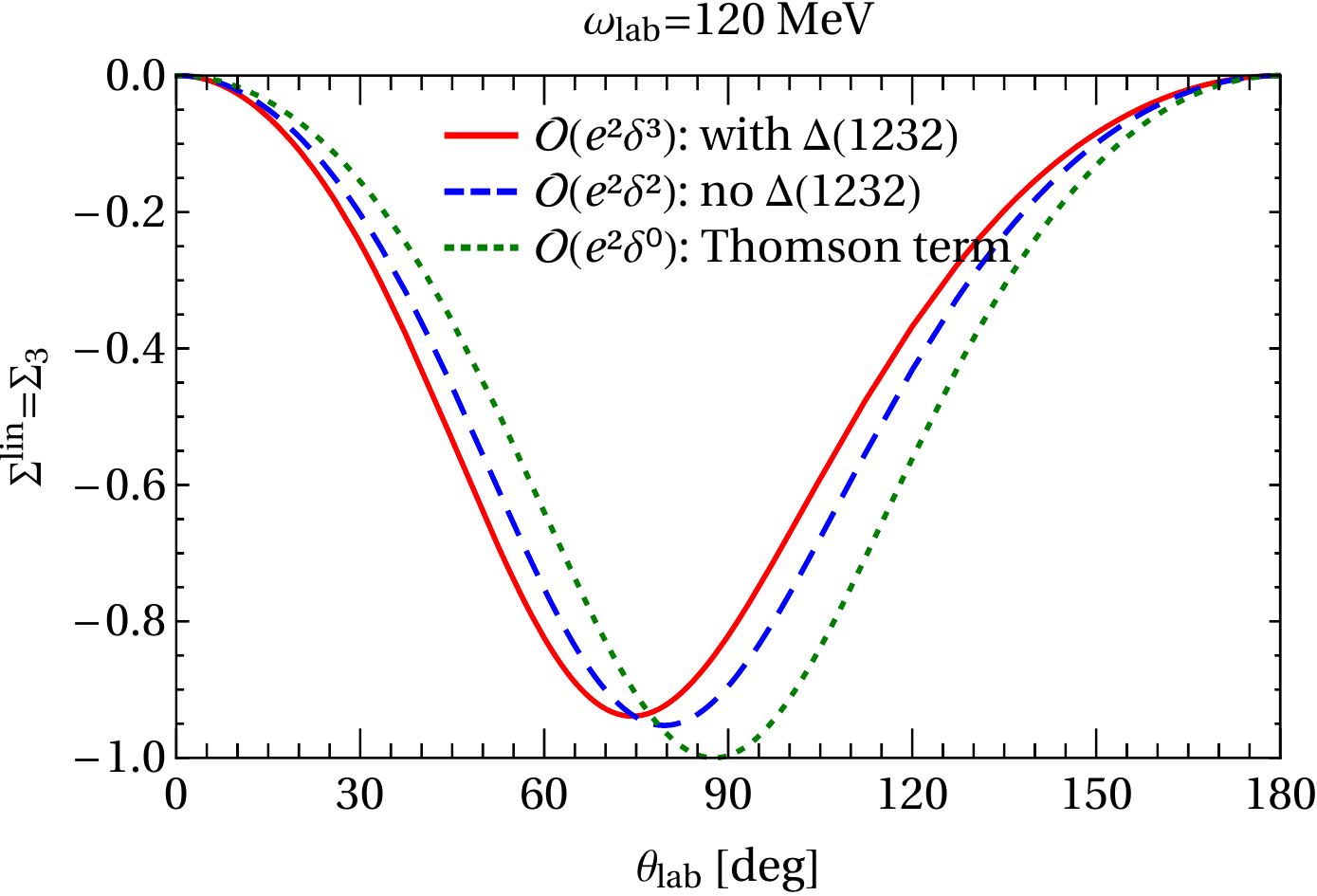}

    \caption{(Colour on-line) The beam asymmetry
      $\Sigma^\mathrm{lin}=\Sigma_3$ in the lab frame. Top: Energies between
      $\omegalab=50\;\MeV$ and $120\;\MeV$ in $10\;\MeV$ steps, for the
      asymmetry (left) and the corresponding rate difference (right).
      Bottom: order-by-order convergence at $\omegalab=60\;\MeV$ (left) and
      $120\;\MeV$ (right), with notation as in
      fig.~\ref{fig:crosssect-convergence}.}
\label{fig:beamasym-many}
\end{center}
\end{figure}

We start our discussion of the observables for polarised beams and/or targets
with the beam asymmetry $\Sigma^\mathrm{lin}=\Sigma_3$; see
eq.~\eqref{eq:beamasym}. As shown in the top row of
fig.~\ref{fig:beamasym-many}, its magnitude and shape do not change
significantly between $\omegalab=50\;\MeV$ and $120\;\MeV$.  The rate
difference $\Delta_3$ declines steadily: at $\omegalab=120\;\MeV$ it is one
third of its value at $50\;\MeV$. This is also by far the largest asymmetry,
near-saturating the extreme value of $-1$ around $\thetalab=90^\circ$.
%
%
As the second row of fig.~\ref{fig:beamasym-many} shows, these features follow
from the fact that the beam asymmetry is dominated by the single-nucleon Thomson term. This is
also why it is the only asymmetry that is nonzero for $\omega=0$: Compton
scattering on a charged point-like particle leads to the
$(\cos^2\theta-1)/(\cos^2\theta+1)$ shape at zero energy which is well-known
from Classical Electrodynamics~\cite{Jackson}.  Even at the highest energies
considered here, the nucleons' magnetic moments and structure change the asymmetry 
by less than $10\%$, indicating that this observable converges rapidly
in \ChiEFT. All this makes it unsurprising that the wave function dependence
is very small, and so we do not display it here.

\begin{figure}[!htbp]
  \begin{center}
    \includegraphics[width=0.85\linewidth] {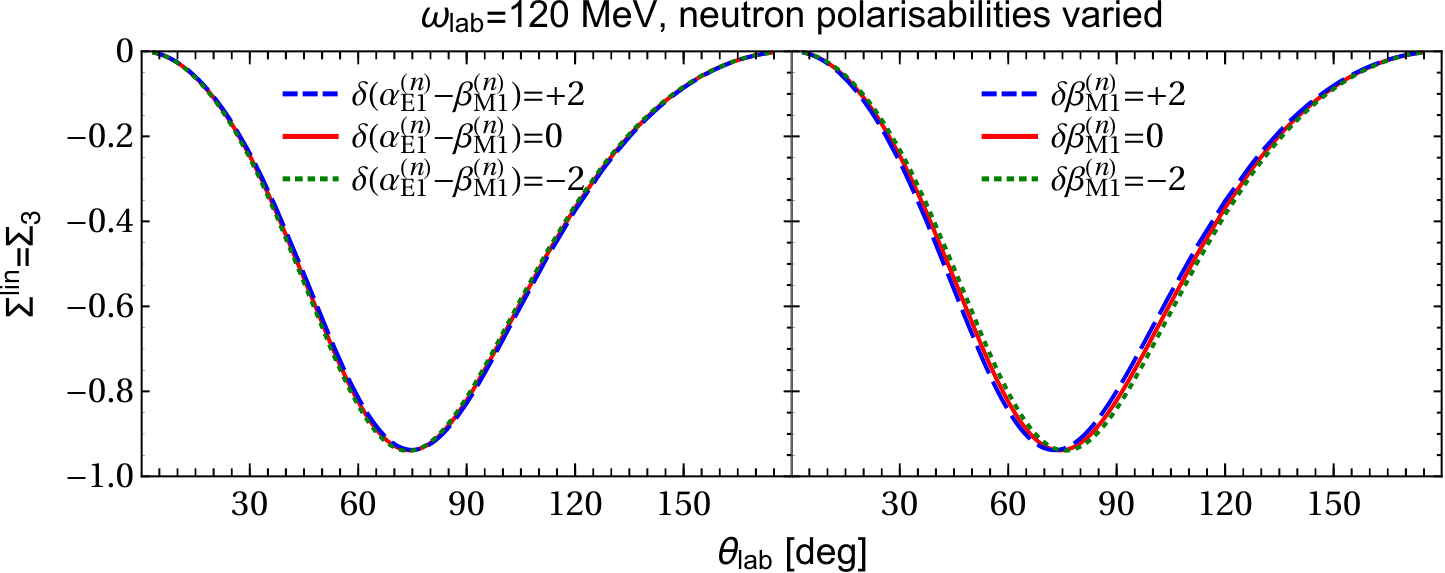}

    \caption{(Colour on-line) Sensitivity of $\Sigma^\mathrm{lin}=\Sigma_3$ to
      varying the scalar polarisabilities of the neutron $\alphaen-\betamn$
      (left) and $\betamn$ (right) about their central values
      (\textcolor{red}{\protect\solid}) of eqs.~\eqref{eq:LundPRL} by $+2$
      (\textcolor{blue}{\protect\longdashed}) and $-2$
      (\textcolor{green}{\protect\dotted}) units, at
      $\omegalab=120\;\MeV$. The effect of spin-polarisability variations is
      considerably exceeded by the thickness of the lines.}
    \label{fig:beamasym-scalarpols}
  \end{center}
\end{figure}

Given the persistence of this point-like behaviour, it should come as no
surprise that there is very little sensitivity to polarisabilities. In
fig.~\ref{fig:beamasym-scalarpols} we show what variation of
$\alphaen-\betamn$ and $\betamn$ do to $\Sigma^\mathrm{lin}=\Sigma_3$. Even
where the effect is largest, at $120\;\MeV$, it is still slight.  Varying the
electric scalar polarisabilities or the spin ones impacts the beam symmetry
even in extreme cases by hardly more than the thickness of the lines
(cf.~ref.~\cite{Choudhury:2004yz} for analytic arguments for this behaviour in
the deuteron case).

\subsection{Circularly Polarised Beam on Transversely Polarised Target}
\label{sec:Tcirc11}

\begin{figure}[!htbp]
  \begin{center}
    \includegraphics[height=0.3\linewidth]
{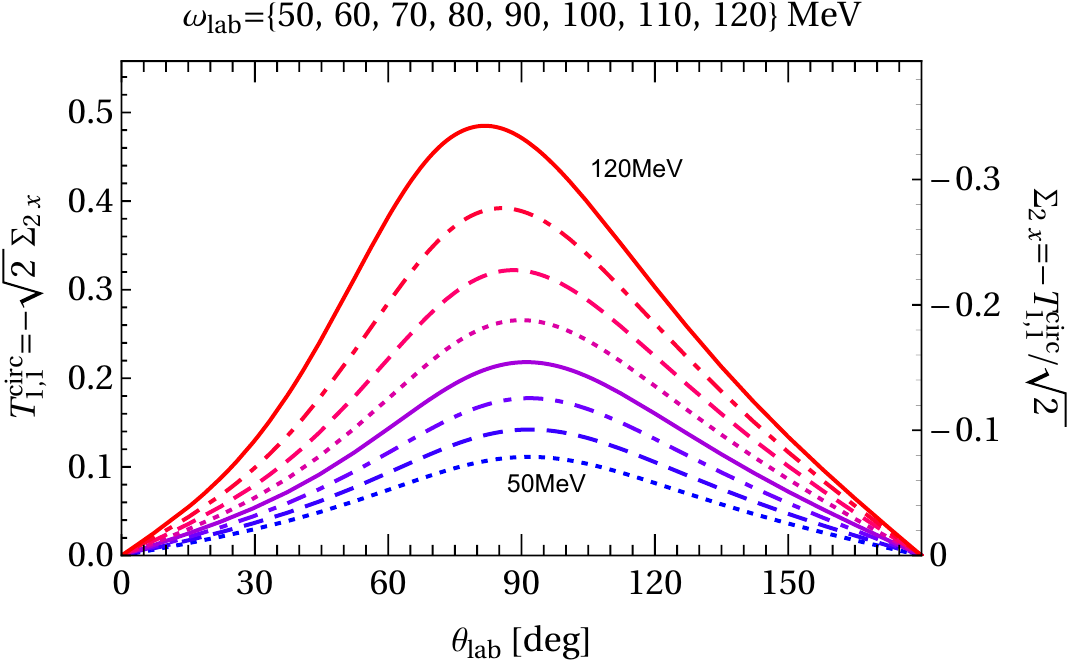}\hqq
    \includegraphics[height=0.3\linewidth]
{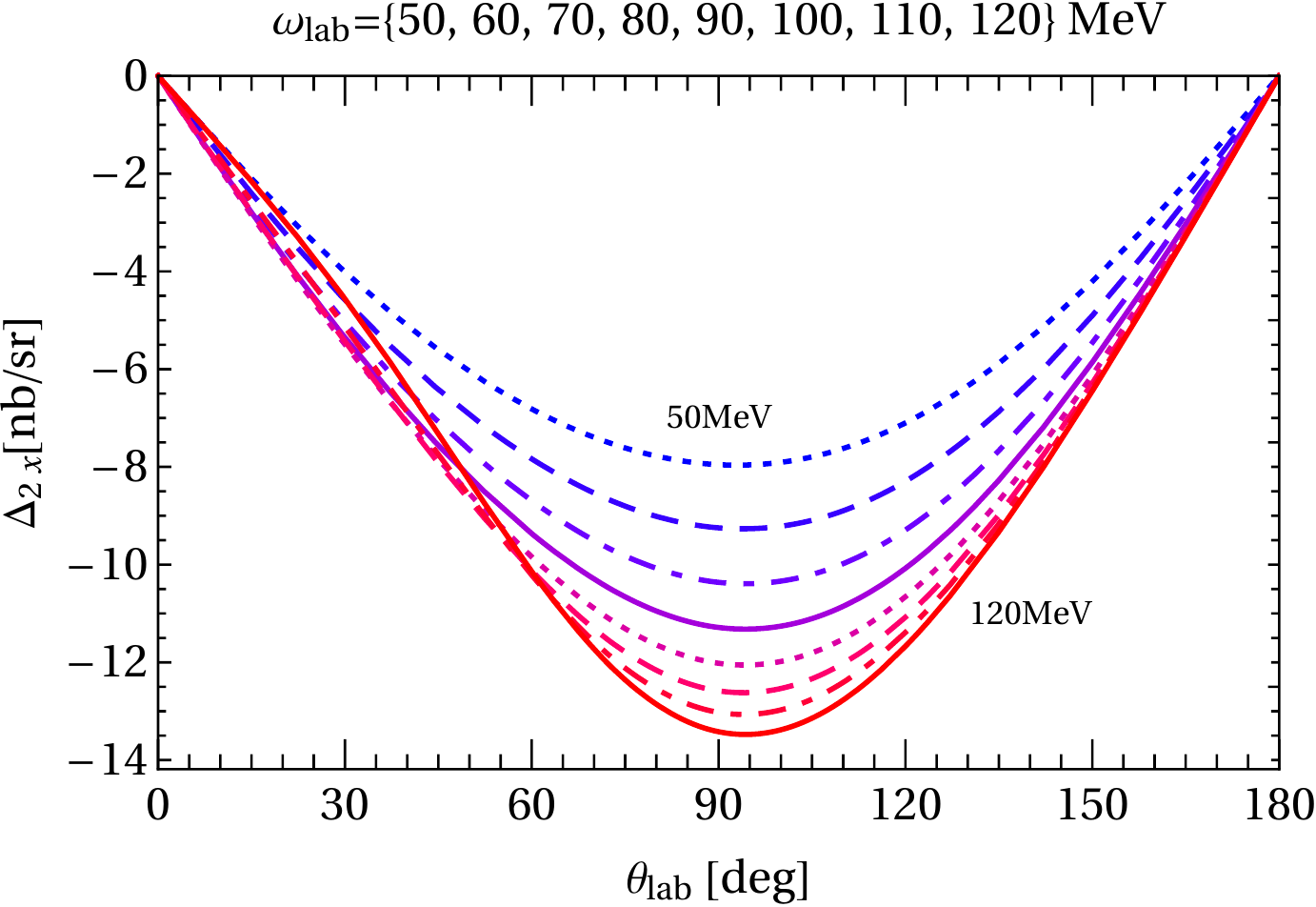}\hspace*{4.ex}
\\[2ex]
    \includegraphics[height=0.3\linewidth]
    {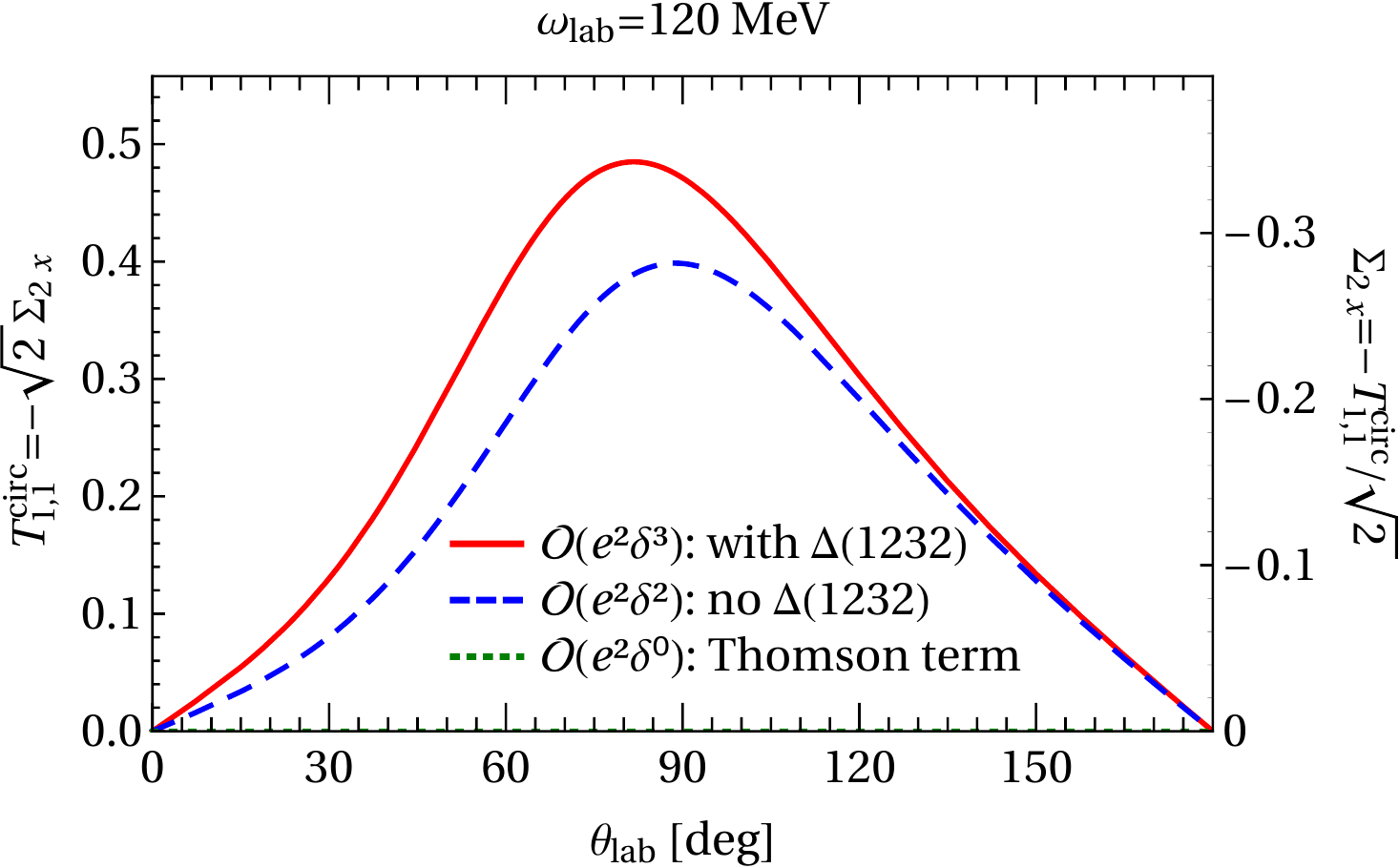}
    \hq
    \includegraphics[height=0.3\linewidth]
    {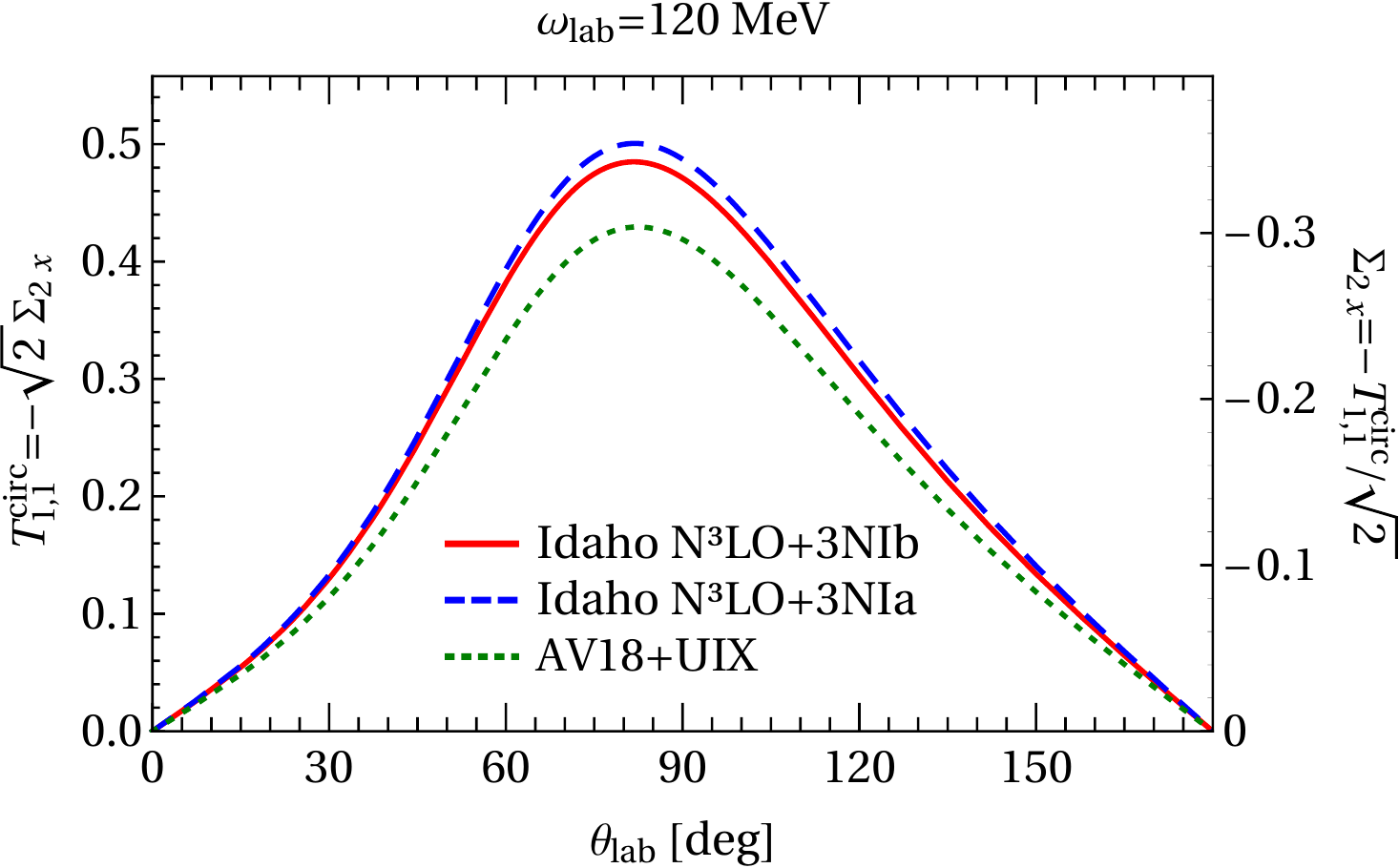}

    \caption{(Colour on-line) The double asymmetry
      $T_{11}^\mathrm{circ}=-\sqrt{2}\;\Sigma_{2x}$ in the lab frame. Top: Energies between
      $\omegalab=50\;\MeV$ and $120\;\MeV$ in $10\;\MeV$ steps, for the asymmetry (left) and rate difference (right).
Bottom left: order-by-order convergence at $\omegalab=120\;\MeV$, with
notation as in fig.~\ref{fig:crosssect-convergence}; the LO result is
identically zero.
Bottom right: wave function dependence at $\omegalab=120\;\MeV$, with notation
as in fig.~\ref{fig:crosssect-wfdep}. }
\label{fig:T11circ-many}
\end{center}
\end{figure}

Figure~\ref{fig:T11circ-many} shows results for the double asymmetry
$T_{11}^\mathrm{circ}=-\sqrt{2}\;\Sigma_{2x}$ of eq.~\eqref{eq:T1Xcirc}, which
is formed by considering scattering of a circularly polarised beam on a target
polarised in the $+x$ vs.~$-x$ direction.  Like the beam asymmetry, it is zero
at $\theta=0^\circ,180^\circ$; unlike $\Sigma^\text{lin}$, this asymmetry vanishes
as $\omegalab \rightarrow 0$, and is consequently quite small (maximum value
of $0.1$) at $\omegalab=50\;\MeV$. It grows faster than linearly with
$\omegalab$, with a maximum value at $120\;\MeV$ of $0.5$.  The corresponding
rate difference $\Delta_{2x}$ increases more slowly, by about $50\%$ between
$50\;\MeV$ and $120\;\MeV$, because the cross section decreases with
energy. Nevertheless, the figure-of-merit for the asymmetry measurement is
larger at higher energies---and so, of course, is the sensitivity of the
asymmetry to spin polarisabilities.  Even though the asymmetry is zero at LO
[$\calO(e^2\delta^0)$, single-nucleon Thomson term], the \ChiEFT result
converges well (see third panel of fig.~\ref{fig:T11circ-many}).  The
correction from including the Delta at $\calO(e^2\delta^3)$ is $25\%$ even at
$\omegalab=120\;\MeV$, and less than $5\%$ at $\omegalab=60\;\MeV$.  The wave
function dependence---displayed at $120\;\MeV$ in the fourth panel of
fig.~\ref{fig:T11circ-many}---is about $\pm10\%$ and essentially
energy-independent.  We observe that---as for the differential cross section
and $\Sigma^\mathrm{lin}=\Sigma_3$---all wave functions predict the same
shape.

\begin{figure}[!htbp]
  \begin{center}
    \includegraphics[width=0.94\linewidth]
{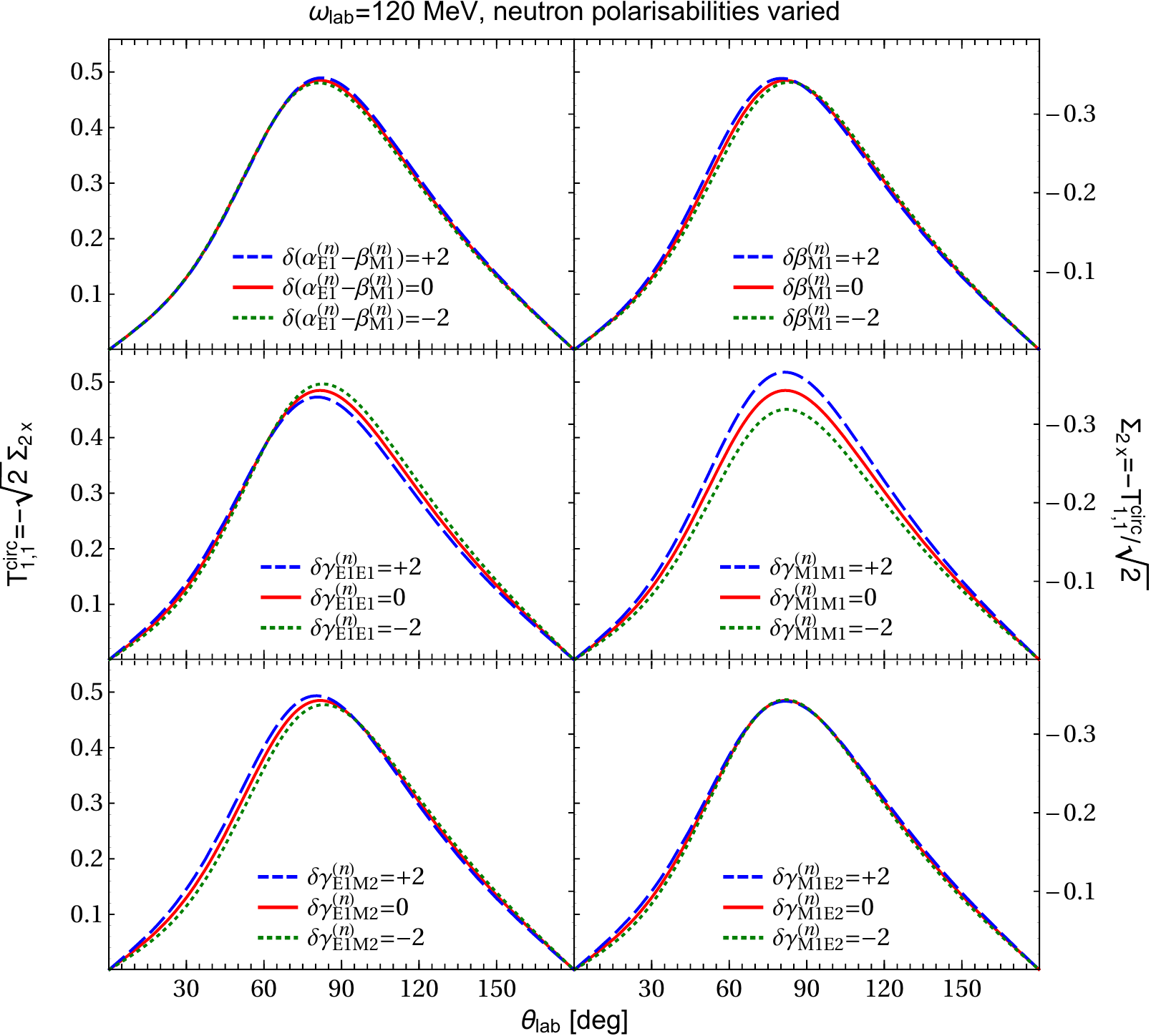}

\caption{(Colour on-line) Sensitivity of
  $T_{11}^\mathrm{circ}=-\sqrt{2}\;\Sigma_{2x}$ to variations of the
   neutron polarisabilities $\alphaen-\betamn$ (top left),
  $\alphaen$ (top right), $\gammaeen$ (centre left), $\gammammn$
  (centre right), $\gammaemn$ (bottom left), and $\gammamen$ (bottom
  right), about their central values by $\pm2$ units, at
  $\omegalab=120\;\MeV$. Notation as in fig.~\ref{fig:beamasym-scalarpols}.}
    \label{fig:T11circ-spinpols}
  \end{center}
\end{figure}

Figure~\ref{fig:T11circ-spinpols} shows the effect of varying the
polarisabilities.  The upper two panels of fig.~\ref{fig:T11circ-spinpols}
show that, even at $\omegalab=120\;\MeV$, there is little sensitivity to the
scalar polarisabilities, especially if the Baldin sum rule constraint on
$\alphaen + \betamn$ is imposed.

The largest sensitivity is to the spin polarisability $\gammammn$. Changing it
by $\pm2$ units affects $T_{11}^\mathrm{circ}$ by about $\pm8\%$ of its peak
value at $\thetalab\approx90^\circ$ and $\omegalab=120\;\MeV$. Changes in the
other spin polarisabilities produce noticeably smaller effects. Varying
$\gammaemn$ alters $T_{11}^\mathrm{circ}$ more at $\thetalab<90^\circ$;
varying $\gammaeen$ affects it at $\thetalab>90^\circ$; the sensitivity to
$\gammamen$ vanishes at $90^\circ$.  Therefore, a measurement around
$\thetalab\approx90^\circ$ provides a good opportunity to extract $\gammammn$
with little contamination from the other polarisabilities.

For this observable, the multipole basis for the spin polarisabilities
exhibits a predominant sensitivity to a single spin polarisability. Figure 15
of Shukla \etal~\cite{Shukla:2008zc} shows noticeable effects when two
of the spin polarisabilities in the ``Ragusa'' basis, namely
$\gamma_4=\gammamm$ and $\gamma_1=-\gammaee-\gammaem$, were varied by $\pm 100\%$. 
The substantial shifts on varying $\gamma_1$ largely reflected the
fact that it is the largest to start with at ${\cal O}(e^2\delta^2)$. In
contrast, our work considers variations of $\pm 2$ units.  Allowing
for this, the results of ref.~\cite{Shukla:2008zc} and the current work are in
fact consistent and serve to show that the multipole basis is more convenient
for this observable. In that spirit, we also briefly comment---although we do
not show explicit results---that when the forward- and backward-combinations
$\gamma_{0/\pi}$ of the spin-polarisabilities are used as constraints, the
signal is still strong, as is the one for the combination
$\gammammn-\gammamen$ which also appears in the alternative multipole basis
introduced in ref.~\cite{Griesshammer:2017txw}.

Finally, we see that an explicit Delta increases both the magnitude of the
asymmetry (fig.~\ref{fig:T11circ-many}) and its sensitivity to the
polarisabilities, compared to the $\calO(e^2 \delta^2)$-result.

\clearpage

\subsection{Circularly Polarised Beam on Longitudinally Polarised Target}
\label{sec:Tcirc10}

\begin{figure}[!htbp]
  \begin{center}
    \includegraphics[height=0.3\linewidth]
{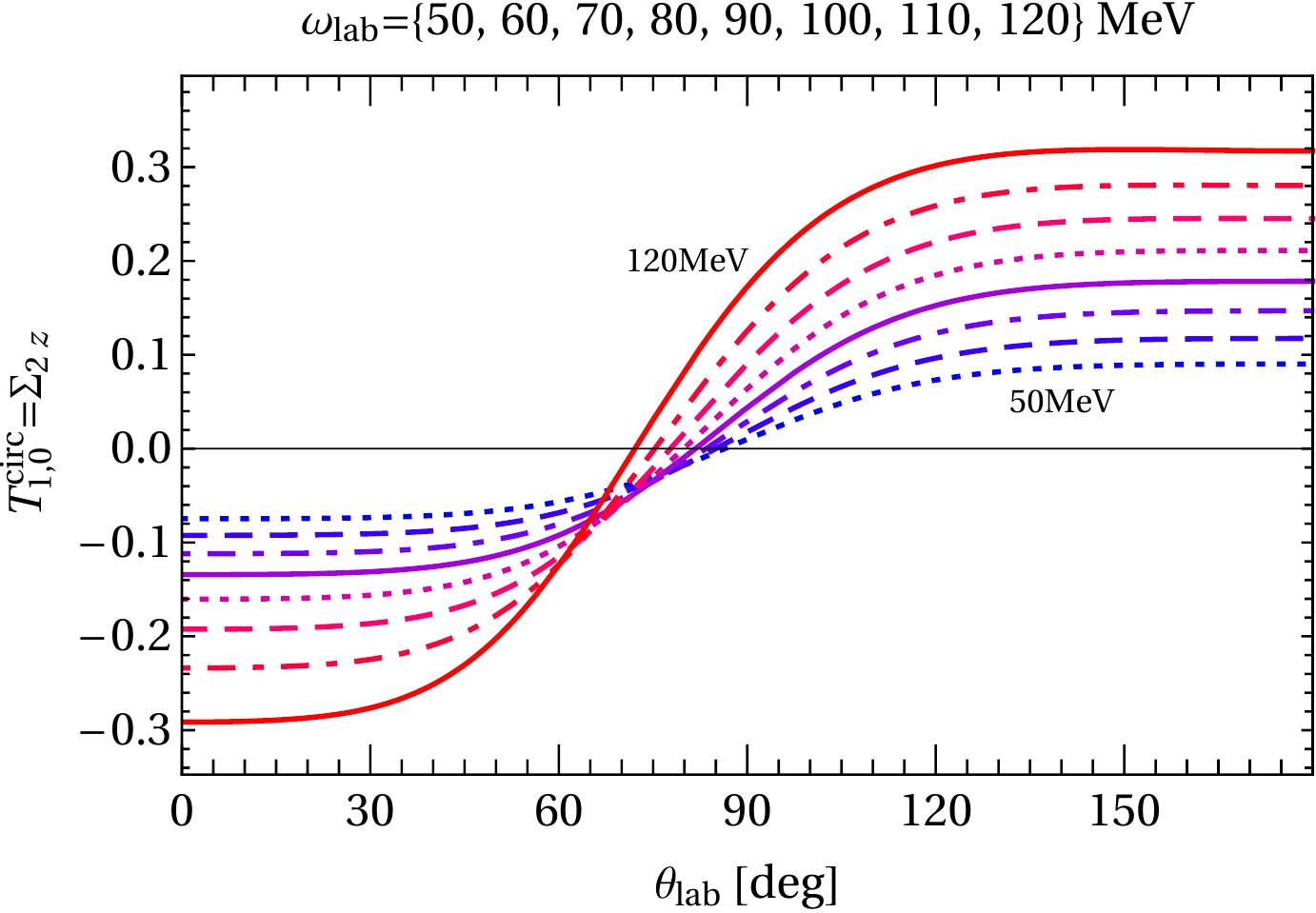}
    \includegraphics[height=0.3\linewidth]
{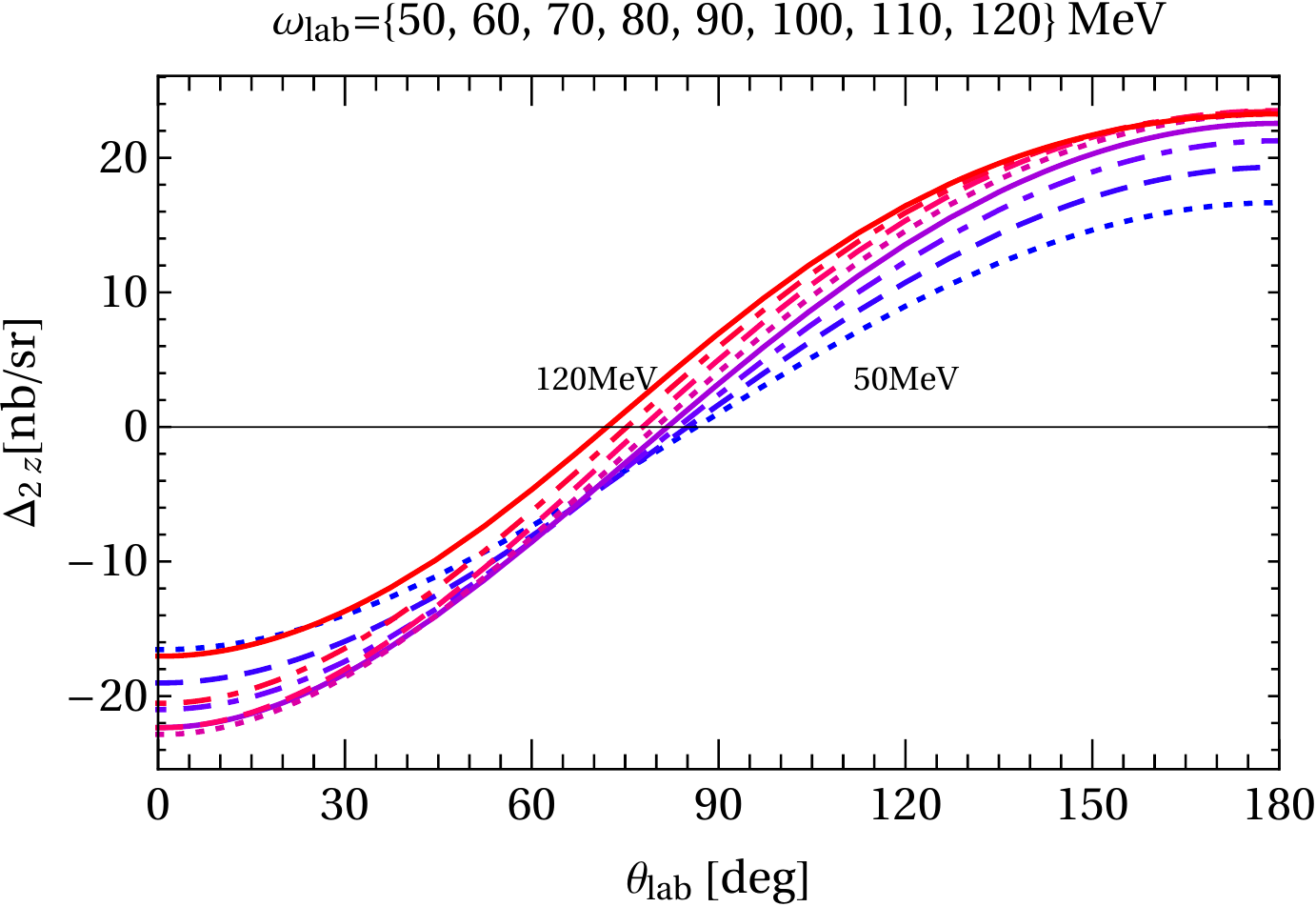}
\\[2ex]
    \includegraphics[height=0.3\linewidth]
    {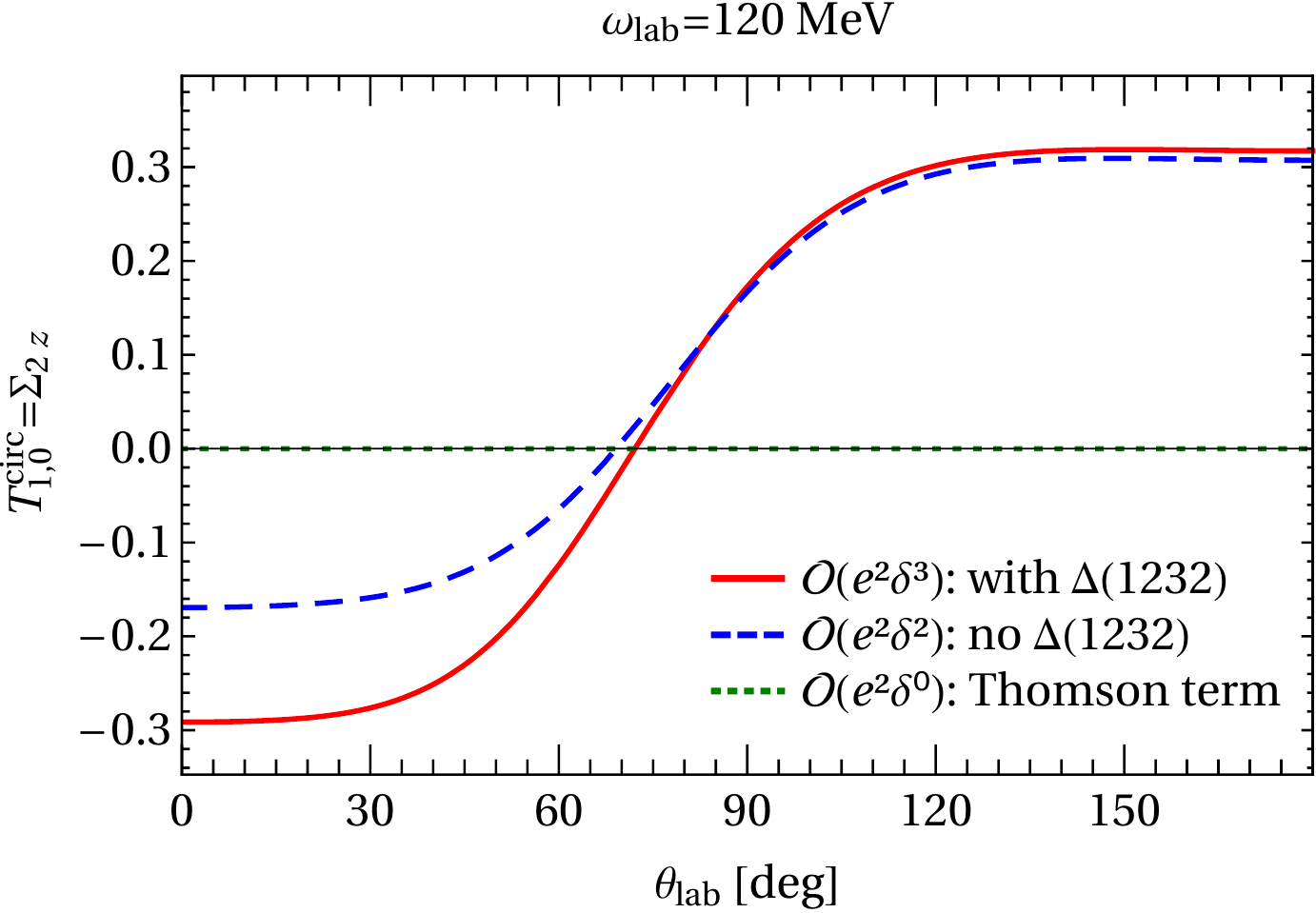}
    \includegraphics[height=0.3\linewidth]
    {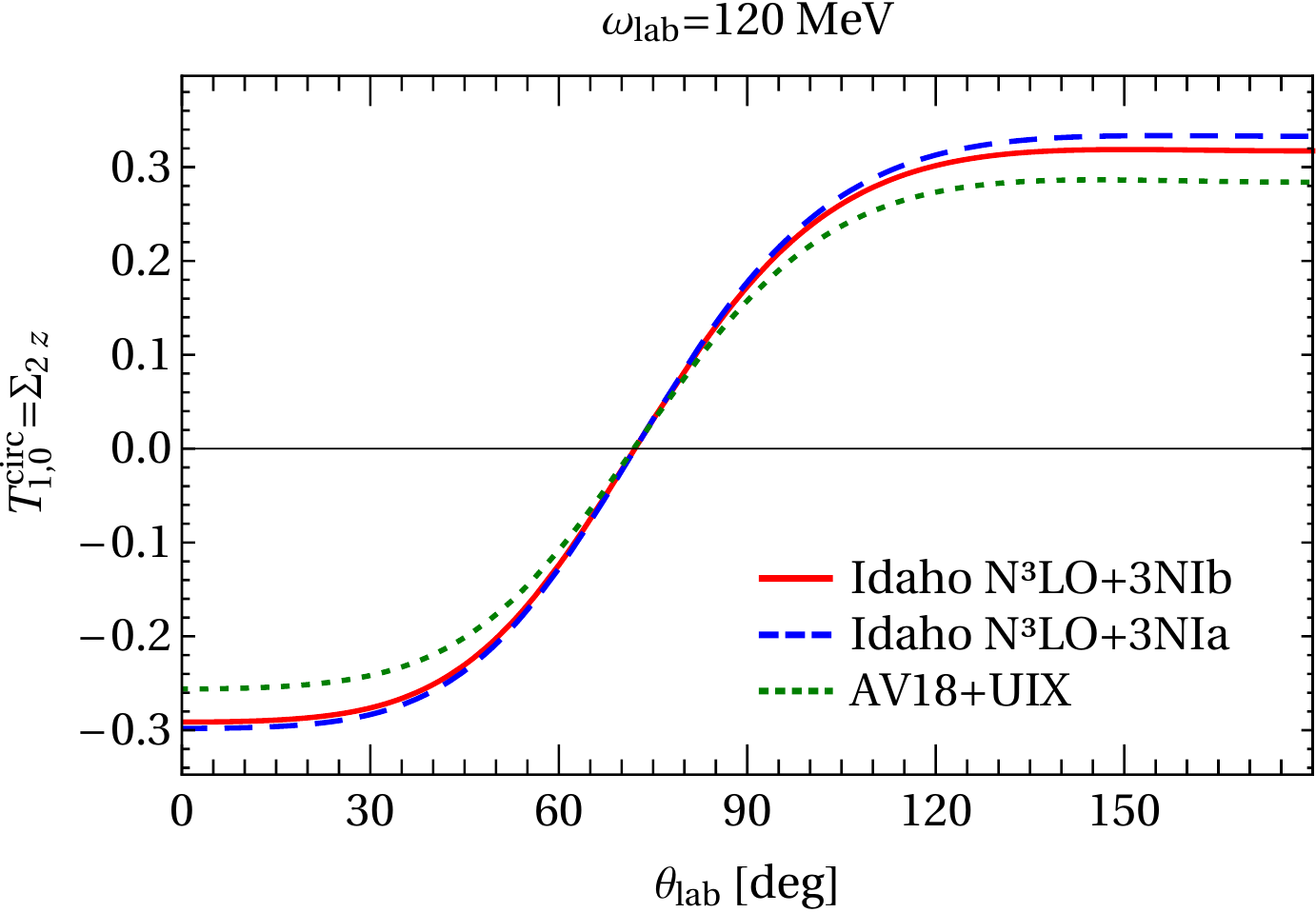} 

    \caption{(Colour on-line) The double asymmetry
      $T_{10}^\mathrm{circ}=\Sigma_{2z}$ in the lab frame. Top: Energies between
      $\omegalab=50\;\MeV$ and $120\;\MeV$ in $10\;\MeV$ steps, for the
      asymmetry (left) and rate difference (right).  Bottom left: order-by-order
      convergence at $\omegalab=120\;\MeV$, with notation as in
      fig.~\ref{fig:crosssect-convergence}; the LO result is identically zero.
      Bottom right: wave function dependence at $\omegalab= 120\;\MeV$, with
      notation as in fig.~\ref{fig:crosssect-wfdep}.}
\label{fig:T10circ-many}
\end{center}
\end{figure}

Finally, we turn to the double asymmetry $T_{10}^\mathrm{circ}=\Sigma_{2z}$ of
a circularly polarised beam on a target which is polarised parallel
vs.~antiparallel to the beam direction; see eq.~\eqref{eq:T1Xcirc}. Our
predictions are shown in fig.~\ref{fig:T10circ-many}. The asymmetry is again
zero at LO [$\calO(e^2\delta^0)$, single-nucleon Thomson term], and is about
the same size as $T_{11}^\mathrm{circ}$. It is also zero for $\omega=0$, but
nonzero at $\theta=0^\circ,180^\circ$. The upper two panels show that it increases
by a factor of about $4$ between $\omegalab=50\;\MeV$ and $120\;\MeV$. However,
the concomitant decrease of the cross section with energy now produces a rate
difference that is essentially constant with photon energy.  At
$\omegalab=50\;\MeV$ the change from $\calO(e^2 \delta^2)$ to
$\calO(e^2 \delta^3)$ is at most 5\%; order-by-order results at
$\omegalab=120\;\MeV$ are displayed in the third panel of
fig.~\ref{fig:T10circ-many}. The \ChiEFT result converges well, showing a
significant change from the $\calO(e^2\delta^2)$ result only at high energies
and forward angles. The wave function dependence is small at
all energies and angles; the fourth panel of fig.~\ref{fig:T10circ-many} is
quite representative.

\begin{figure}[!t]
  \begin{center}
    \includegraphics[width=0.85\linewidth]
{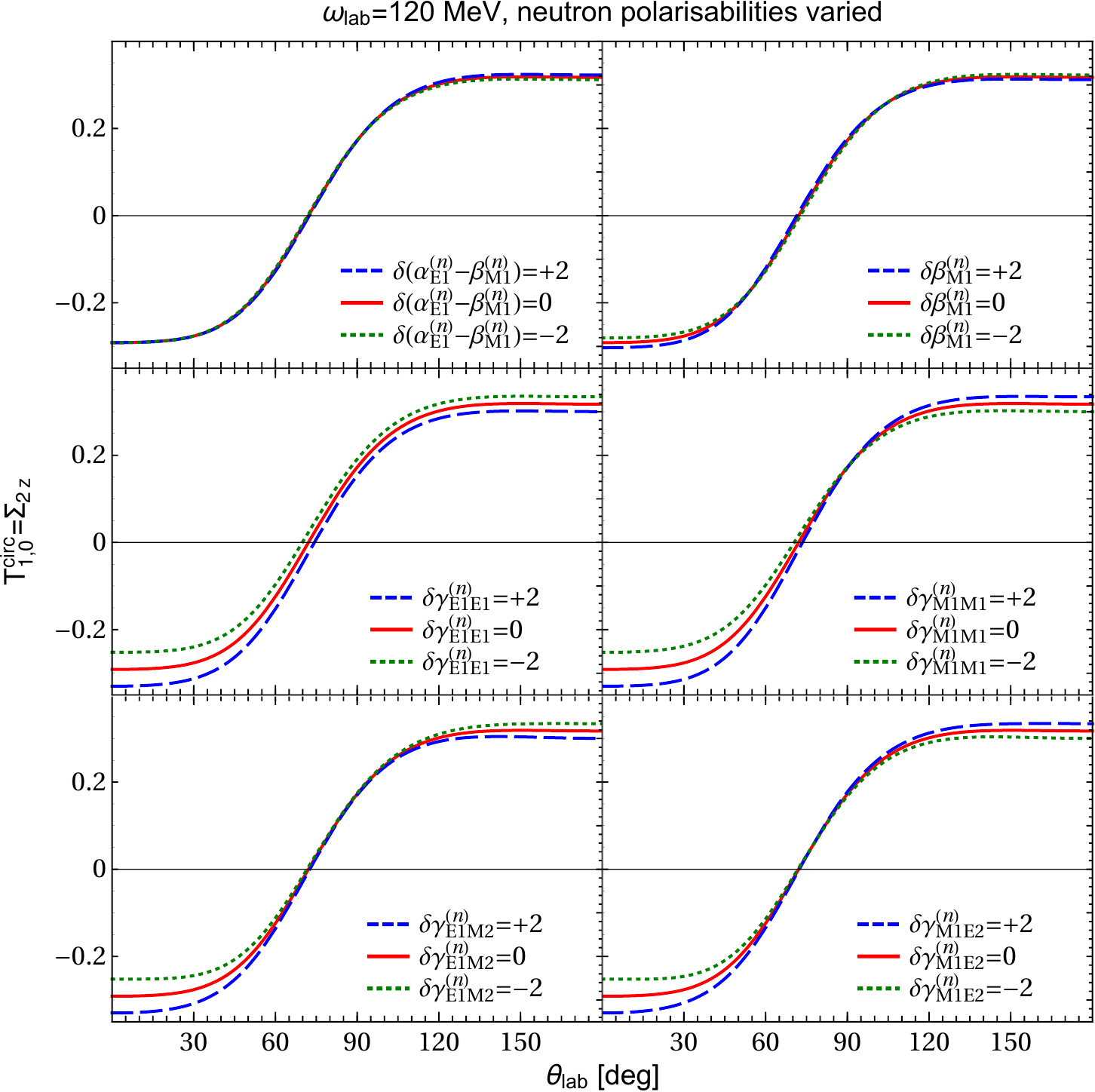}

\caption{(Colour on-line) Sensitivity of $T_{10}^\mathrm{circ}=\Sigma_{2z}$ to
  varying the scalar and spin polarisabilities of the neutron about their
  central values by $\pm2$ units, at $\omegalab=120\;\MeV$. Notation and panels as in
  fig.~\ref{fig:T11circ-spinpols}.}
    \label{fig:T10circ-spinpols}
  \end{center}
\end{figure}

Figure~\ref{fig:T10circ-spinpols} shows that variation of the scalar
polarisabilities around the baseline values of eq.~\eqref{eq:LundPRL} has even
less impact on $T_{10}^\mathrm{circ}$ than on $T_{11}^\mathrm{circ}$.
%
%
To maximise sensitivity to the spin polarisabilities, we again show
results at $\omegalab=120\;\MeV$. This time there is no one
multipole-basis spin polarisability that shows a unique signal: $\gammaeen$
and $\gammaemn$ are essentially degenerate with one another, as are
$\gammammn$ and $\gammamen$. We briefly comment without showing figures that
the results in the alternative multipole basis of the spin polarisabilities
proposed in ref.~\cite{Griesshammer:2017txw} indeed show
cleaner signals for the combinations $\gammaeen-\gammaemn$ and
$\gammammn-\gammamen$.   $\gammaeen$ could possibly be
extracted by measuring the point where $T_{10}^\mathrm{circ}$ is zero, as the effect of other spin
polarisabilities is markedly smaller there. For $\omegalab=120\;\MeV$, the
zero-crossing at $\thetalab\approx84^\circ$ varies by $\pm2.5^\circ$ for
$\delta\gammaeen=\pm2$.

\section{Summary, Observations and Outlook}
\label{sec:conclusion}

We presented \ChiEFT predictions with a dynamical Delta degree of freedom at
\NXLO{3} [$\calO(e^2 \delta^3)$] for Compton scattering from \threeHe for
photon lab energies between $50$ and $120\;\MeV$. We showed results for the
differential cross section, for the beam asymmetry
$\Sigma^\mathrm{lin}=\Sigma_3$, and for the two double asymmetries with
circularly polarised photons and transversely or longitudinally polarised
targets, $T_{11}^\mathrm{circ}=-\sqrt{2}\,\Sigma_{2x}$ and
$T_{10}^\mathrm{circ}=\Sigma_{2z}$. These are the only observables that are
non-zero below pion-production threshold in our formulation.  We also
corrected previous results in refs.~\cite{Choudhury:2007bh,Shukla:2008zc,
  ShuklaPhD} for these observables at \NXLO{2} [$\calO(e^2 \delta^2=Q^3)$,
without dynamical Delta] (see concurrent erratum~\cite{Shukla:2017}). As
expected, the dynamical effects of the Delta do not enter at low energies, but
they are visible in all observables at the upper end of the energy range. In
particular, they markedly reduce the forward-backward asymmetry of the cross
section and increase the magnitude of the double asymmetries and their
sensitivity to spin polarisabilities, echoing similar findings for the
deuteron~\cite{Hildebrandt:2005ix, Hildebrandt:2005iw, Griesshammer:2012we,
  Griesshammer:2013vga, erratum2}. 

We showed that the convergence of the chiral expansion in this energy range is
quite good. The dependence of the results on the choice of \threeHe wave
function is not large, either, and can usually be distinguished from the
effects of neutron polarisabilities by its different angular dependence. We
found that $\alphaen - \betamn$ could be extracted from the cross section, and
$T_{11}^\mathrm{circ}=-\sqrt{2}\;\Sigma_{2x}$ has a non-degenerate sensitivity
to $\gammammn$ around $90^\circ$. $T_{10}^\mathrm{circ}=\Sigma_{2z}$ is
sensitive to $\gammaeen$ and $\gammaemn$. $\Sigma^\mathrm{lin}=\Sigma_3$ is a
useful cross-check on the accuracy of the \ChiEFT amplitudes, but is dominated
by the single-nucleon Thomson term to surprisingly high energies, and so
measurements of this may not be especially useful for polarisability
determinations.

Ultimately, the most accurate values of polarisabilities will be inferred from
a data set that includes all four observables. For the spin polarisabilities,
the upper end of our energy range will be crucial, as the sensitivities at
$\omegalab\lesssim100\;\MeV$ are so small that discriminating between
different spin-polarisability values would be very challenging.

As in ref.~\cite{Griesshammer:2017txw}, we argue that our results can be
considered quite robust, \ie~that varying the single-nucleon amplitudes of
complementary theoretical approaches like dispersion relations will lead to
sensitivities which are hardly discernible from ours. We did not quantify
residual theoretical uncertainties in detail, as this presentation is meant to
provide an exploratory study of magnitudes and sensitivities of observables to
the nucleon polarisabilities. Once data are available, a polarisability
extraction will of course need to address residual theoretical uncertainties
with more diligence, as was already done for the proton and deuteron in
refs.~\cite{Griesshammer:2012we, McGovern:2012ew, Myers:2014ace,
  Myers:2015aba, Griesshammer:2015ahu}. 

Our exploration of different observables is therefore sufficient to assess
experimental feasibilities, and we see it as part of an ongoing dialogue on
the best kinematics and observables for future experiments to obtain
information on neutron polarisabilities~\cite{Weller:2013zta, Annand:2015,
  Ahmed:2016, HIGSPAC}.  To facilitate that dialogue, our results are
available as an interactive \emph{Mathematica} notebook from
\texttt{hgrie@gwu.edu}, where cross sections, rates and asymmetries are
explored when the scalar and spin polarisabilities are varied, including
variations constrained by sum rules.

\begin{figure}[!htbp] \thisfloatpagestyle{empty}
  \begin{center}
\pbox{\linewidth}{
  \hq\includegraphics[height=0.28\linewidth]
{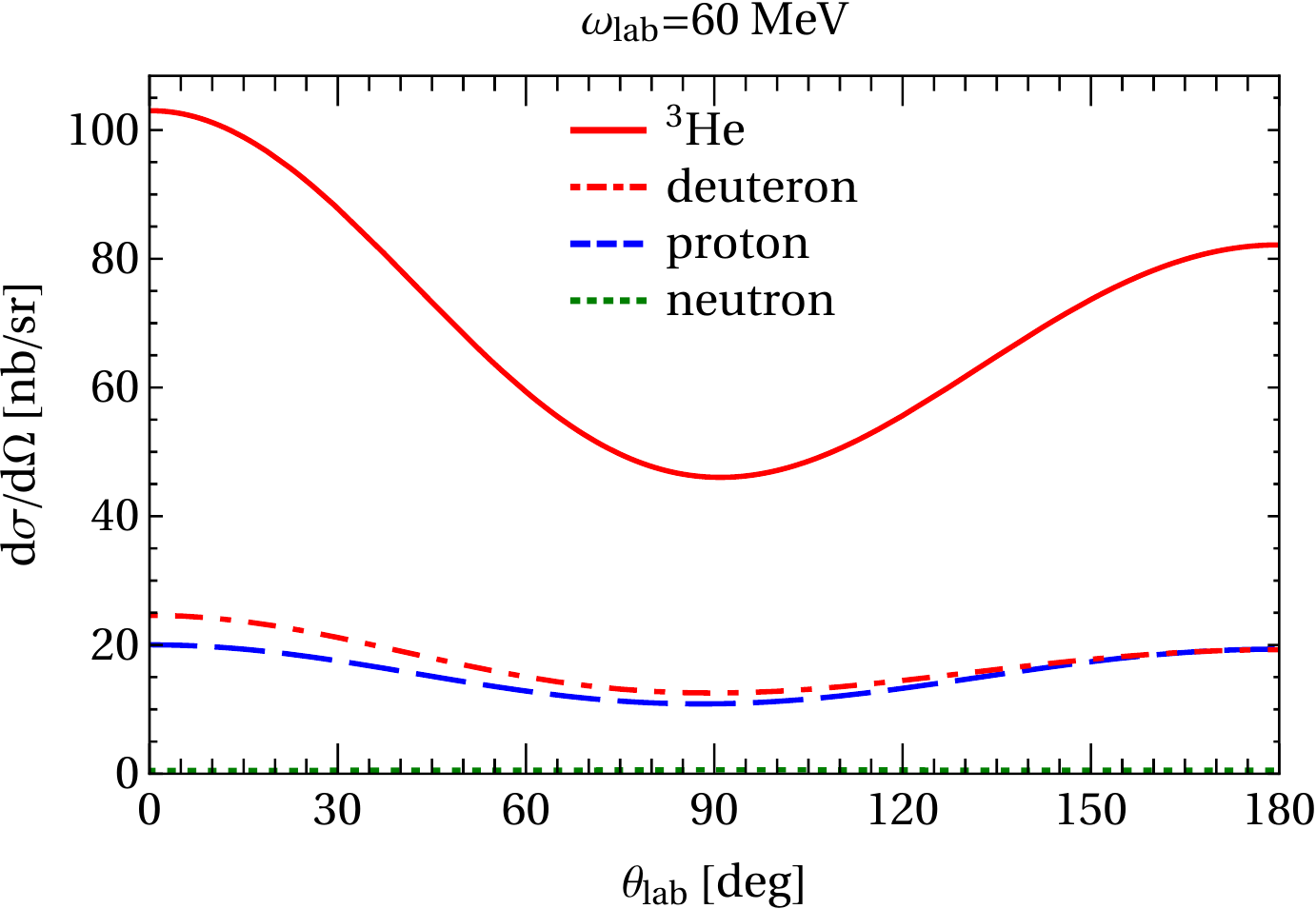}\hspace*{9.4ex}
    \includegraphics[height=0.28\linewidth]
{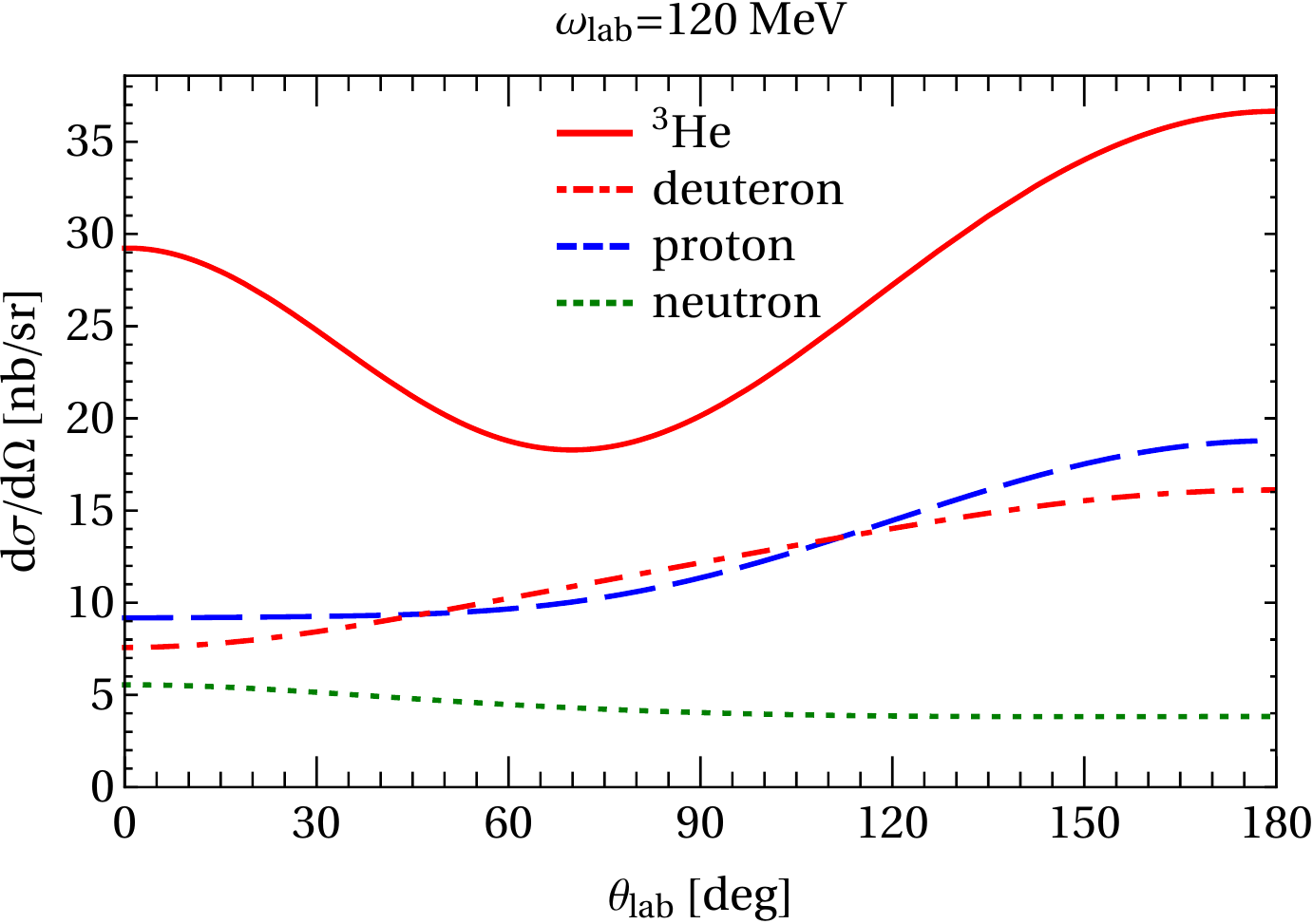}\\[2ex] 
\includegraphics[height=0.28\linewidth]
{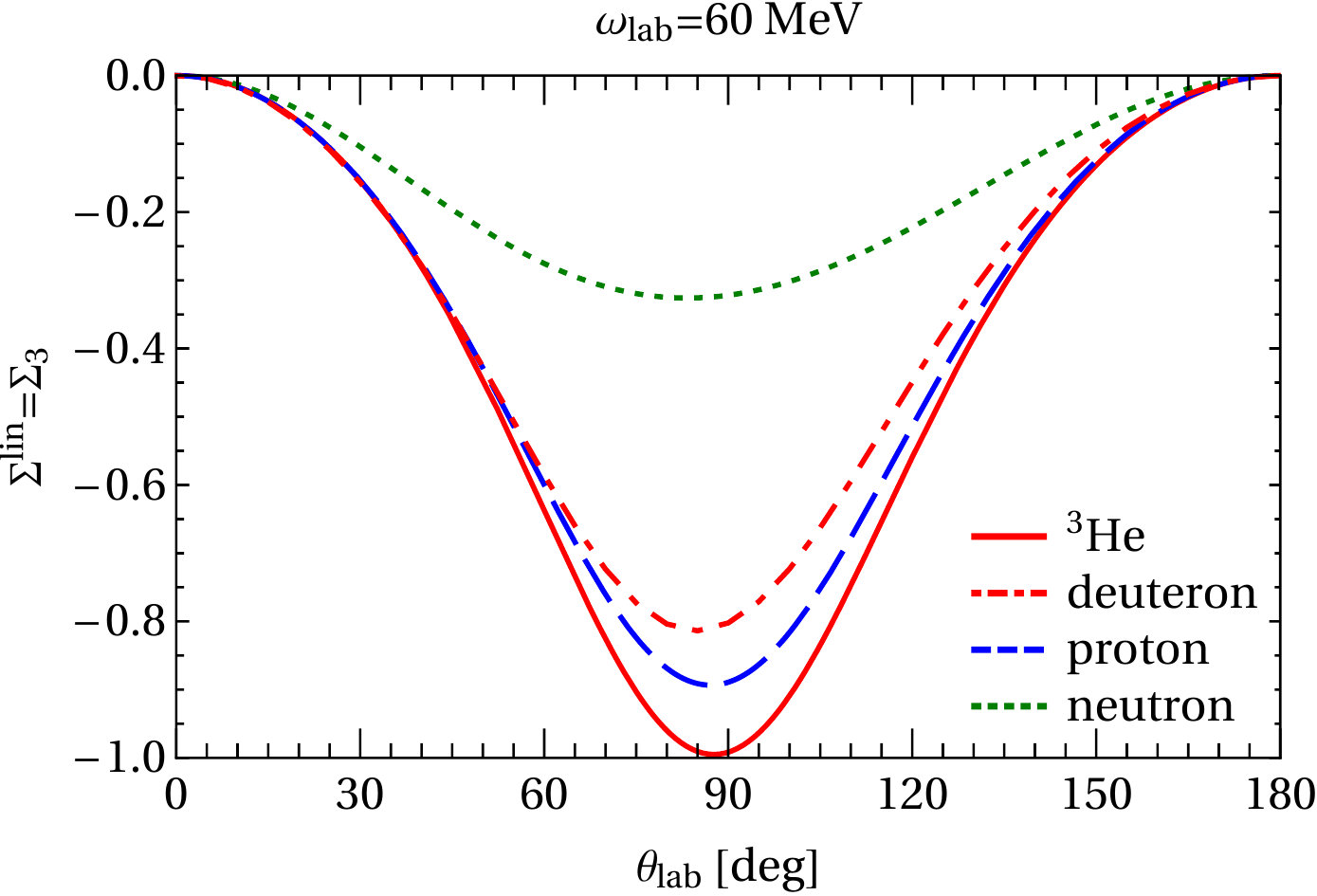}\hspace*{8.7ex}
\includegraphics[height=0.28\linewidth]
{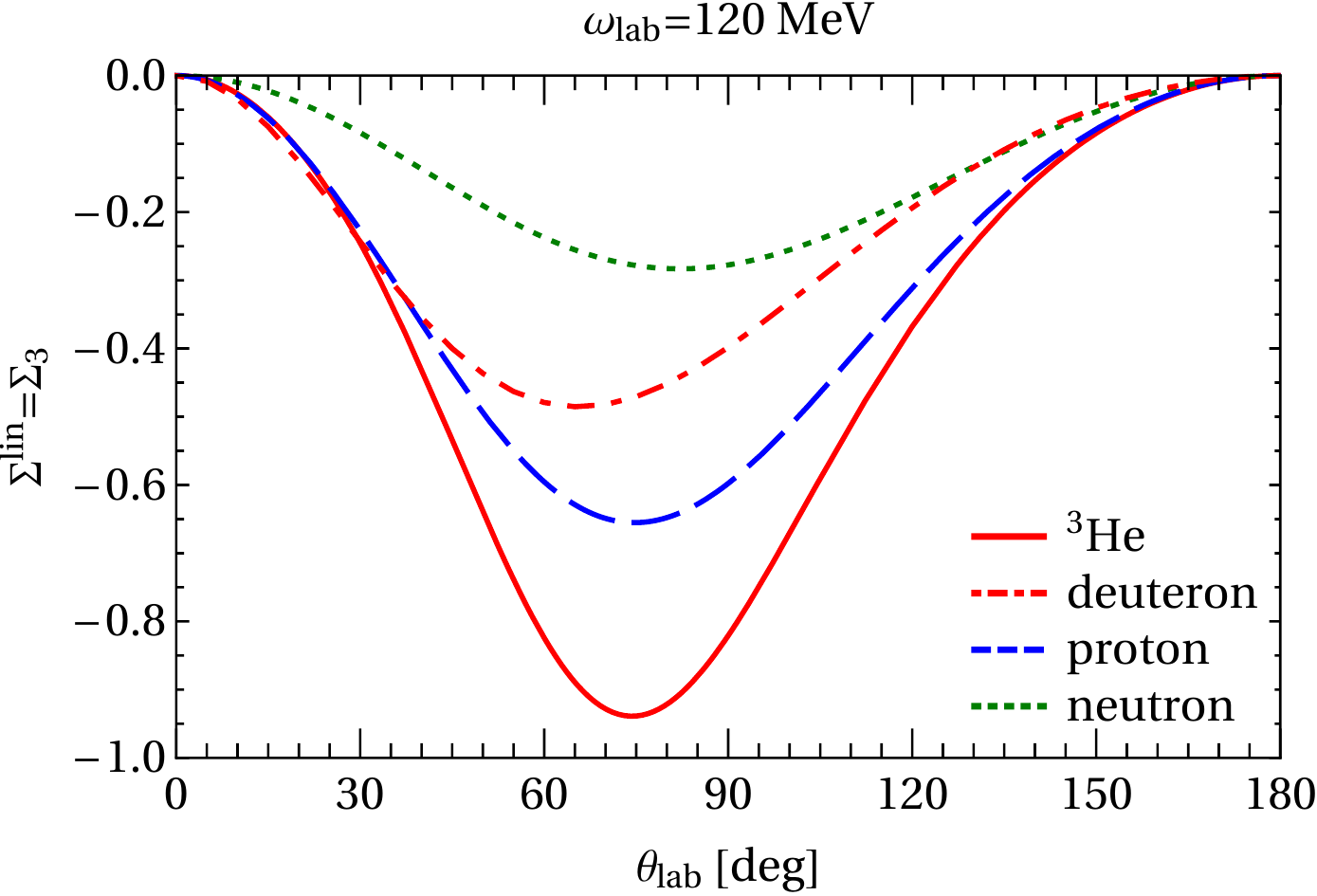}\\[2ex] 
  \includegraphics[height=0.28\linewidth]
{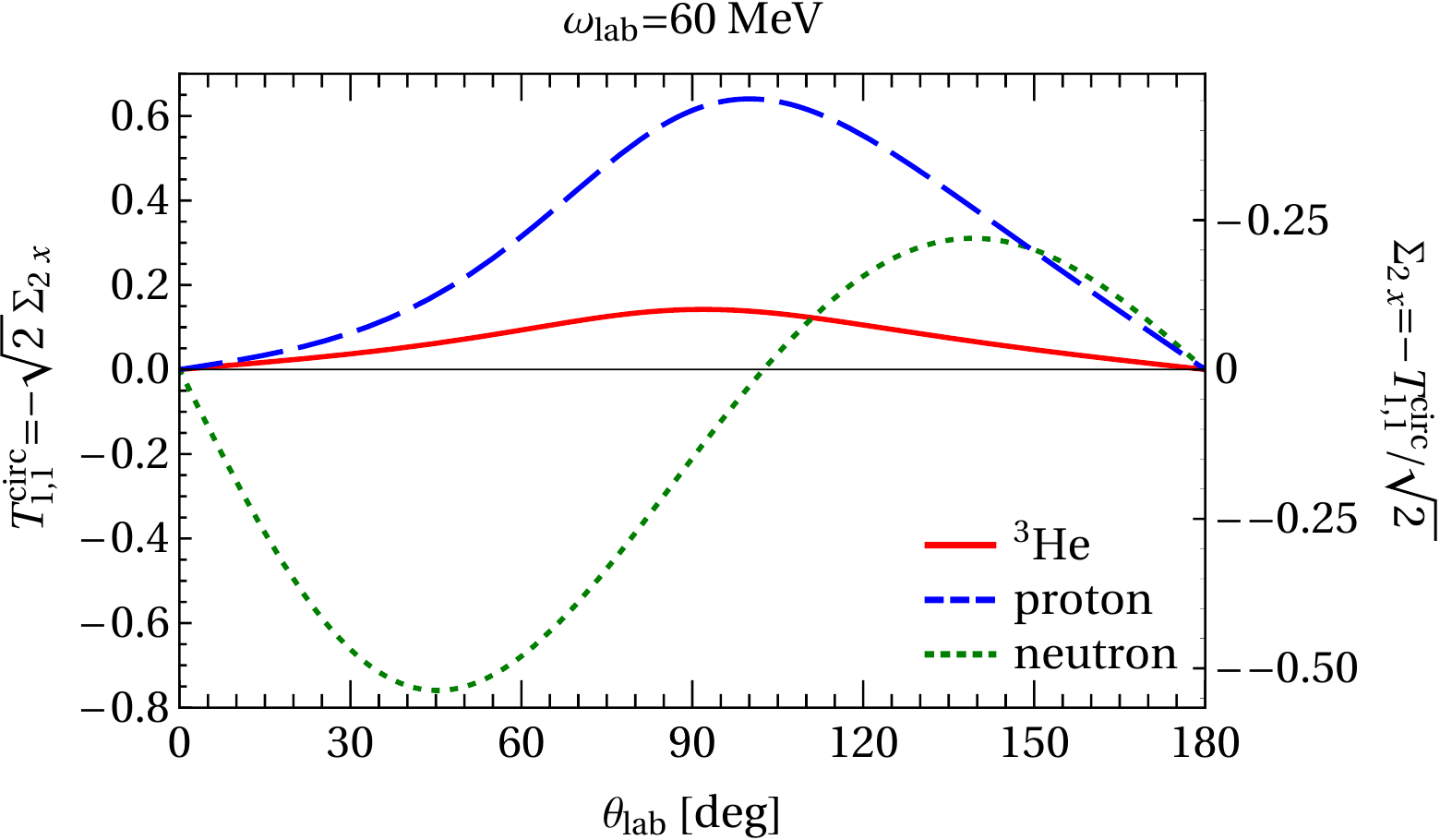}\hqq\hq
    \includegraphics[height=0.28\linewidth]
{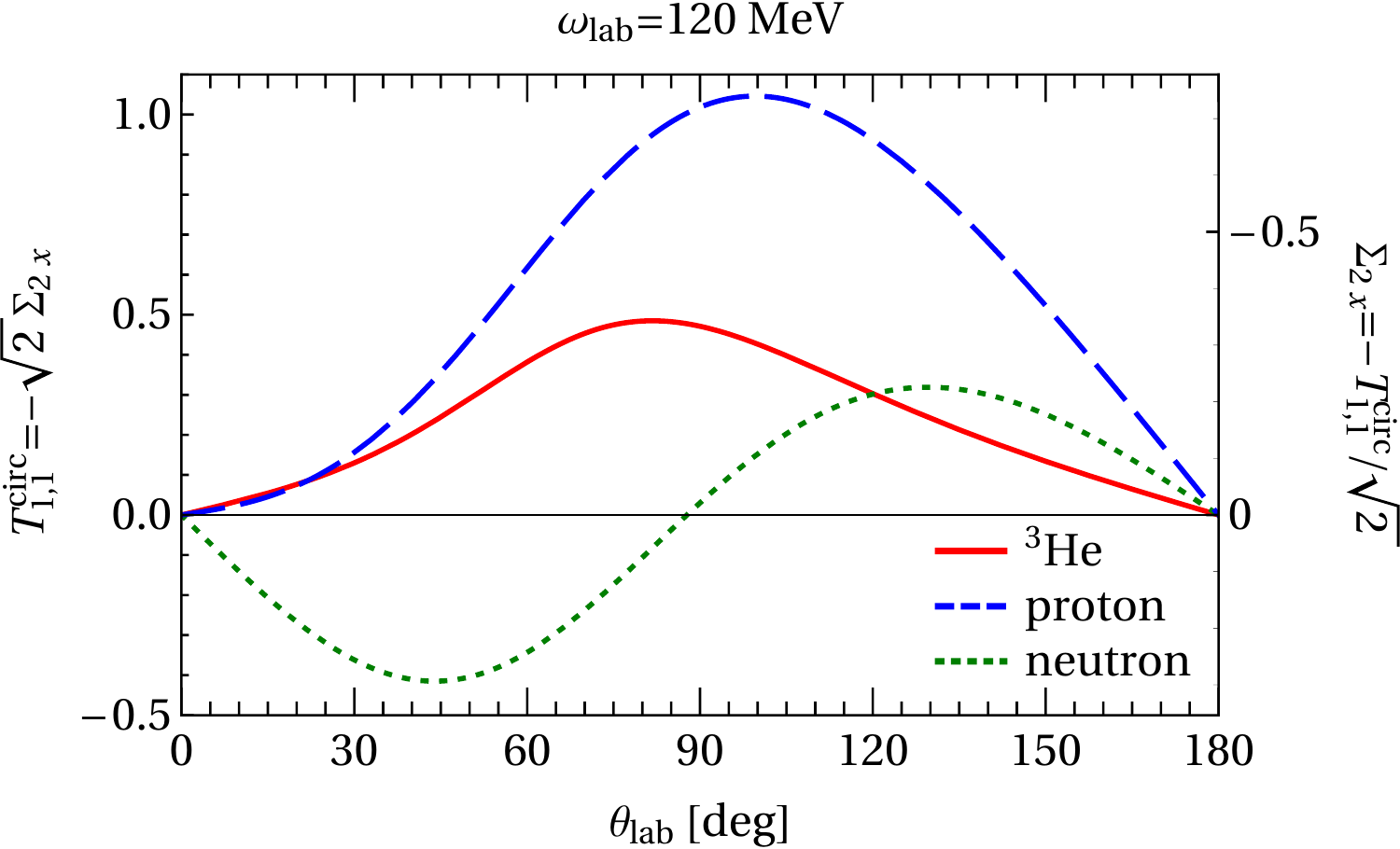}\\[2ex] 
\hqq \includegraphics[height=0.28\linewidth]
{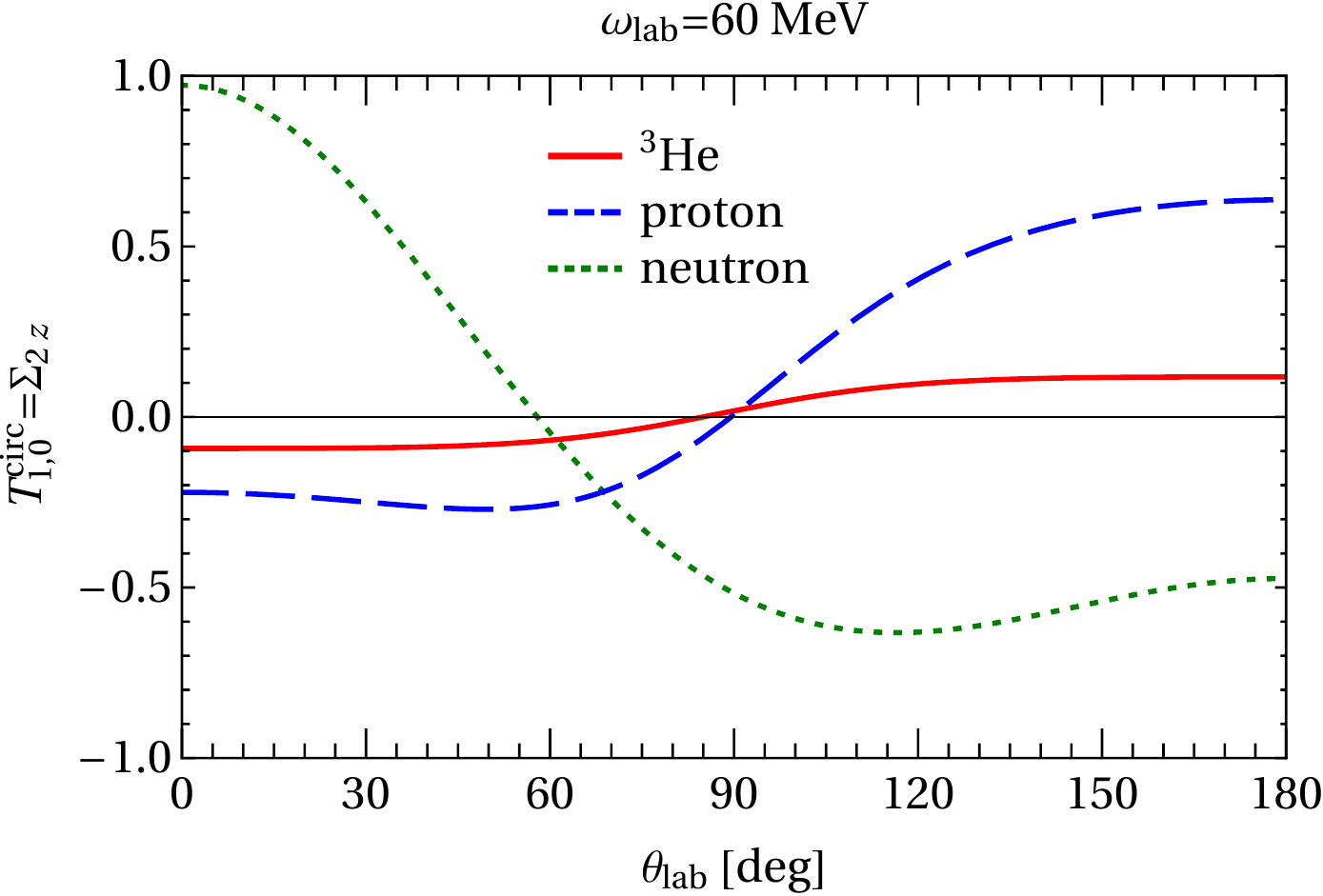}\hspace*{8.4ex}
    \includegraphics[height=0.28\linewidth]
{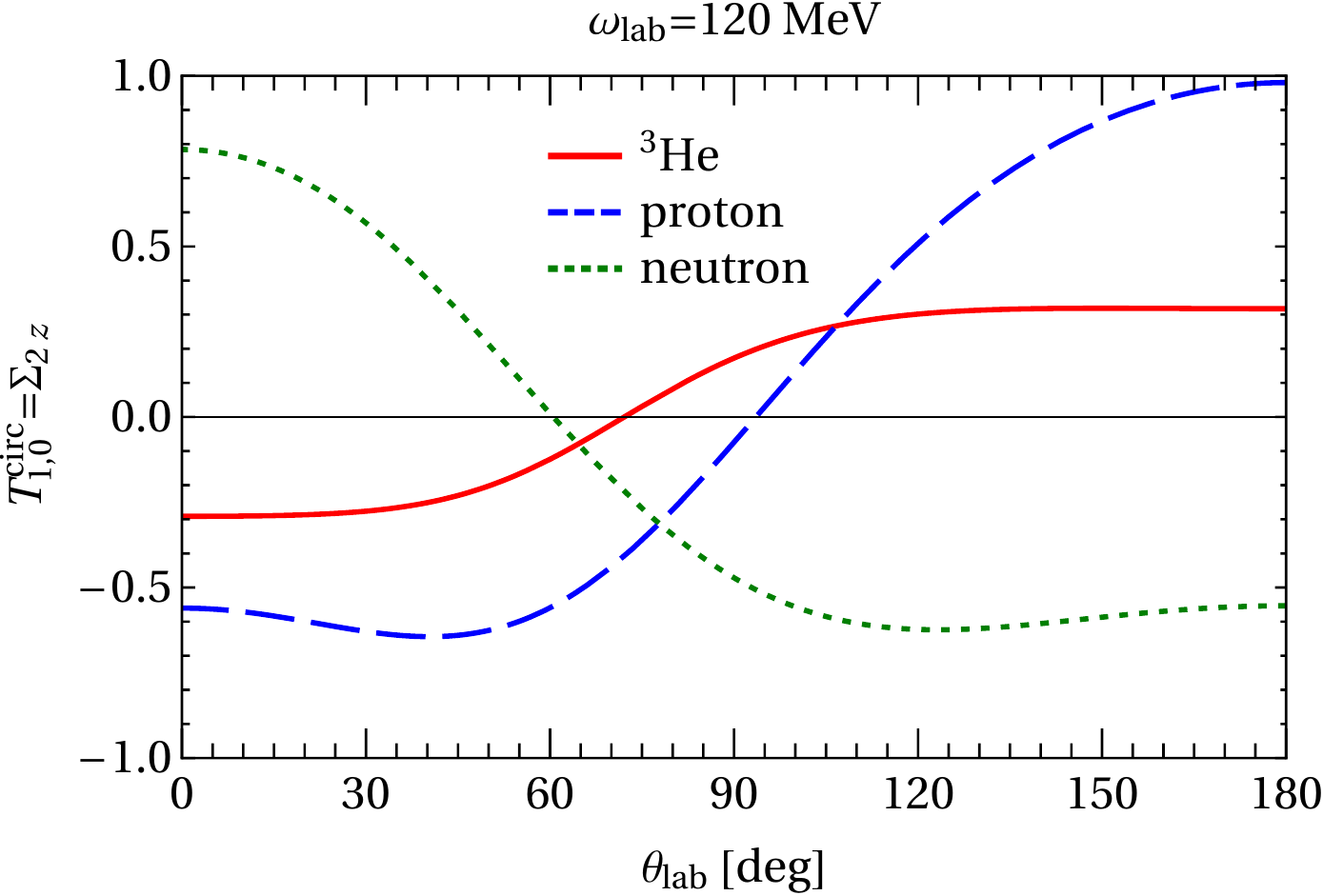}
}

\caption{(Colour on-line) Observables at $\calO(e^2\delta^3)$ (\ie with
  Delta) for $\omegalab=60\;\MeV$ (left column) and $120\;\MeV$
  (right column) on $4$ targets: \threeHe (red
  solid, this work); the deuteron (red dot-dashed, described in
  refs.~\cite{Hildebrandt:2005ix, Hildebrandt:2005iw, Griesshammer:2012we}); the proton (blue
  dashed) and neutron (green dotted) (both from
  ref.~\cite{Griesshammer:2017txw}).
  Top two rows: cross section (top) and beam asymmetry
  $\Sigma^\text{lin}=\Sigma_3$ (second). 
  Bottom two rows: The double asymmetries
  $T^\text{circ}_{11}=-\sqrt{2}\;\Sigma_{2x}$ (top) and
  $T^\text{circ}_{10}=\Sigma_{2z}$ (bottom) on the three spin-$\half$ targets:
  proton, neutron and \threeHe. Notice the difference of the scales in rows 1 and
  3.}
    \label{fig:targets}
  \end{center}
\end{figure}

To put \threeHe Compton scattering in context, we compare predictions of
observables with those for the proton, neutron and deuteron in
fig.~\ref{fig:targets}.  Clearly, measurements on each target are sensitive to
different combinations of nucleon polarisabilities: the proton and neutron
only to their respective polarisabilities; the deuteron only to the isoscalar
components $\alphaep+\alphaen$ etc.; and \threeHe roughly to
$2\alphaep+\alphaen$, $2\betamp+\betamn$ and to the spin polarisabilities of
the neutron but not of the proton  (see discussion below). Each target thus offers
complementary linear combinations. Here we focus on comparisons of the
magnitudes of observables, important for planning experiments.

For the cross section at $\omegalab=60\;\MeV$, the ratio between the \threeHe
and the proton varies between $4$ and $5$; but at $120\;\MeV$, it drops to $3$
at forward and $1.5$ at backward angles. The \threeHe-to-deuteron ratio is $4$
at $60\;\MeV$ and the angular dependence is very similar there. At
$120\;\MeV$, however, it drops from nearly $5$ at forward angles to about $2$
at backward angles. Obviously, \threeHe Compton scattering will
produce higher rates than either $\gamma$-deuteron or $\gamma$-proton elastic
scattering and probe a new linear combination of proton and neutron scalar
polarisabilities. In addition though, the
sensitivity to neutron scalar polarisabilities is also larger (in absolute
terms) for \threeHe than for the deuteron. Qualitatively, this is because the neutron
polarisabilities interfere with the Thomson term of two protons in \threeHe,
while they only interfere with that of one proton in the
deuteron. Quantitative comparison shows that the sensitivity enhancement over the deuteron case is around
2.2--3 for $\omegalab$ between $60$ and $120\;\MeV$, with a mild energy
and angle dependence.

The beam asymmetry of all targets (except of course the neutron) is dominated by the Thomson limit of
scattering on a charged point-particle at low energies. Figure~\ref{fig:targets} shows that this effect is
more prominent for \threeHe and survives to higher energies than for the other
targets. This means that there is no significant sensitivity of $\Sigma_3$ to
variations of the polarisabilities.

Figure~\ref{fig:targets} also shows double asymmetries (without the
deuteron, which is not a spin-half target). There is some qualitative resemblance between the results for the
proton and for \threeHe, but the \threeHe results are much smaller. The
neutron results are driven by the structure parts of the amplitudes. 
Due to the absence of pieces involving the overall particle charge, the
double asymmetries of the neutron have a very different angular behaviour than those
of the \threeHe or proton: for $\T_{11}^\text{circ}=-\sqrt{2}\;\Sigma_{2x}$
there is a clear valley-and-mountain shape instead of the single ``hump'';
$\T_{10}^\text{circ}=\Sigma_{2z}$ even has an inverted angular dependence.

The results presented above included the sensitivities of observables to
neutron polarisabilities.  We have also computed their sensitivity to proton
polarisabilities, and found that---to a good approximation---the
relevant combinations of proton and neutron scalar polarisabilities are
$2 \alphaep + \alphaen$ and $2 \betamp + \betamn$, just as expected from the
na\"ive picture of Compton scattering from $\threeHe$.  \ChiEFT 
corrections to this result never amount to more than $\pm5\%$.
Similarly, the \ChiEFT calculations also demonstrate that the double-asymmetries are $10-20$
times more sensitive to the spin polarisabilities of the neutron than of the
proton, with the precise ratio again depending on energy and angle. However,
the results of Fig.~\ref{fig:targets} emphasise that there is no energy where
polarised $\threeHe$ really does act as a ``free neutron-spin target''. The
\emph{sensitivities} of $T_{10}^{\rm circ}$ and $T_{11}^{\rm circ}$ to neutron
spin polarisabilities closely mimic those of the free-neutron observables. But
their \emph{magnitudes} do not.

In conclusion, \ChiEFT allows one to quantify the important angle- and
energy-dependent corrective to the na\"ive picture of \threeHe as having a
response that is the sum of that for two protons with antiparallel spins and
one neutron. In particular, we stress that the cross section predicted by the
impulse approximation for $\gamma\,\threeHe$ scattering is far too small; see
fig.~\ref{fig:crosssect-convergence}. The impulse approximation omits the
charged pion-exchange currents that are a key mechanism for elastic
$\gamma\,\threeHe$ scattering at the energies considered here.  Neglecting
pion-exchange currents distorts extractions of nucleon polarisabilities.

\ChiEFT provides both a power-counting argument that these effects are at
least as large as the sought-after nucleon-structure effects, and a
quantitative prediction for the two-body currents, with a reliable assessment
of theoretical uncertainties. Indeed, detailed work to check the
convergence of the expansion for exchange currents and the other pieces of the
\threeHe-Compton amplitude by performing a \NXLO{4} [$\calO(e^2 \delta^4)$]
calculation and extending the applicable energy range is essential if these
theory studies are to move from exploratory into the realm of high-accuracy
extractions of polarisabilities from data. Work along these lines, using the
same \ChiEFT framework for the nucleon, deuteron and \threeHe, is in
progress~\cite{future3He}.


\section*{Acknowledgements}

Andreas Nogga's assistance in providing \threeHe wave functions, and answering
questions about their implementation was invaluable. Ch.~Hanhart directed us
to the Lebedev-Laikov method for solid-angle integrations which significantly
sped up the code.
We gratefully acknowledge discussions with J.~R.~M.~Annand, M.~W.~Ahmed,
E.~J.~Downie and M.~Sikora. We are particularly grateful to the organisers and
participants of the workshop \textsc{Lattice Nuclei, Nuclear Physics and QCD -
  Bridging the Gap} at the ECT* (Trento), and of the US DOE-supported
\textsc{Workshop on Next Generation Laser-Compton Gamma-Ray Source}, and for
hospitality at KITP (Santa Barbara; supported in part by the US National
Science Foundation under Grant No.~NSF PHY-1125915) and KPH (Mainz). HWG is
indebted to the kind hospitality and financial support of the Institut f\"ur
Theoretische Physik (T39) of the Physik-Department at TU M\"unchen and of the
Physics Department of the University of Manchester. AM is grateful to his
thesis advisor, R.~P.~Springer, for her advice and support during the
completion of this project.
This work was supported in part by UK Science and Technology Facilities
Council grants ST/L0057071/1 and ST/L0050727/1 (BS), as well as ST/L005794/1
and ST/P004423/1 (JMcG), by the US Department of Energy under contracts
DE-FG02-05ER41368, DE-FG02-06ER41422 and DE-SC0016581 (all AM), DE-SC0015393
(HWG) and DE-FG02-93ER-40756 (DRP), by the Scottish Universities Physics
Alliance Prize Studentship (BS), and by the Dean's Research Chair programme of
the Columbian College of Arts and Sciences of The George Washington University
(HWG).




\end{document}